\documentclass[structabstract,usenatbib]{aa}  
\usepackage{longtable,lscape}
\usepackage[figuresright]{rotating}
\usepackage{graphicx}
\usepackage[varg]{txfonts}
\usepackage{natbib}
\usepackage{amssymb}

\newcommand{\hii}{\relax \ifmmode {\mbox H\,{\scshape II}}\else H\,{\scshape II}\fi}
\newcommand{\mi}{\relax \ifmmode {\mu{\mbox m}}\else $\mu$m\fi}
\newcommand{\ha}{\relax \ifmmode {\mbox H}\alpha\else H$\alpha$\fi}
\newcommand{\hb}{\relax \ifmmode {\mbox H}\beta\else H$\beta$\fi}
\newcommand{\hg}{\relax \ifmmode {\mbox H}\beta\else H$\gamma$\fi}
\newcommand{\hd}{\relax \ifmmode {\mbox H}\beta\else H$\delta$\fi}

\newcommand{\sii}{\relax \ifmmode {\mbox S\,{\scshape II}}\else S\,{\scshape II}\fi}
\newcommand{\siii}{\relax \ifmmode {\mbox S\,{\scshape III}}\else S\,{\scshape III}\fi}
\newcommand{\nii}{\relax \ifmmode {\mbox N\,{\scshape II}}\else N\,{\scshape II}\fi}
\newcommand{\oii}{\relax \ifmmode {\mbox O\,{\scshape II}}\else O\,{\scshape II}\fi}
\newcommand{\oiii}{\relax \ifmmode {\mbox O\,{\scshape III}}\else O\,{\scshape III}\fi}
\newcommand{\neiii}{\relax \ifmmode {\mbox Ne\,{\scshape III}}\else Ne\,{\scshape III}\fi}
 
 \newcommand{\rdostres}{\relax \ifmmode {\,\mbox{R}}_{\rm 23}\else \,\mbox{R}$_{\rm 23}$\fi} 
\newcommand{\metal}{\mbox{12$+$log(O/H)}}
\newcommand{\NDOS}{\mbox{\it N2}}
\newcommand{\RDOSTRES}{\mbox{\it R23}}

\begin{document}

   \title{Extreme emission-line galaxies out to z$\sim$\,1 in zCOSMOS}
   \subtitle{I. Sample and characterization of global properties}

\author{
R. Amor\'in\inst{1,2}
\and
E.~P\'erez-Montero\inst{2,3,4}
\and
T.~Contini\inst{3,4}
\and
J.M. V\'ilchez\inst{2}
\and
M.~Bolzonella\inst{10}
\and
L.~A.~M.~Tasca\inst{6}
\and
F.~Lamareille\inst{3,4}
\and 
G.~Zamorani\inst{10}
\and
C.~Maier\inst{5,16}
\and
C.~M.~Carollo\inst{5}
\and
J.-P.~Kneib\inst{6}
\and
O.~Le~F\`{e}vre\inst{6}
\and
S.~Lilly\inst{5}
\and
V.~Mainieri\inst{7}
\and
A.~Renzini\inst{8}
\and
M.~Scodeggio\inst{9}
\and
S.~Bardelli\inst{10}
\and 
A.~Bongiorno\inst{11}
\and
K.~Caputi\inst{21}
\and
O.~Cucciati\inst{13}
\and
S.~de~la~Torre\inst{12}
\and
L.~de~Ravel\inst{12}
\and
P.~Franzetti\inst{9}
\and
B.~Garilli\inst{9,6}
\and
A.~Iovino\inst{14}
\and
P.~Kampczyk\inst{5}
\and 
C.~Knobel\inst{5}
\and
K.~Kova\v{c}\inst{5,15}
\and
J.-F.~Le~Borgne\inst{3,4}
\and
V.~Le~Brun\inst{6}
 \and
M.~Mignoli\inst{10}
\and 
R.~Pell\`o\inst{3,4}
\and
Y.~Peng\inst{5}
\and
V.~Presotto \inst{17,14}
\and
E.~Ricciardelli\inst{18}
\and
J.~D.~Silverman\inst{19}
\and
M.~Tanaka\inst{19}
\and
L.~Tresse\inst{6}
\and
D.~Vergani\inst{10,20}
\and
E.~Zucca\inst{10}
}

 \offprints{R. Amor\'in \email{ricardo.amorin@oa-roma.inaf.it}}

  \institute{
{INAF -- Osservatorio Astronomico di Roma, via Frascati 33, 00040  Monteporzio Catone, Roma, Italy}  
\and
{Instituto de Astrofis\'ica de Andaluc\'ia, CSIC, 18008 Granada, Spain}  
\and
{IRAP, Universit{\'e} de Toulouse, UPS-OMP, Toulouse, France} 
\and
{Institut de Recherche en Astrophysique et Plan{\'e}tologie, CNRS, 14, avenue Edouard Belin, F-31400 Toulouse, France} 
\and
{Institute of Astronomy, ETH Zurich, CH-8093, Z\"urich, Switzerland}   
\and
{Laboratoire d'Astrophysique de Marseille, CNRS-Universit{\'e} d'Aix-Marseille, 38 rue Frederic Joliot Curie, F-13388 Marseille, France}  
\and
{European Southern Observatory, Karl-Schwarzschild-Strasse 2, Garching b. Muenchen, D-85748, Germany}  
\and
{Dipartimento di Astronomia, Universit\'a di Padova, vicolo Osservatorio 3, I-35122 Padova, Italy}  
\and
{INAF - IASF Milano, Via Bassini 15, I-20133, Milano, Italy}   
\and
{INAF - Osservatorio Astronomico di Bologna, via Ranzani 1, I-40127 Bologna, Italy}  
\and 
{Max-Planck-Institut f\"ur extraterrestrische Physik, D-84571 Garching b. Muenchen, D-85748, Germany}  
\and 
{SUPA Institute for Astronomy, The University of Edinburgh, Royal Observatory, Edinburgh, EH9 3HJ}  
\and
{INAF - Osservatorio Astronomico di Trieste, Via Tiepolo, 11, I-34143 Trieste, Italy}  
\and
{INAF - Osservatorio Astronomico di Brera, Via Brera, 28, I-20159 Milano, Italy}  
\and
{MPA - Max Planck Institut f\"ur Astrophysik, Karl-Schwarzschild-Str. 1,  85741 Garching, Germany}  
\and 
{University of Vienna, Department of Astronomy, Tuerkenschanzstrasse 17, 1180 Vienna, Austria} 
\and
{Universit\'a degli Studi dell'Insubria, Via Valleggio 11, 22100 Como, Italy} 
\and 
{Instituto de Astrof\'isica de Canarias, V\'ia Lactea s/n, E-38200 La Laguna, Tenerife, Spain} 
\and
{IPMU, Institute for the Physics and Mathematics of the Universe, 5-1-5 Kashiwanoha, Kashiwa, 277-8583, Japan}  
\and
{INAF-IASFBO, Via P. Gobetti 101, I-40129, Bologna, Italy}  
\and
{Kapteyn Astronomical Institute, University of Groningen, 9700 AV Groningen,
The Netherlands} 
  }

   \date{}  

 \abstract
{
The {study} of large and representative samples of low-metallicity star-forming galaxies  
at different cosmic epochs is of great interest to the detailed understanding of the assembly history and evolution of low-mass galaxies.
}
{
We present {a thorough characterization of} a large sample of 183 extreme emission-line galaxies (EELGs) at redshift $0.11 \leq z \leq 0.93$ selected from the 20k zCOSMOS bright survey because of their unusually large emission line equivalent widths. 
}
{
{We use multiwavelength COSMOS photometry, HST-ACS {\it I}-band imaging, 
and optical zCOSMOS spectroscopy to derive the main global properties of star-forming 
EELGs, such as sizes, stellar masses, star formation rates (SFR), and reliable oxygen abundances 
using both ``direct'' and ``strong-line'' methods. }
} 
{
The EELGs are extremely compact ($r_{50} \sim 1.3$\,kpc), low-mass
(M$_*$\,$\sim$\,10$^{7}$-10$^{10}$\,M$_{\odot}$) galaxies 
forming stars at unusually high {specific star formation} rates (sSFR\,$\equiv$\,SFR/M$_{\star}$ up to 10$^{-7}$\,Gyr$^{-1}$) 
compared to main sequence star-forming galaxies of the same stellar 
mass {and redshift}. 
{At rest-frame UV wavelengths}, the EELGs are luminous and {show high surface brightness} 
and include strong Ly$\alpha$ emitters, as revealed by GALEX spectroscopy. 
We show that zCOSMOS EELGs are {high-ionization}, low-metallicity 
systems, {with median} 12$+\log$(O/H)$=$8.16$\pm$0.21 
(0.2\,$Z_{\odot}$) {including a handful of extremely metal-deficient 
($<$\,0.1\,$Z_{\odot}$) EELGs}. 
{While} $\sim$80\% of the EELGs show non-axisymmetric morphologies, 
including clumpy and cometary or tadpole galaxies, we find that {$\sim$29\%} of {them} show additional low-surface-brightness 
features, {which} strongly {suggests} recent or ongoing interactions. 
As star-forming dwarfs in the local Universe, EELGs are most often 
found in relative isolation. 
While only very few EELGs belong to compact groups, almost one third of 
them are found in spectroscopically confirmed {loose} pairs or triplets.   
}
{
The zCOSMOS EELGs are galaxies caught in a
  transient and probably early period of their evolution, where they are 
  efficiently building up a significant fraction of their
  present-day stellar mass in {an ongoing}, galaxy-wide starburst. 
Therefore, the EELGs constitute an ideal benchmark for comparison
studies between low- and high-redshift low-mass 
star-forming galaxies.
} 


\titlerunning {Extreme emission-line galaxies in zCOSMOS}        
\authorrunning{R. Amor\'in et al.}
  \maketitle
 
%

\section{Introduction}

Low-mass galaxies undergoing vigorous bursts of star formation over
galaxy-wide scales provide unique laboratories for understanding 
{galaxy} mass assembly and chemical evolution over cosmic times. 
{In the local Universe, these systems are often referred to 
as H{\sc II} galaxies \citep{Terlevich1991} and Blue Compact Dwarfs 
\citep[BCDs;][]{Thuan1981}, depending on the observational technique or
the selection criteria \citep[see][for a review]{KunthOstlin2000}. 
In spectroscopic surveys, they are generally recognized by their high-excitation emission lines with unusually large equivalent widths 
(EW)\footnote{We use the convention of positive equivalent widths for 
emission lines.}, as a product of the photoionization 
of gas by hot massive stars in a young burst of star formation 
\citep{Sargent1970}.} 

Over the last decade, the advent of all-sky optical and UV surveys
such as the {\it Sloan Digital Sky Survey}
\citep[SDSS;][]{Abazajian2003} and the {\it Galaxy Evolution Explorer} 
\citep[GALEX;][]{Martin2005}, along with other smaller surveys, have
allowed us to systematically search and characterize relatively large
samples of extreme emission-line galaxies (EELGs) out to the
frontiers of the local Universe \citep[$z \la 0.3$,
e.g.,][]{Kniazev2004,Kakazu2007,Overzier2008,Salzer2009,Cardamone2009,
Cowie2010,Izotov2011,Shim2013}. 
{This has made} it possible to discover  an
increasing number of extremely compact, low-metallicity galaxies with
unusually high {specific} star formation rates (SFR, 
sSFR$=$SFR/M$_{\star} \sim$1-100\,Gyr$^{-1}$),  
 such as the {\it \emph{green peas}}  \citep{Cardamone2009,Amorin2010}
and a handful of {extremely metal-poor galaxies 
\citep[XMPs; $Z$\,$\la$\,0.1\,$Z_{\odot}$][]{KunthOstlin2000}} at 
$0.1$\,$\la$$z$$\la$\,0.4 
\citep[e.g.,][]{Kakazu2007,Hu2009,Cowie2010}.  

Similarly to nearby H{\sc II} galaxies and some BCDs, EELGs 
{are probably} the youngest and chemically least evolved 
population of low$-z$ star-forming galaxies 
\citep[SFGs, e.g.,][]{Searle1972,Rosa2007,Jaskot2013}. 
These properties make them unique probes with which to study the details of
chemical enrichment, massive star formation, and feedback processes in
galaxies with physical properties (i.e., size, mass, SFR,
metallicity, gas, and dust relative content) most closely resembling
those prevailing at high redshift, e.g., Lyman-break galaxies and
Lyman-$\alpha$ emitters \citep[e.g.,][]{Pettini2001,Finkelstein2011}. 
{Furthermore, increasing observational evidence point to EELGs 
as} the likely environments to host the progenitors of long-duration gamma-ray bursts  
\citep[e.g.,][]{Christensen2004,Kewley2007,Savaglio2009,Guseva2011} 
and the most luminous supernovae \citep{Chen2013,Lunnan2013,Leloudas2014,Toene2014}. 

In order to  understand comprehensively  the properties of
EELGs as a class, to select best case studies for detailed analysis, and
{to} provide a valuable benchmark for comparative studies at higher
redshifts, large and representative samples of EELGs must be assembled. 
{Although} EELGs are generally rare 
among local low-mass galaxies \citep[$<$0.5\% of galaxies in SDSS;]
[]{Kniazev2004}, their frequency and significance 
 in the context of the cosmic star formation rate density is expected 
 to increase out to $z\sim$\,1 \citep{Guzman1997,Kakazu2007}. 
However, because of their faintness and compactness, studying EELGs at
these intermediate redshifts requires a great deal of observational
effort. Thus, pioneering studies have been limited to relatively small 
samples of intrinsically luminous objects \citep[e.g.,][]{Koo1995,Phillips1997}.   

In this context, recent deep multiwavelength surveys have offered 
a new avenue for studying chemical enrichment and starburst activity 
and its associated feedback processes in strongly star-forming EELGs out 
to $z\sim$\,1 and beyond 
\citep[see, e.g,][]{Hoyos2005,vanderWel2011,Atek2011,Trump2011,Xia2012,
Henry2013,Ly2014,Amorin2014a,Amorin2014b,Masters2014,Maseda2014}.
This is the case of the COSMOS survey \citep{Scoville2007} and one of
its spectroscopic follow-ups, the zCOSMOS 20k bright survey
\citep{Lilly2007}.
In particular, the wealth of {high-quality photometric and 
spectroscopic data} provided by these surveys allow {us to perform} 
a thorough and systematic characterization of a large probe of faint 
($I_{\rm AB}\la$\,22.5 mag) EELGs out to $z\sim$\,1. 

While the large collection of deep broad- and narrowband photometric 
measurements provided by COSMOS allows  luminosities 
and reliable stellar masses to be derived, HST-ACS $I-$band imaging provides the
spatial resolution {required} to study morphological properties. 
Moreover, zCOSMOS provides the {high signal-to-noise (S/N) 
spectroscopy required to properly measure the emission lines 
used to derive reliable gas-phase metallicities.}    
Remarkably, and despite {the challenge of measuring 
temperature sensitive emission line ratios (e.g., [\oiii]\,5007/4363), 
zCOSMOS spectroscopy allows  gas-phase metallicity 
to be derived using the so-called} {\it direct} ($T_{\rm e}$) 
an unprecedentedly large number of EELGs at intermediate redshifts. 
Thus, our survey offers the opportunity of identifying a relatively 
large number of extremely metal-deficient 
($\la$\,0.1 $Z_{\odot}$) galaxy candidates.

{This is the first of a series of papers aimed at investigating 
the formation history and evolution of low-mass star-forming galaxies  
over cosmological time scales using deep multiwavelength surveys. 
Here, we present the largest spectroscopic sample of galaxies 
with extreme nebular emission in the range 0.1\,$\la$$z$$\la$\,1 assembled 
so far. We characterize more than 150 EELGs selected from the zCOSMOS 20k 
bright survey, based on different key properties, namely size, stellar 
mass, metallicity, and SFR, which are discussed as a function 
of morphology and environment. }
The {derived properties will be used in a companion paper 
(Amor\'in et al., in prep.; Paper\,{\sc II}) to discuss possible 
evolutionary scenarios based on their position in scaling relations 
involving mass, size, metallicity, and SFR.} 

Our paper is organized as follows. In Section\,\ref{sect:sample} we 
describe the parent sample, our dataset, and the selection criteria 
adopted to {compile} the sample of EELGs. 
In Section\,\ref{sect:properties} {we present the main physical 
properties of the sample. We describe the methodology used to derive 
stellar masses, star formation rates and UV properties, and gas-phase metallicities. 
As part of the analysis, we also present an alternative 
method aimed at obtaining $T_{\rm e}-$based metallicities in those 
EELGs without available measurements of the [\oii]\,3727,3729  
doublet. We finish Section\,\ref{sect:properties} by studying the morphological and environmental properties of EELGs.} Later, in {Sections\,\ref{sect:discussion1}-\ref{sect:discussion3}, 
we highlight the discovery of a number of extremely metal-poor galaxy candidates, 
discuss the connection between EELGs and Ly$\alpha$ emitters, and compare   
the zCOSMOS EELGs with other previous samples.   
Finally, Section\,\ref{sect:conclusions} summarizes our main results and 
conclusions. }

Throughout this paper we adopt the standard $\Lambda$-CDM cosmology, 
{\em \emph{i.e.}}, $h$ = 0.7, $\Omega_m$ = 0.3, and $\Omega_\Lambda$ = 0.7 
(Spergel et al., 2007) {and a solar metallicity value of 12$+\log$(O/H)$=$8.69 
\citep{AllendePrieto2001}. Magnitudes are given in the AB system. }

\section{Sample and data}
\label{sect:sample}

\subsection{The parent zCOSMOS 20k bright sample}

{COSMOS} is a large HST-ACS survey, with $I$-band exposures
down to $I_{\rm AB}=28$ on a field of {1.7\,deg$^2$} 
\citep{Scoville2007}.  The COSMOS field has been the object of
extensive multiwavelength ground- and space-based observations
spanning the entire spectrum: X-ray, UV, optical/NIR, mid-infrared,
mm/submillimeter, and radio, providing {photometry over 30 bands} 
\citep{Hasinger2007,Taniguchi2007,Capak2007,Lilly2007,
  Sanders2007,Bertoldi2007,Schinnerer2007,Koekemoer2007,McCracken2010}.
   \begin{figure}[t!]
   \centering
   \includegraphics[angle=0,width=8.cm]{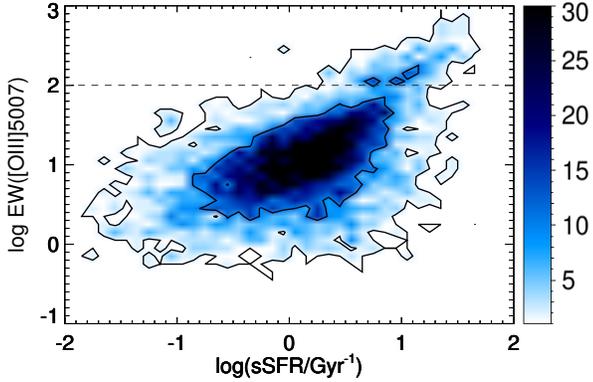}
     \caption{[\oiii]\,$\lambda$\,5007 equivalent width as a
       function of specific SFR for {5056 galaxies at 
       redshift 0.1$<$\,$z$\,$<$\,0.94 in the SFG-20k sample of  
       \citet{Perez-Montero2013}. The inner and outer contours 
       enclose  68\% and 99\% of the sample, respectively.   
       The black dashed line delimits our selection threshold, 
       EW(\oiii)$\geq$\,100\AA, above which galaxies in the 
       zCOSMOS 20k sample are considered to be EELGs.      
       }
              }
         \label{selection}
   \end{figure}

The zCOSMOS survey \citep{Lilly2007} is a large spectroscopic follow-up 
undertaken in the COSMOS field, which {used} about 600\,h of 
ESO observing time with the VIMOS multi-object spectrograph 
\citep{LeFevre2003} mounted on the Melipal 8m telescope of the VLT. 
The survey was divided in two parts, zCOSMOS-bright and zCOSMOS-deep. 
The zCOSMOS-deep observed $\sim 10\,000$ galaxies selected through color 
criteria to have $1.4 \la z\la 3.0$ on the central 1\,deg$^2$ of 
the COSMOS field. 
The zCOSMOS-bright survey is purely magnitude-limited in 
$I-$band and covered the whole area of 1.7 deg$^2$ of the COSMOS field. 
The zCOSMOS-bright survey provides redshifts for $\sim 20\,000$ 
galaxies down to $I_{\rm AB} \leq 22.5$ as measured from the HST-ACS imaging.  
The success rate in redshift measurements is very high, 95\% in the 
redshift range $0.5 < z < 0.8$, and the velocity accuracy is 
$\sim 100$ km/s \citep{Lilly2009}. 
Each observed object has been assigned a flag according to the reliability
of its measured redshift. Classes 3.x and 4.x redshifts, plus Classes
1.5, 2.4, 2.5, 9.3, and 9.5 are considered a secure set, with an
overall reliability of 99\% \citep[see for details][]{Lilly2009}.

The current work is based on the zCOSMOS-bright survey final release, 
the  20k-bright {sample}. 
{This catalog} consists of about $20\,000$ spectra for galaxies 
with $z\leq 2$ and secure redshifts according to the above flag classification. 
{The} zCOSMOS-bright data were acquired with 1$''$ slits and 
the medium-resolution ($R = 600$) grism of VIMOS, providing spectra 
sampled at $\sim$\,2.5 \AA\,pixel$^{-1}$ over a wavelength range of 
approximately $5550-9650$\AA. This spectral range enables important diagnostic emission lines to be followed in order to compute metallicity up to 
redshift $z\sim 1.5$. 
The observations were acquired with a seeing lower than 1.2\arcsec. 
The total integration time was set to $1$ hour to secure redshifts 
with a high success rate. Detailed information about target selection, 
observations, and data reduction can be found in \citet{Lilly2009}.
   \begin{figure}[t!]
   \centering
   \includegraphics[angle=0,width=9.0cm]{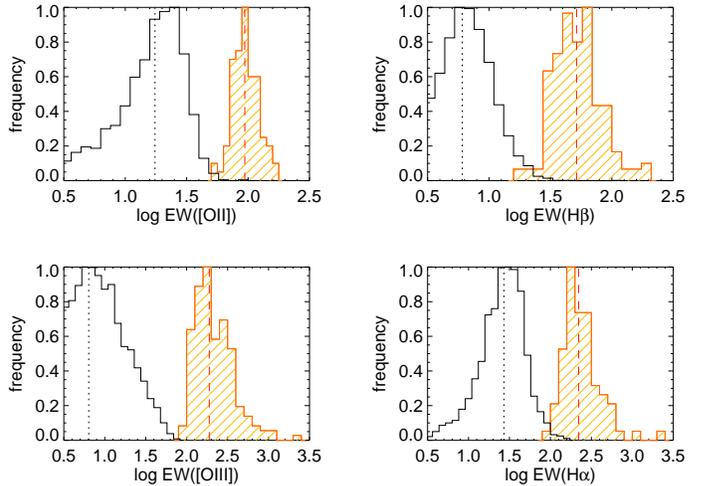}
     \caption{{Distribution of rest-frame equivalent widths for the most luminous 
       hydrogen and oxygen emission lines. Black open histograms correspond to 
       the SFG-20k sample, while the red lined histograms correspond to the EELG 
       sample. Dashed and dotted lines indicate median values. }
       {The adopted EELG selection criteria include galaxies with 
       the highest EWs in all the observed strong emission lines.} 
              }
         \label{histograms}
   \end{figure}

Spectroscopic measurements (emission and absorption lines
fluxes, and equivalent widths) in zCOSMOS were {performed with} the 
automated pipeline {\it platefit-vimos} 
(Lamareille et al., in preparation) similar to those performed
on SDSS \citep[e.g.,][]{Tremonti2004} and VVDS spectra
\citep{Lamareille2009}. This routine fits the stellar component 
of galaxy spectra as a combination of 30 single stellar population 
(SSP) templates, with different ages and metallicities from the 
library of \citet{Bruzual2003}. 
The best-fit synthetic spectrum is used to remove the stellar 
component. Emission lines are then {fit} as a single 
nebular spectrum made of a sum of Gaussians at specified 
wavelengths. Further details can be found in 
\citet{Lamareille2006,Lamareille2009}. 

\subsection{The EELG sample selection}
   \begin{figure}[t!]
   \centering
   \includegraphics[angle=0,width=8.5cm]{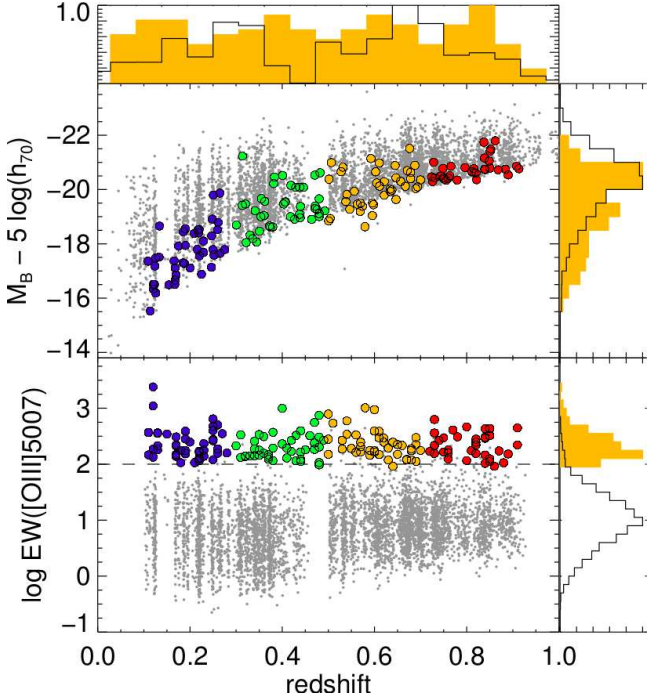}
     \caption{{The two large panels show the rest-frame $B-$band absolute 
     magnitude (\textit{above}) and rest-frame [\oiii] equivalent width 
     (\textit{below}) of the EELGs (\textit{large circles}) and SFGs 
     (\textit{small dots}) in zCOSMOS, as a function of redshift. 
     The upper panel and the two  small panels on the right show the normalized distribution 
     of EELGs (\textit{yellow histogram}) and SFG (\textit{black histogram}) in 
     zCOSMOS for these three quantities. 
     The color code denotes bins of redshift: $0.11 \leq z \leq 0.30$ (\textit{purple}), 
     $0.30 < z \leq 0.50$ (\textit{green}), $0.50 < z \leq 0.70$ ({yellow}), and 
     $0.50 < z \leq 0.93$ (\textit{red}). The black dashed line delimits our selection 
     threshold, EW(\oiii)$\geq$\,100\AA , above which galaxies in the zCOSMOS 20k sample 
     are considered to be EELGs.  }
               }
         \label{z_distribution}
   \end{figure}

In \citet{Perez-Montero2013} we selected a large subsample of more
than 5300 star-forming galaxies at redshift $z=0-1.3$ from the 
{20k-bright} sample to study their physical properties and chemical evolution. 
In order to define our sample of EELGs, {we} repeat the same 
procedure, discarding all broad-line AGNs and selecting only galaxies with 
S/N$>$2 for all the emission-lines automatically measured by {\it platefit-vimos}, 
and involved in the derivation of the oxygen abundance.   
We limit the sample to $\sim$\,5000 galaxies in the redshift range 
$0.11 \leq z  \leq 0.93$ 
to keep only galaxies with [\oiii]\,$\lambda$\,5007 still included in the 
observed spectral range. 
Finally, from {this subset} we select $\sim$\,200 galaxies with the 
largest {rest-frame} equivalent widths in [\oiii]\,$\lambda$\,5007,  EW([\oiii])$\geq$100\AA. {The remaining galaxies -- hereafter referred 
to as the SFG-20k sample -- are used as a comparison set.}  

There are several reasons for the [\oiii]-based selection of EELGs. 
Given the spectral range of our VIMOS data and the rest 
wavelength of [\oiii], a selection {criterion} based on this line 
is more convenient compared to other strong lines (e.g., \ha\ or [\oii]) if 
one intends to maximize the redshift range to be explored {using zCOSMOS data. 
Thus, we are able to collect EELGs on a large redshift range, 
$0.11$\,$\leq$$z$$\leq$\,0.93.}

{In Figure~\ref{selection} we show the relation between the 
{\it \emph{specific}} SFR and the rest-frame [\oiii] equivalent widths of the
SFG-20k  sample of \citet{Perez-Montero2013}. 
Despite the relatively large scatter, we find a clear trend with large
sSFR galaxies showing higher EWs. 
Therefore, our selection limit in the EW[\oiii] for the EELGs guarantees that all 
these EELGs are among the most efficient SFGs out to $z\sim$1. }

{Moreover, it is worth noting that our} [\oiii] criterion
{also leads us to} select galaxies with   strong oxygen
lines, and with unusually strong hydrogen emission lines and extremely
faint and flat continuum. 
This is shown in Fig.~\ref{histograms}, where we compare the 
\ha, \hb, [\oii], and [\oiii] EW distribution of both EELG and SFG-20k samples. 
Clearly, our limit in EW([\oiii])$\geq$\,100\AA\ also leads to the selection of galaxies 
with very high EW(\ha)$\ga$\,100\AA\ and EW(\hb)$\ga$\,20\AA.  
{According to population synthesis models, these limits are an indication of 
young star formation ($<$\,10 Myr) \citep{Leitherer1999} and they have been considered  
in the literature as a powerful tool with which to select young metal-poor starbursts 
\citep[e.g., H{\sc ii} galaxies;][]{SanchezAlmeida2012}}. 
{An} alternative selection {criterion} based on other strong
emission lines would have the drawback that we could only  
select a smaller number of galaxies over a smaller redshift range,
e.g., H$\alpha$ emitters can be selected at $z\la$\,0.5. 

Finally, the choice of a sample {of strong [\oiii] emitters has 
also been motivated } by the aim of collecting a representative 
and statistically significant sample of star-forming galaxies that would be 
easily detectable at {higher} redshifts ($z \sim$\,2-3) in 
deep wide-field NIR surveys 
\citep[e.g.,][]{vanderWel2011,Atek2011,Xia2012,Guaita2013,Maseda2013,Maseda2014}.  
Since they would be affected by similar biases, our sample is intended
to offer a valuable benchmark for future direct comparison with 
other probes of strong emission-line galaxies at higher redshifts. 

Most of the selected galaxies are faint, with 
$I_{\rm AB}$ magnitudes of about $\sim$\,22 mag. 
This can make the measurement of their continuum and faint
emission-lines relatively uncertain when done with 
automatic procedures. 
In order to double check the {fluxes and EWs of} our EELG sample 
we have re-measured by hand (using the {\sl splot} task in IRAF) all 
their emission lines to be sure of their values. 
{Uncertainties} on the line fluxes have been computed 
following \citet{Perez-Montero2003}. 
After discarding a few spurious cases (i.e., extremely noisy spectra or with 
some defects) we finally define a total sample of 183 EELGs. 

\begin{figure*}[t!]
   \centering
\includegraphics[angle=0,width=8.7cm]{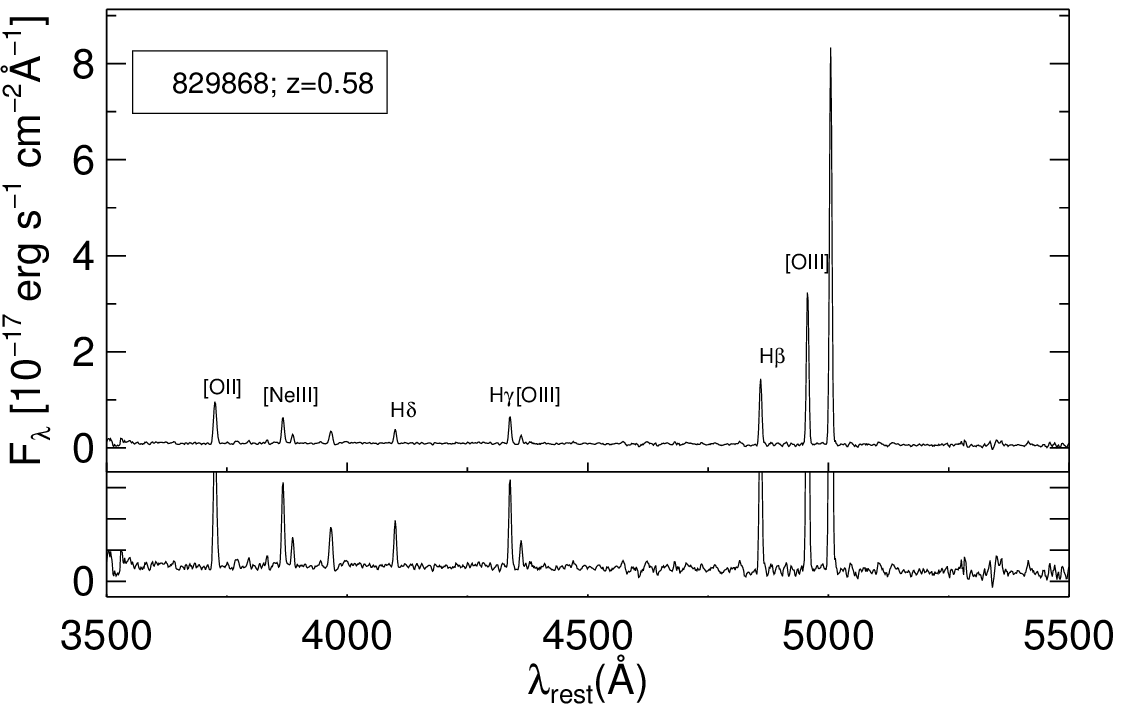}\hspace{8mm}
\includegraphics[angle=0,width=8.45cm]{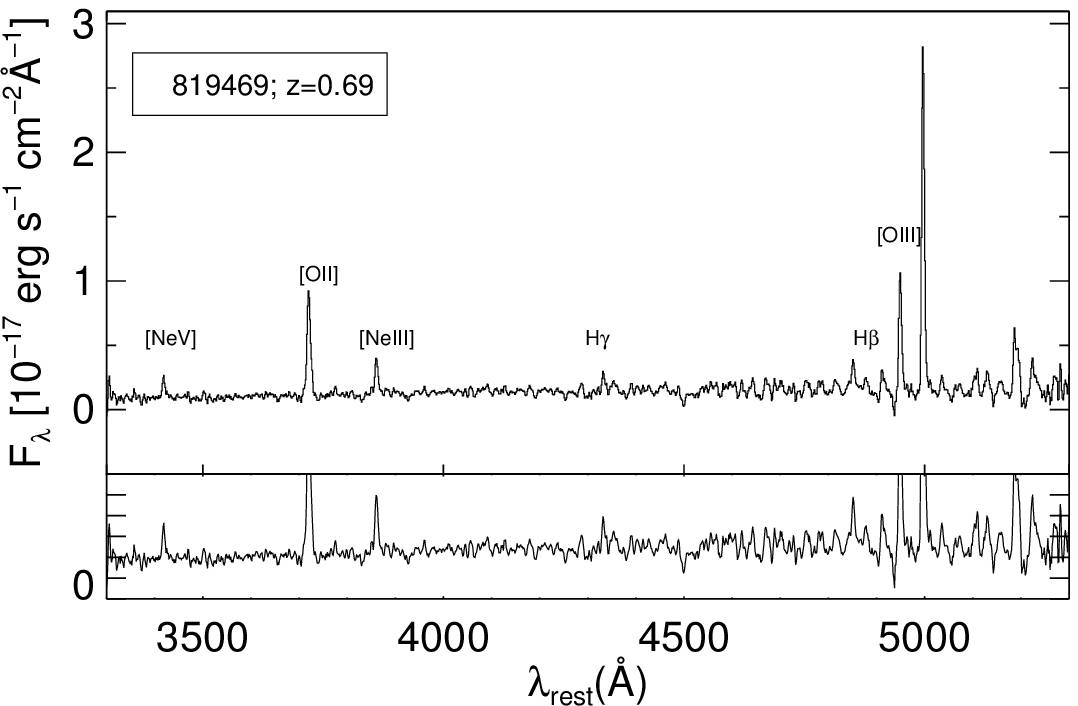}
     \caption{VIMOS spectrum of a purely star-forming 
      (\textit{left}) and a NL-AGN candidate {(\textit{right})}. 
     The spectra have been smoothed by a two-pixel {boxcar filter. 
     The zCOSMOS ID number, the spectroscopic redshift, and the main emission lines 
     are labeled}.
     }
         \label{spectrum}
   \end{figure*}

{Figure~\ref{z_distribution} shows the distribution of rest-frame 
absolute magnitude $M_{B}$ and [\oiii] EW with redshift for both the EELG 
and SFG-20k samples. 
The selected EELGs are {approximately} uniformly distributed in 
redshift out to $z\sim1$ and, by construction, they have the largest EWs. 
Figure~\ref{z_distribution} shows that this property also leads us to 
preferentially select low-luminosity galaxies {in the $B$ band}, including 
most of the less luminous SFGs in zCOSMOS. 
The $B$-band luminosity of EELGs spans a wide range, 
$-16$\,$\la$\,$M_B$\,$\la$\,$-21.5$, and increases with redshift following 
the same trend {as} the SFG-20k  sample. 

\subsection{Identification of AGNs: diagnostic diagrams}

In order to distinguish between purely and non-purely star-forming galaxies 
in the EELGs sample, we  need to identify narrow-line (NL) AGNs (Seyfert 2 
and LINERs) because broad-line AGNs were previously excluded from the sample 
in the selection process.  
{To that end}, we use the combination of {four} empirical diagnostic 
diagrams based on different bright emission-line ratios, {which are presented 
in Fig.~\ref{BPT}. In addition}, we cross-correlate our sample galaxies with the 
XMM and Chandra X-rays catalogs from COSMOS \citep{Hasinger2007,Elvis2009}. 
   \begin{figure*}[t!]
   \centering
   \includegraphics[angle=0,width=7.2cm]{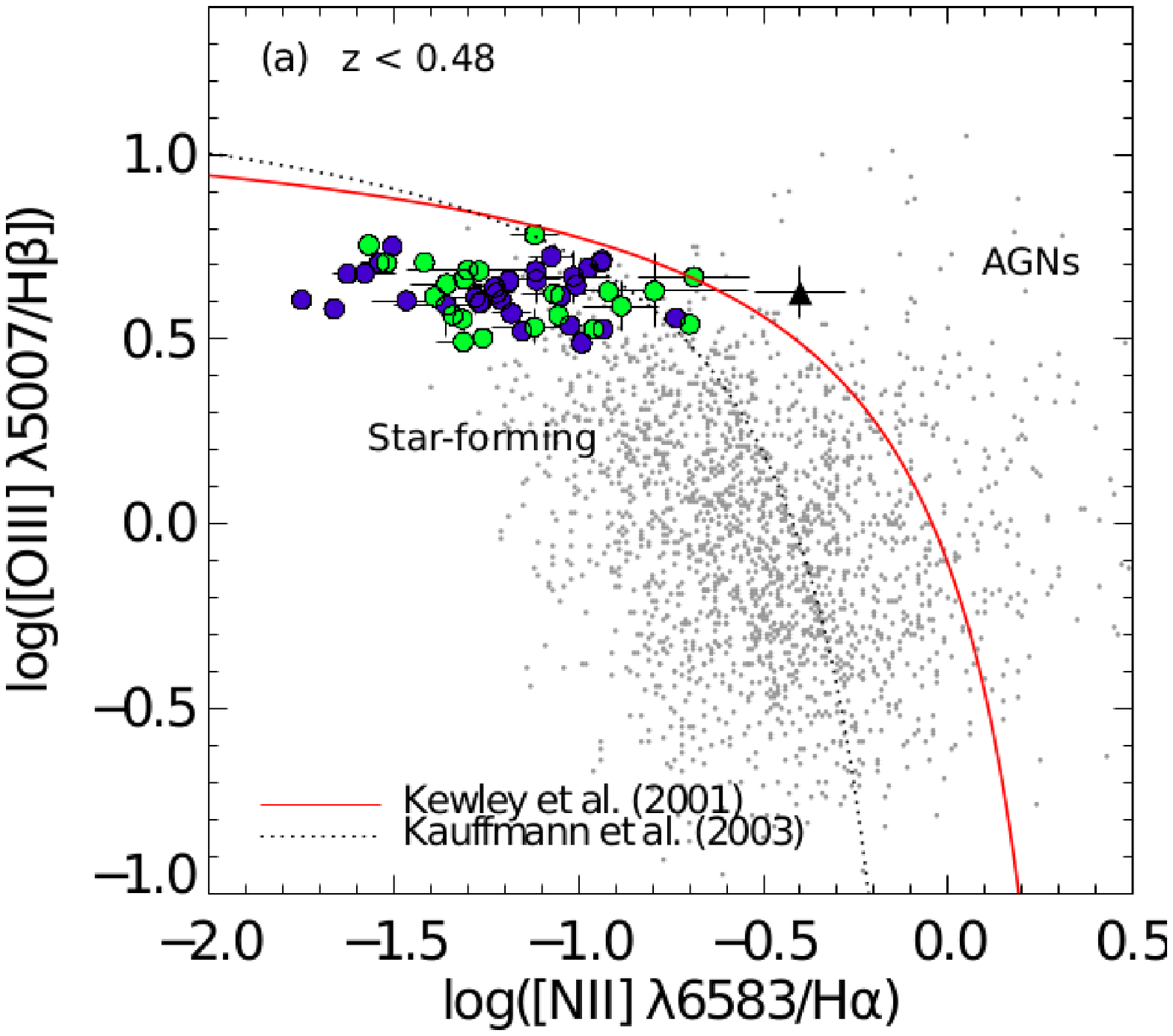} \hspace{2pt}
  \includegraphics[angle=0,width=7.cm] {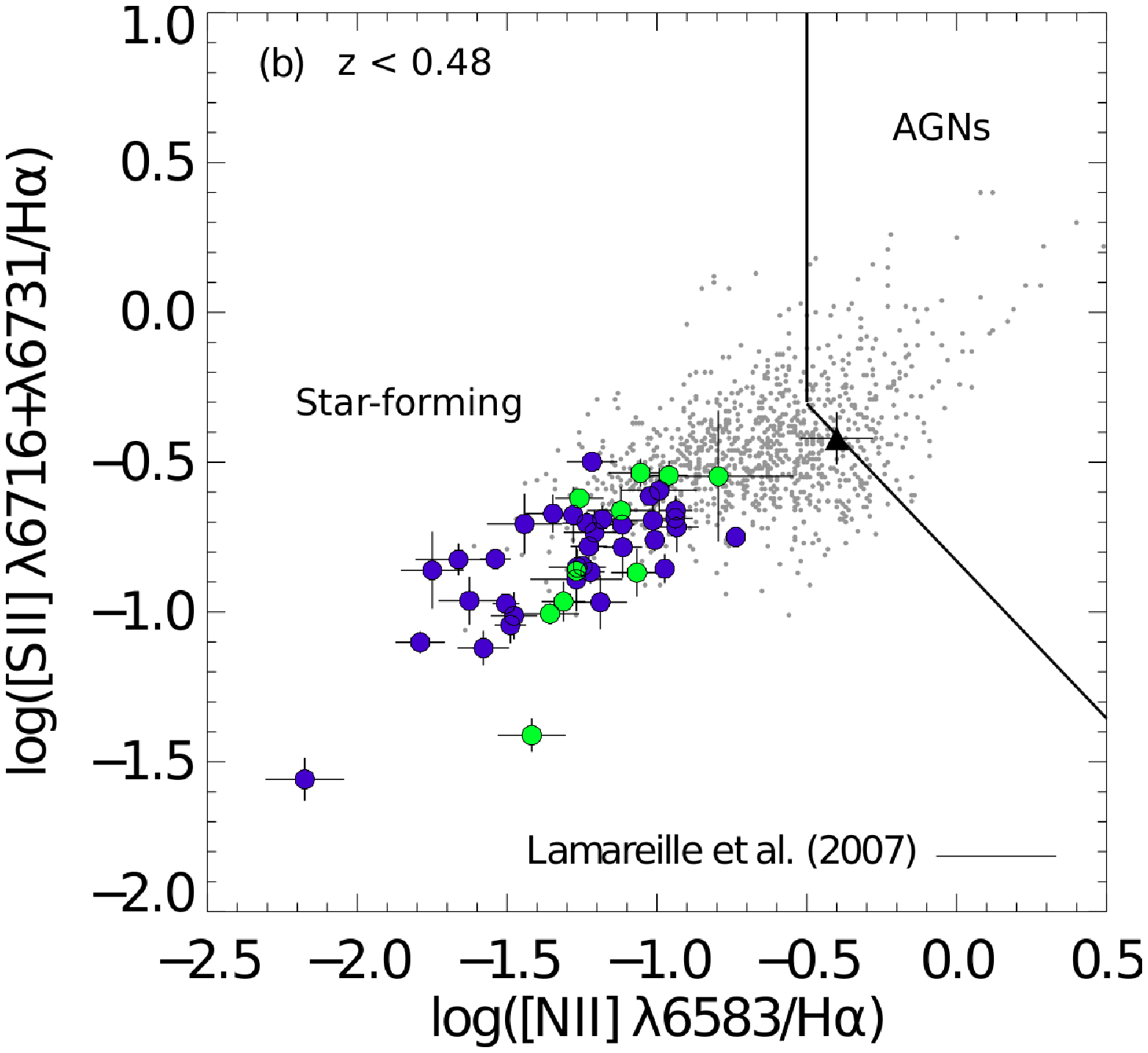} \\[2pt]
   \includegraphics[angle=0,width=7.cm]{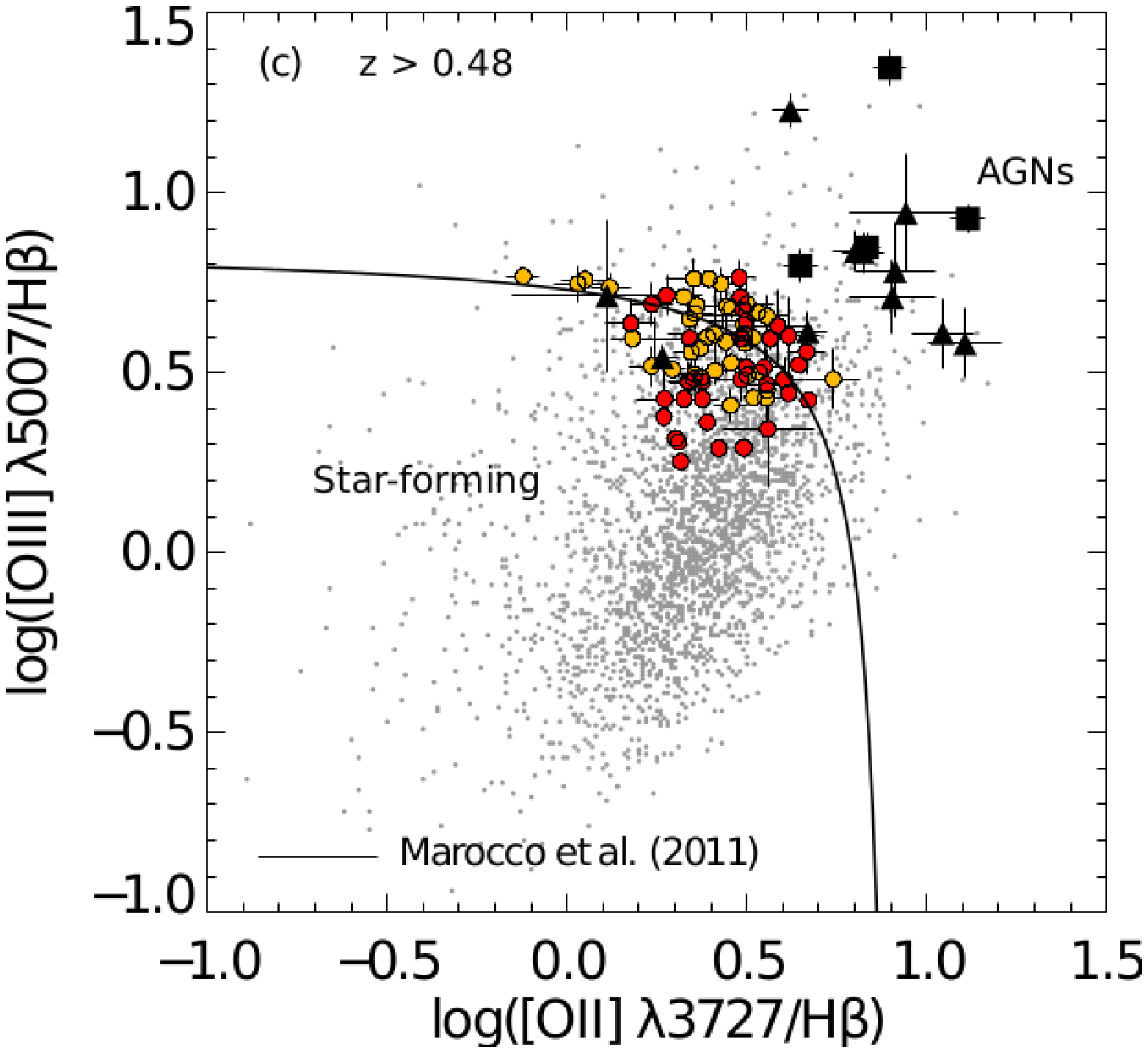} \hspace{2pt}
   \includegraphics[angle=0,width=7.cm]{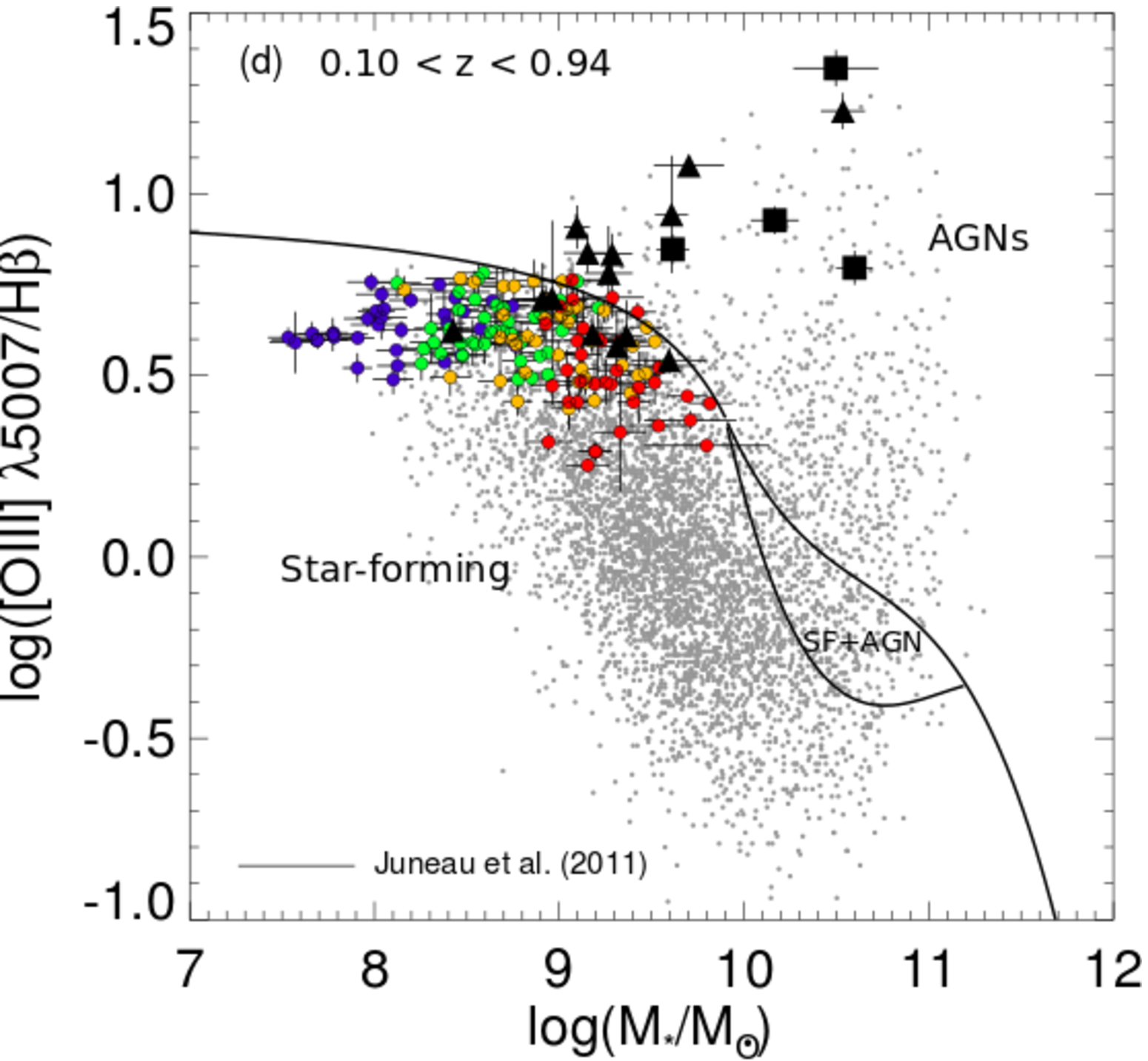}
    \caption{Diagnostic diagrams {for EELGs. Large symbols indicate purely 
    star-forming systems (\textit{colored}) and AGN candidates (\textit{black}). 
    Small gray dots show galaxies from the SFG-20k parent sample. Colors are as in 
    Fig.~\ref{z_distribution} and labels indicate the redshift of 
    each subset of galaxies. 
    AGN candidates include both galaxies with detected 
    X-ray counterparts (\textit{squares}) and galaxies with broad Balmer line 
    components and/or high-ionization emission lines (\textit{triangles}). 
    Lines show the empirical separation between SFGs and AGNs.} 
              }
         \label{BPT}
   \end{figure*}

For {63 EELGs} at $z$\,$\la$\,0.48, and depending on the set of lines 
with {available and} reliable measurements, we use both the well-known 
diagnostic diagram \citep[e.g.,][]{Baldwin1981,Veilleux1987} based 
on the line ratios [\oiii]/\hb\ and [\nii]/\ha\ (Fig.~\ref{BPT}\,$a$), 
and the \ha\ classification proposed by \citet{Lamareille2007} based on 
[\nii], [\sii], and \ha\ emission-line ratios (Fig.~\ref{BPT}\,$b$). 
For {95 EELGs} galaxies with $z$\,$>$\,0.48, the \ha,
[\nii]\,6584, and [\sii]\,6717, 6731 emission lines are no longer visible  
in the zCOSMOS {VIMOS} spectra and therefore the above diagnostics 
cannot be used. Instead,  {for these galaxies} we use the diagnostic 
diagram defined {in} \citet{Lamareille2004}, involving the [\oiii]/\hb\ and 
[\oii]/\hb\ emission-line ratios, {as shown Fig.~\ref{BPT}\,$c$. 
This diagnostic diagram includes the corrections proposed by 
\citet{Perez-Montero2013} to minimize the impact of reddening effects 
due to the long wavelength baseline between [\oii]\,3727 and \hb. 
Finally, for the entire EELG sample we use the empirical MEx diagram 
\citep{Juneau2011}, where SFGs and AGNs are distinguished by their stellar 
mass and excitation level (Fig.~\ref{BPT}\,$d$). 
Galaxies clearly located above the empirical limits shown in 
Fig.~\ref{BPT} should be considered  AGN candidates.} 

Overall, the agreement between {the first three diagnostics and the MEx 
diagram is good. 
Similar results, although with slightly larger dispersion, are found 
using alternative diagnostics, such as excitation vs. $U-B$ 
color \citep{Yan2011}.
Nonetheless, the above empirical limits typically have 1$\sigma$ 
uncertainties of about 0.2 dex and many EELGs are located very close to  
these boundaries. This can make the distinction between SFGs and AGNs 
somewhat tricky for some objects. 
As an additional test to select AGN candidates} we have checked them 
one-by-one for the presence of {X-ray counterparts and} 
bright high-ionization emission lines (e.g., [Ne{\sc V}]), and/or very
extended Balmer line components, which can be indicative of the presence of AGNs. 

Only  four EELGs, all of them at $z$\,$>$\,0.47, are confirmed X-ray sources 
(zCOSMOS IDs 819469, 825103, 839230, and 841281). 
All of them are also clear AGN candidates in the optical diagnostics. The remaining} AGN
candidates without X-ray counterparts{}  
show high-ionization lines such as [Ne{\sc V}] and He{\sc II} 
{or unusually broad components in their Balmer lines, 
and they are typically  redder  than the rest of the EELGs.} 
In Fig.~\ref{spectrum} we present one example of a VIMOS spectrum for both a 
purely star-forming EELG and a NL-AGN candidate with an X-ray counterpart.  

{To summarize}, using the above criteria, our analysis finds 165 purely 
star-forming EELGs (90\%) and 18 EELGs (10\%) with likely NL-AGN contribution. 

\section{The properties of extreme emission-line galaxies in zCOSMOS}
\label{sect:properties}
\begin{sidewaystable*}
\caption{{Emission-line fluxes}}
\label{Tab0}
\centering
\footnotesize
\begin{tabular}{lccccccccccccc}
\noalign{\smallskip}
\hline\hline
\noalign{\smallskip}
zCOSMOS\,ID & $\alpha (J2000)$ & $\delta (J2000)$ & $z$ & [\oii]\,3727 &  H$\gamma$ & [\oiii]\,4363 & H$\beta$ & [\oiii]\,4958 & [\oiii]\,5007 & H$\alpha$ & [\nii]\,6584 &  [\sii]\,6716 & [\sii]\,6730 \\
\hline
\noalign{\smallskip}
\vspace{0.15mm}
 700882 & 150.349518 & 2.275816 & 0.464 & ... & 1.0$\pm$0.3 & ... & 1.9$\pm$0.4 & 3.5$\pm$0.5 & 9.5$\pm$0.6 & ... & ... & ... & ... \\
 701051 & 149.856964 & 2.245983 & 0.345 & ... & 2.3$\pm$0.3 & ... & 6.1$\pm$0.5 & 8.0$\pm$0.6 & 23.9$\pm$0.6 & 26.5$\pm$0.4 & ... & ... & ... \\ 
 701741 & 150.393982 & 2.578904 & 0.504 & 9.2$\pm$1.0 & 2.2$\pm$0.5 & 0.6$\pm$0.2 & 4.5$\pm$0.3 & 5.5$\pm$0.5 & 14.1$\pm$0.9 & ... & ... & ... & ... \\
 800984 & 150.286469 & 1.623921 & 0.595 & 7.5$\pm$0.5 & 3.4$\pm$0.3 & 0.9$\pm$0.2 & 9.5$\pm$1.0 & 20.5$\pm$0.8 & 53.0$\pm$1.0 & ... & ... & ... & ...      \\
 801094 & 150.242004 & 1.610143 & 0.546 & 6.0$\pm$0.3 & 1.3$\pm$0.2 & ... & 2.5$\pm$0.5 & 2.9$\pm$0.3 & 9.6$\pm$0.3 & ... & ... & ... & ... \\
 802275 & 149.711777 & 1.616209 & 0.635 & 5.1$\pm$0.2 & 0.7$\pm$0.1 & ... & 2.0$\pm$0.1 & 2.6$\pm$0.1 & 6.1$\pm$0.2 & ... & ... & ... & ... \\
 803226 & 150.705399 & 1.716969 & 0.570 & 8.5$\pm$0.8 & 2.3$\pm$0.1 & 0.6$\pm$0.2 & 4.5$\pm$0.5 & 8.8$\pm$0.6 & 25.7$\pm$0.8 & ... & ... & ... & ...       \\
 803892 & 150.526794 & 1.787344 & 0.439 & ... & 2.0$\pm$0.3 & ... & 3.9$\pm$0.3 & 4.5$\pm$0.4 & 15.0$\pm$0.2 & 6.8$\pm$1.0 & ... & ... & ... \\
 804130 & 150.452408 & 1.632364 & 0.429 & ... & 1.3$\pm$0.2 & ... & 5.4$\pm$0.3 & 9.3$\pm$0.2 & 25.2$\pm$0.3 & 5.9$\pm$0.5 & 1.2$\pm$0.4 & ... & ...  \\
 804791 & 150.286530 & 1.633338 & 0.603 & 13.9$\pm$0.3 & 2.7$\pm$0.2 & ... & 5.7$\pm$0.4 & 7.0$\pm$0.3 & 18.2$\pm$0.5 & ... & ... & ... & ... \\
\noalign{\smallskip}                                                       
\hline                                                                          
\hline
\end{tabular}                                                                   
\begin{list}{}{}
\item {Notes: 
{Measured emission-line fluxes are given in units of 10$^{-17}$\,erg s$^{-1}$ cm$^{-2}$. 
Flux errors have been derived following \citet{Perez-Montero2003}. 
No extinction correction has been applied to these fluxes. 
(The entire version of this table for the full sample of EELGs is 
available {\it On-line}).}}
\end{list}
\end{sidewaystable*}

{In Table~\ref{Tab0}\footnote{\label{note0}A preview table is shown. 
A complete version of this table is available {\it \emph{online.}}} 
we present the sample of 165 star-forming EELGs, including fluxes and 
uncertainties for the most relevant emission lines. 
These quantities, along with an exquisite multiwavelength dataset, have been 
used to derive their main properties. 
In this section we will describe the methodology and briefly 
discuss our results. }
{A catalog of the most relevant properties} for each galaxy 
{is presented in} Table 2\textasciicircum 2. 
{The sample has been divided into four redshift bins to further examine 
possible trends in their main properties with redshift.}
These bins are almost equally populated and {they can be distinguished 
by color in our figures. Finally, in} Table~\ref{Tab2} {we} show 
median values and standard deviations of the main properties of EELGs 
according to {the defined redshift bins, their optical morphology 
and their environment. 
The median properties of the NL-AGN candidates are also included 
in Table~\ref{Tab2}. However, the subset of NL-AGN candidates is not 
considered for the subsequent  analysis.}  

   \begin{figure}[t!]
   \centering
   \includegraphics[angle=0,width=8.32cm]{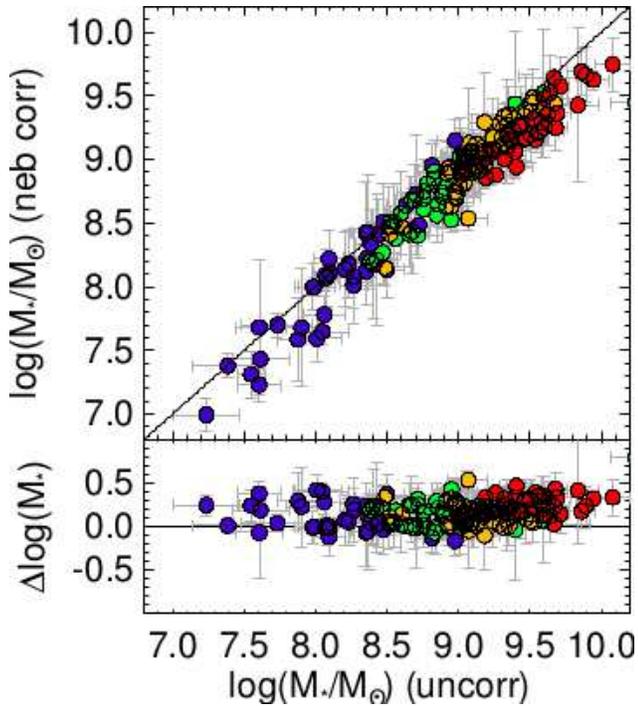}\hspace{2mm}
   \caption{Comparison of stellar masses derived from SED fitting before {(uncorr, $x-$axis) and after (nebcorr, $y-$axis)} removing the contribution of emission line fluxes to the broadband photometry. {The bottom panel shows the difference in stellar mass (uncorr--nebcorr)  on the $y$-axis. Symbols and colors are as in Fig.~3.} 
              }
         \label{mass}
   \end{figure}

\subsection{The low stellar masses of EELGs}

Total  stellar masses, M$_*$, for {SFGs in the zCOSMOS 20k} sample 
are taken from \citet{Bolzonella2010}. They were derived by fitting stellar 
population synthesis models to both the broadband  optical/near-infrared 
\citep[CFHT: u, i, Ks; Subaru: B, V, g, r, i, z][]{Capak2007} 
and infrared \citep[Spitzer/IRAC: 3.6$\mu$m, 4.5$\mu$m][]{Sanders2007} 
photometry using a chi-square minimization for each galaxy.
The different methods used to compute stellar masses, based on different 
assumptions about the population synthesis models and the star formation 
histories, are described in detail in \citet{Bolzonella2010}. 
The accuracy of the photometric stellar masses is satisfactory overall, 
with typical dispersions due to statistical uncertainties and 
degeneracies of the order of 0.2 dex. 
The addition of secondary bursts to a continuous
star formation history produces systematically higher (up to
40\%) stellar masses, while population synthesis models taking
into account the TP-AGB stellar phase \citep{Maraston2005} produces 
systematically lower M$_*$ by  0.10 dex. 
The uncertainty on the absolute value of M$_*$ due to assumptions
on the initial mass function (IMF) is within a factor of 2 for the
typical IMFs usually adopted in the literature.
In this paper, we have adopted stellar masses calculated on the basis 
of a \citet{Chabrier2003} IMF and the stellar population models of 
\citet{Bruzual2003}, with the addition of secondary bursts to the 
standard declining exponential star formation history. 

For strong emission line galaxies a significant contribution to the 
broadband flux densities from nebular emission is {superimposed on} the
stellar spectral energy distribution (SED). 
Since standard stellar population synthesis models do not include nebular 
emission this may have an impact on the SED fitting and, in particular, on 
the computed total stellar masses \citep[e.g.,][]{Krueger1995,Papaderos1998,Schaerer2009,Atek2011,Curtis-Lake2013,Stark2013,Castellano2014,Santini2014,Pacifici2015}. 
In order to overcome this potential systematic effect for the sample of EELGs, 
an additional set of fits were computed after removing the contribution of 
emission line fluxes to the observed broadband magnitudes.  
For the models we follow the prescriptions described above for the standard 
zCOSMOS SED fitting \citep[][]{Bolzonella2010}, but fixing {the 
metallicity of the} stellar population models 
(from $Z=[0.02, 0.2, 0.4, 1]\,Z_{\odot}$) to the nearest {available} value 
to the observed gas-phase metallicity of each galaxy (see Section 3.6). 
In those cases where the gas-phase metallicity was not {measured we 
adopted the median metallicity of the full sample as a} reference value. 
   \begin{figure*}[t!]
   \centering
   \includegraphics[angle=0,width=7.5cm]{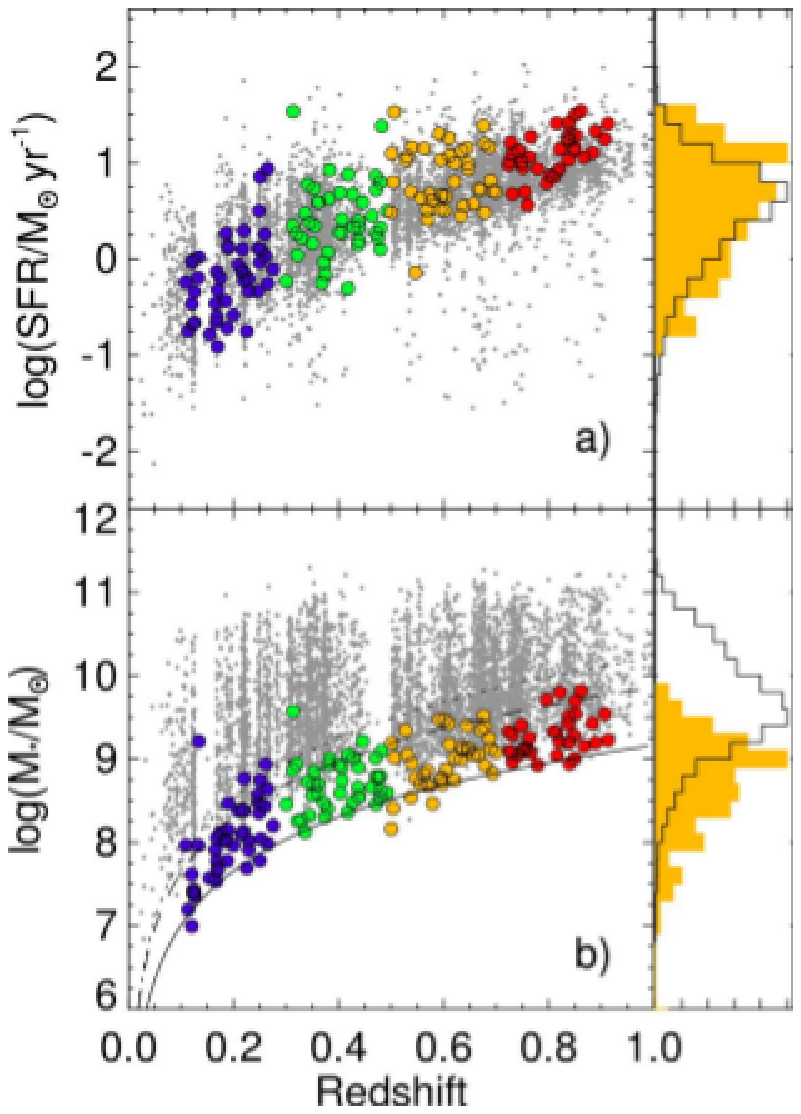}\hspace{5 mm}
   \includegraphics[angle=0,width=7.55cm]{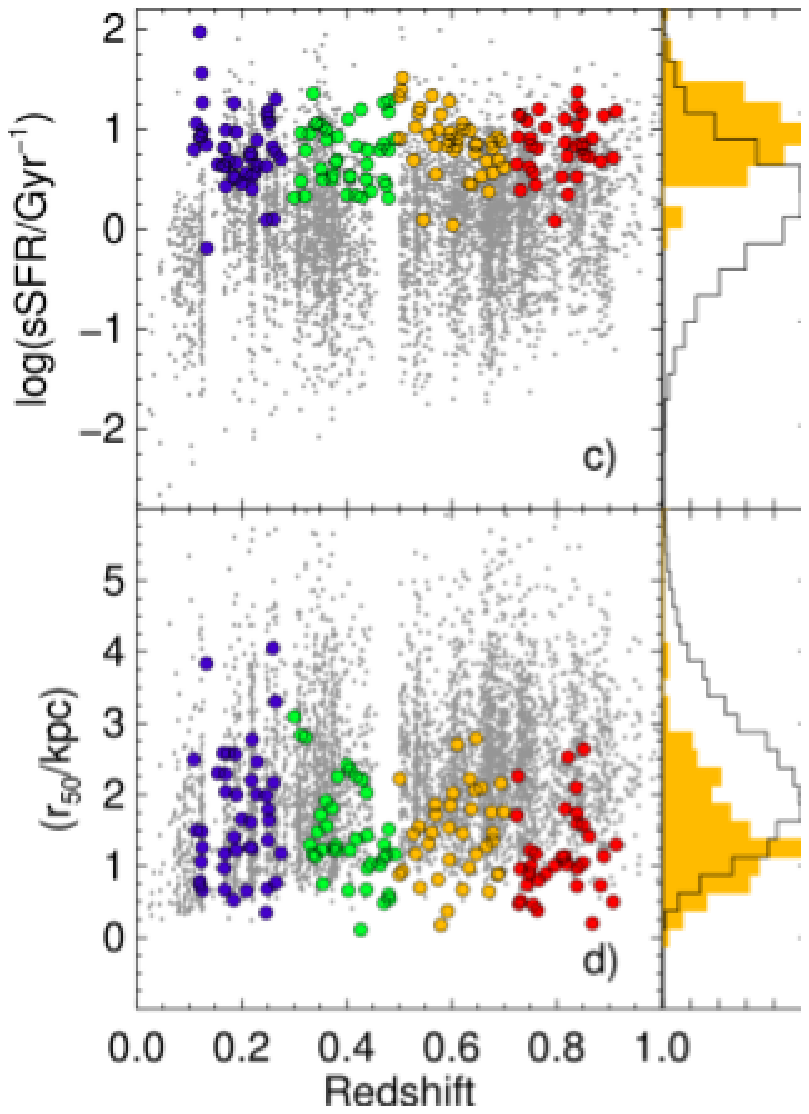}
     \caption{Redshift distributions of SFR ($a$), stellar mass ($b$), specific SFR ($c$), and {HST-ACS $I$-band} half-light radii ($d$). Symbols and colors are the same as in Fig.~\ref{z_distribution}.  {The normalized distribution of EELGs ($filled$) and SFGs ($black$) for each property are shown in the  histograms to the right of the panels}.
     Solid, dashed, and dotted lines in ($b$) show the logarithmic fitting to the limiting masses of the star-forming sample for levels 25\%, 50\%, and 75\% of completeness, respectively. {The EELGs are small galaxies forming the low end of stellar mass and the high end of sSFR distributions of SFGs in zCOSMOS up to $z\sim$\,1}.
                        }
         \label{prop_distribution}
   \end{figure*}

{In Figure~\ref{mass} we show a comparison between the stellar masses 
derived before and after removing the contribution of the 
strong emission lines. 
We find the stellar masses derived from SED fitting using uncorrected magnitudes  systematically offset to higher values ($\sim$\,0.25 dex in the median) compared to 
the masses derived from SED fitting after correction for nebular emission. 
In the most extreme cases (i.e., very high EWs and very low metallicity)} neglecting 
nebular emission in SED fitting may lead to an overestimation of the stellar mass 
of up to a factor of $\sim$\,3-5. 
This result is in good agreement with previous findings for strong emission line 
galaxies at low and high redshift \citep[e.g.,][]{Atek2011,Curtis-Lake2013}. 

In Table~\ref{Tab1} we include the stellar masses {corrected for 
nebular emission, which are adopted for the subsequent analysis. 
Median values for each redshift bin are also listed in} Table~\ref{Tab2}. 
{Finally, in Fig.~\ref{prop_distribution}\,($b$) we show the 
redshift distribution of M$_*$ for EELGs and SFGs in zCOSMOS}; 
Fig.~\ref{prop_distribution}\,($b$) {also includes} the limiting 
mass for the {SFG-20k} sample as derived by \citet{Perez-Montero2013}. 
{We note that EELGs are clearly among the less massive SFGs in} 
zCOSMOS. {Their stellar masses are found to increase slowly with 
redshift}, from $\sim\,$10$^{7}$ at $z\sim$\,0.1 to $\sim$\,10$^{10}$\,M$_{\odot}$ 
at $z\sim$\,0.9. 
{Most EELGs are found between the} 25\% and 75\% {completeness 
limit}. 
Therefore, the EELG sample {is in a range of masses where the 
zCOSMOS 20k sample is not complete}.   

\subsection{Dust extinction and star formation rate of EELGs}

\begin{figure}[t!]
  \centering
  \includegraphics[angle=0,width=4.4cm]{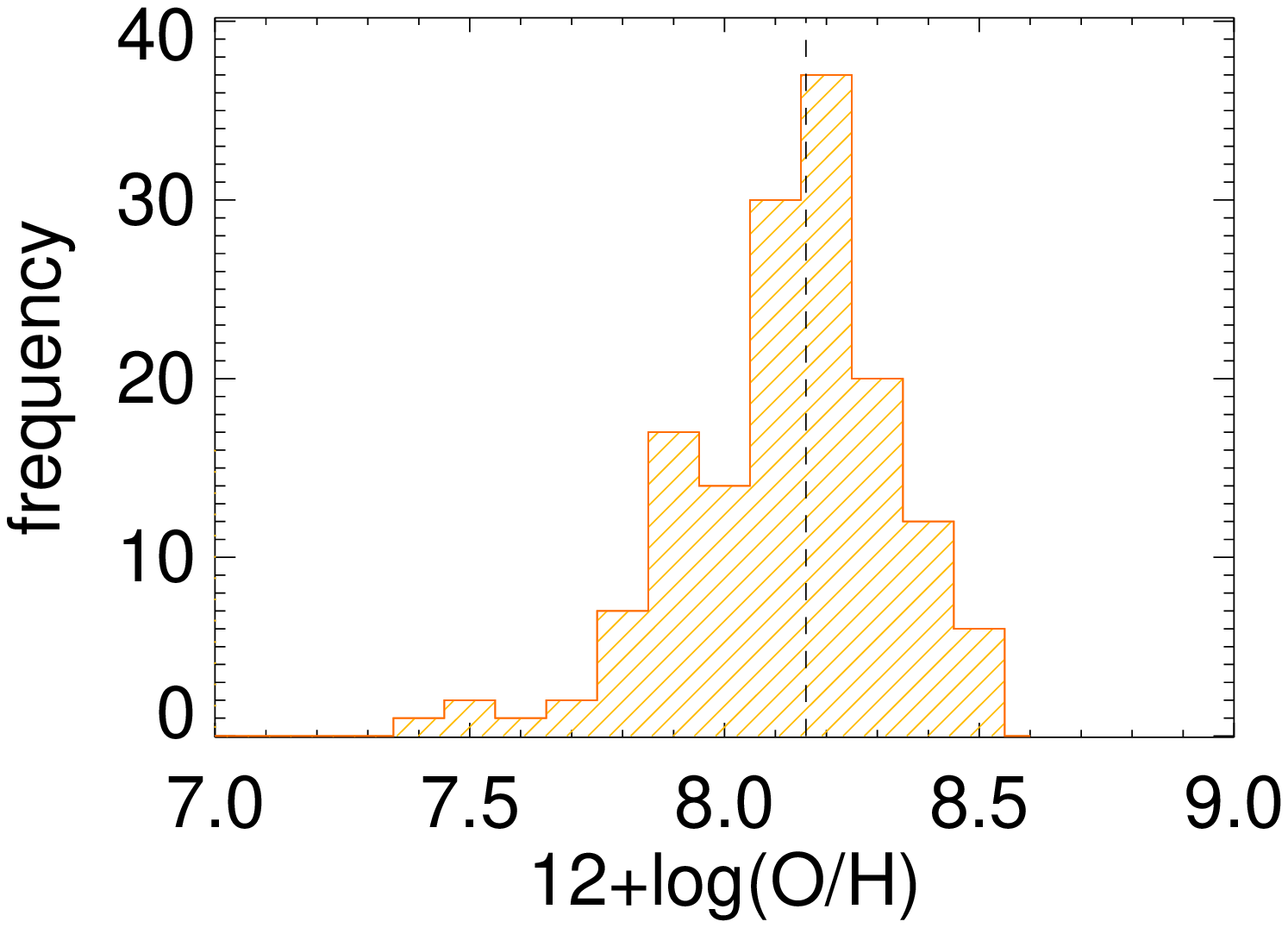}
    \includegraphics[angle=0,width=4.4cm]{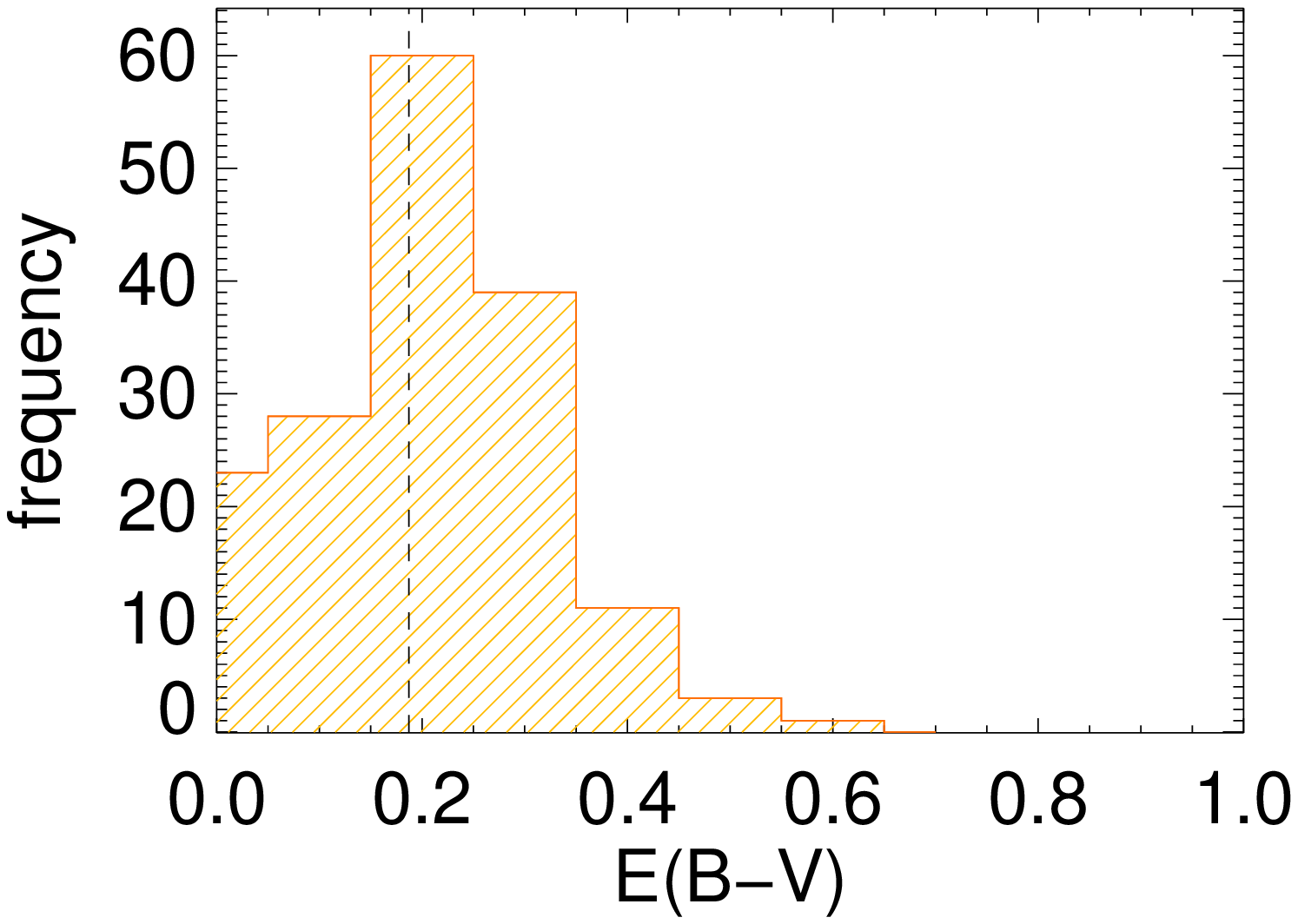}
    \caption{{Distribution of gas-phase metallicity ({\it left}) and reddening ({\it right}) for the sample of EELGs. Vertical dashed lines indicate median values. {The EELGs are low-extinction, metal-poor systems}.}
                 }
        \label{histo_OH}
   \end{figure}

We derive SFRs using the luminosity of the brightest available 
Balmer emission line after correction for aperture effects and reddening. 
Aperture effects were quantified using factors derived from 
photometry\footnote{ACS-HST photometry was used if available, Subaru 
photometry if not.}. {A reddening correction} was carried out using the 
Balmer decrement for those objects with more than one Balmer hydrogen 
recombination line with S/N$>$2 available {(see Tables~\ref{Tab0} and \ref{Tab1})}, and assuming the theoretical ratios at 
standard conditions of temperature and density from \citet{StoreyHummer1995} 
and the \citet{Cardelli1989} extinction law. 
Although gas extinction is preferable to be used whenever possible, for 
a number of galaxies where only one Balmer line is available ($\sim$\,9\% 
of the EELGs and $\sim$\,36\% of the SFG-20k sample), we considered a 
reddening coefficient from the stellar $E(B-V)$ parameter derived 
from the stellar synthesis fitting, assuming that the gas and the stellar 
reddening coefficients are correlated \citep{Calzetti2000}. 
The same rule was applied for $\sim$32\% of the EELGs where the line ratios \ha/\hb\ 
or \hb/\hg\ were lower than their theoretical values, $($\ha/\hb$)_{0}=$\,2.82 
and $($\hg/\hb$)_{0}=$\,0.47, assuming Case B recombination for typical values of 
both electron temperature and density\footnote{\label{note2}We assume the theoretical coefficients of \citet{StoreyHummer1995} for $T_e=$2$\times$10$^4$K and $n_e=$100 cm$^{-3}$}.
{The reddening coefficients, $c($\hb$)$, for the sample of 
EELGs are listed in Table~\ref{Tab1} and its histogram distribution  
is presented in Fig.~\ref{histo_OH}. 
Overall, most} galaxies show relatively low dust extinction, with a median 
{reddening} of $E(B-V)$=0.19 magnitudes.

We derive the {\it \emph{ongoing}} star formation rates from { extinction}-corrected 
H$\alpha$ luminosities and using the { standard} calibration {of}  \citet{Kennicutt1998}, 
\mbox{SFR(\ha)\,$=$\,7.9$\times$\,10$^{-42}$\,$L$(\ha) [erg\,s$^{-1}$]}, 
which assumes a Salpeter IMF from 0.1 to 100 $M_{\odot}$. 
We have scaled down these SFRs by a factor of 1.7 to be consistent with the 
Chabrier IMF used {in} this paper. 
For those galaxies at $z$\,$\ga$\,0.47 for which \ha\ is not observed in the
VIMOS spectra, we derive the expected \ha\ luminosity based on the \hb\ fluxes 
and the {assumed} theoretical ratio$^4$, 
$($\ha/\hb$)_{0}=$\,2.82. 

We  find that EELGs span a large range of SFR$\sim$\,0.1-35 
M$_{\odot}$\,yr$^{-1}$, with median values increasing with redshift, as  
shown in Fig.~\ref{prop_distribution}($a$) and in Table~\ref{Tab2}. 
These high SFRs in combination with their low stellar masses imply that 
EELGs include the most efficient star-forming galaxies of zCOSMOS in 
terms of {\it \emph{specific}} SFR. 
{This is shown in Fig.~\ref{prop_distribution}\,($c$), 
where the sSFR of EELGs does not evolve with redshift and 
shows a median value of log(sSFR)$\sim$\,0.81\,Gyr$^{-1}$.} 
The extremely high sSFRs of EELGs imply that they are rapidly 
building up their stellar components. Their stellar mass {\it \emph{doubling}} 
times (i.e., 1/sSFR, or the time needed to double their total stellar 
mass at their current SFR) are typically  a few hundreds million years. 

\subsubsection{UV properties}

{\it Galaxy Evolution Explorer (GALEX)} data in the FUV 
($\lambda_c$\,$\sim$\,1530\AA) and NUV ($\lambda_c$\,$\sim$\,2315\AA) from 
the COSMOS/GALEX photometric catalog \citep{Schiminovich2007,Zamojski2007} 
were used to derive rest-frame UV luminosities and colors for most of the EELGs. 
We calculate rest-frame FUV absolute magnitudes consistently 
with those in the optical and IR used for the SED fitting and stellar mass 
derivation \citep{Bolzonella2010}. 
In order to account for intrinsic dust attenuation, we have adopted the 
relation between the total dust attenuation and UV spectral slope given by 
\citet{Meurer1999} for a sample of local starbursts with IR and UV 
measurements, which can be expressed as 
A$_{\rm FUV}$\,$=$\,4.43$+$1.99\, $\beta_{\rm UV}$. 
In this equation $\beta_{\rm UV}$\,$=$\,2.32\,($FUV-NUV$)$-$2.0 is the 
photometric {measurement} of the UV spectral slope in rest-frame.  

The resulting dust-corrected FUV luminosities, $L_{\rm FUV}$, for each
galaxy are listed in Table~\ref{Tab1}. 
Median values of $\beta_{\rm UV}$ and $L_{\rm FUV}$ 
{for the various subsets} are also included in Table~\ref{Tab2}. 
Our EELG sample shows typical values of $\beta_{\rm UV}=-1.61$, which 
according to the Meurer formula imply dust attenuations of 
A$_{\rm FUV}$\,$\sim$\,1.23\,mag. 
This value is $\sim$\,30\% lower than the one derived from a mean reddening 
derived from the optical of $E(B-V)\sim$\,0.2 assuming a \citet{Cardelli1989} 
extinction law with $A_{\rm FUV}=$\,8.15\,$E(B-V)$. 
After dust corrections we find for EELGs median luminosities of 
$L_{\rm FUV}$\,$\sim$\,10$^{10.4} L_{\odot}$ 
and FUV surface brightnesses\footnote{$\mu_{\rm FUV} = \frac{0.5 L_{\rm FUV}}{\pi r_{50}^{2}}$, {where $r_{50}$ is the optical half-light radius (see Sect.~3.4.2)} } of $\mu_{\rm FUV}$\,$\ga$\,10$^{9} L_{\odot}$\,kpc$^{-2}$. 
These values {mean} that the EELGs are very compact and luminous 
in the UV continuum. 
   \begin{figure*}[t!]
   \centering
   \includegraphics[angle=0,width=5.5cm]{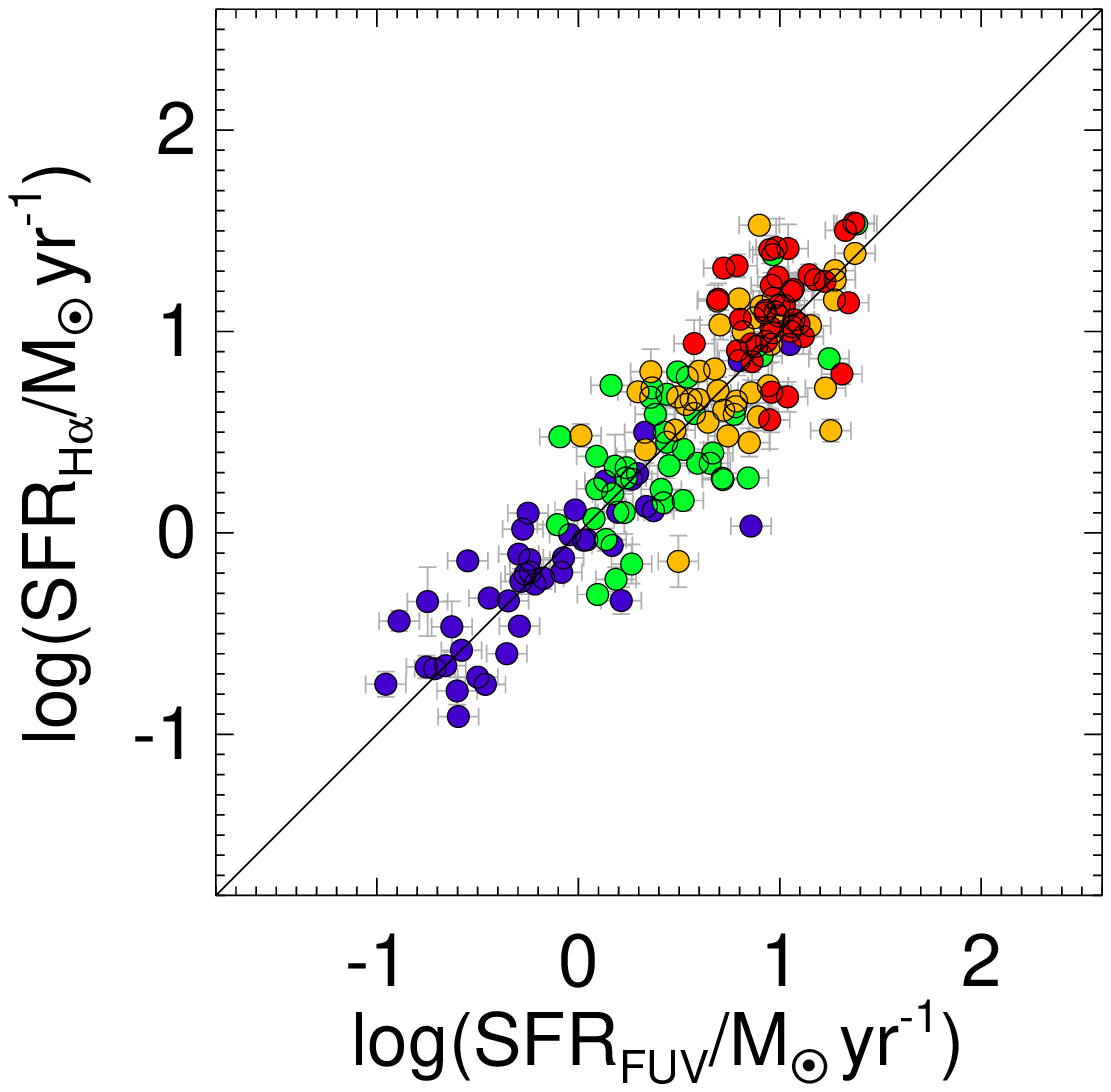}\hspace{2.5mm}
    \includegraphics[angle=0,width=5.5cm]{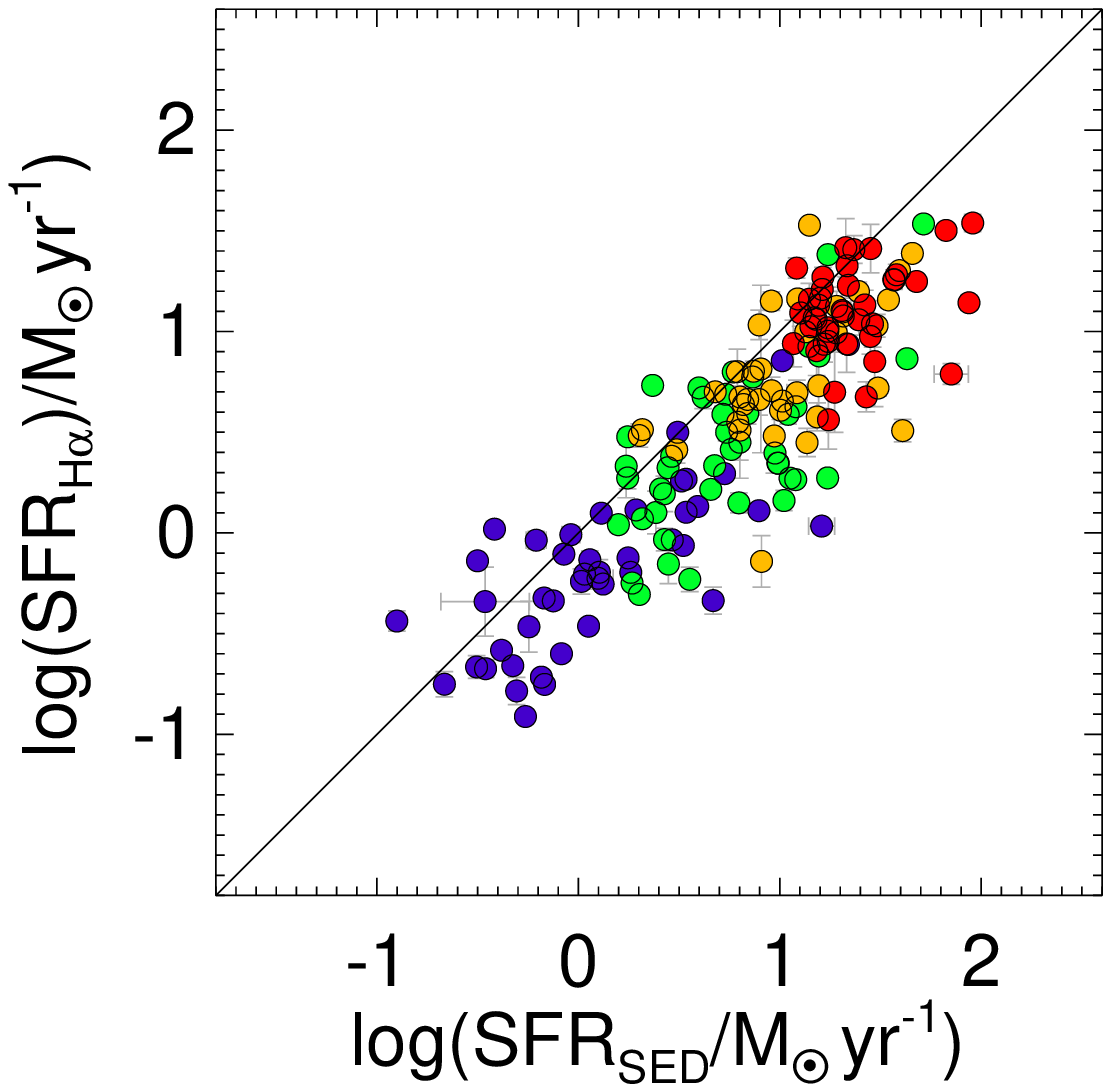}\hspace{2.5mm}
    \includegraphics[angle=0,width=5.5cm]{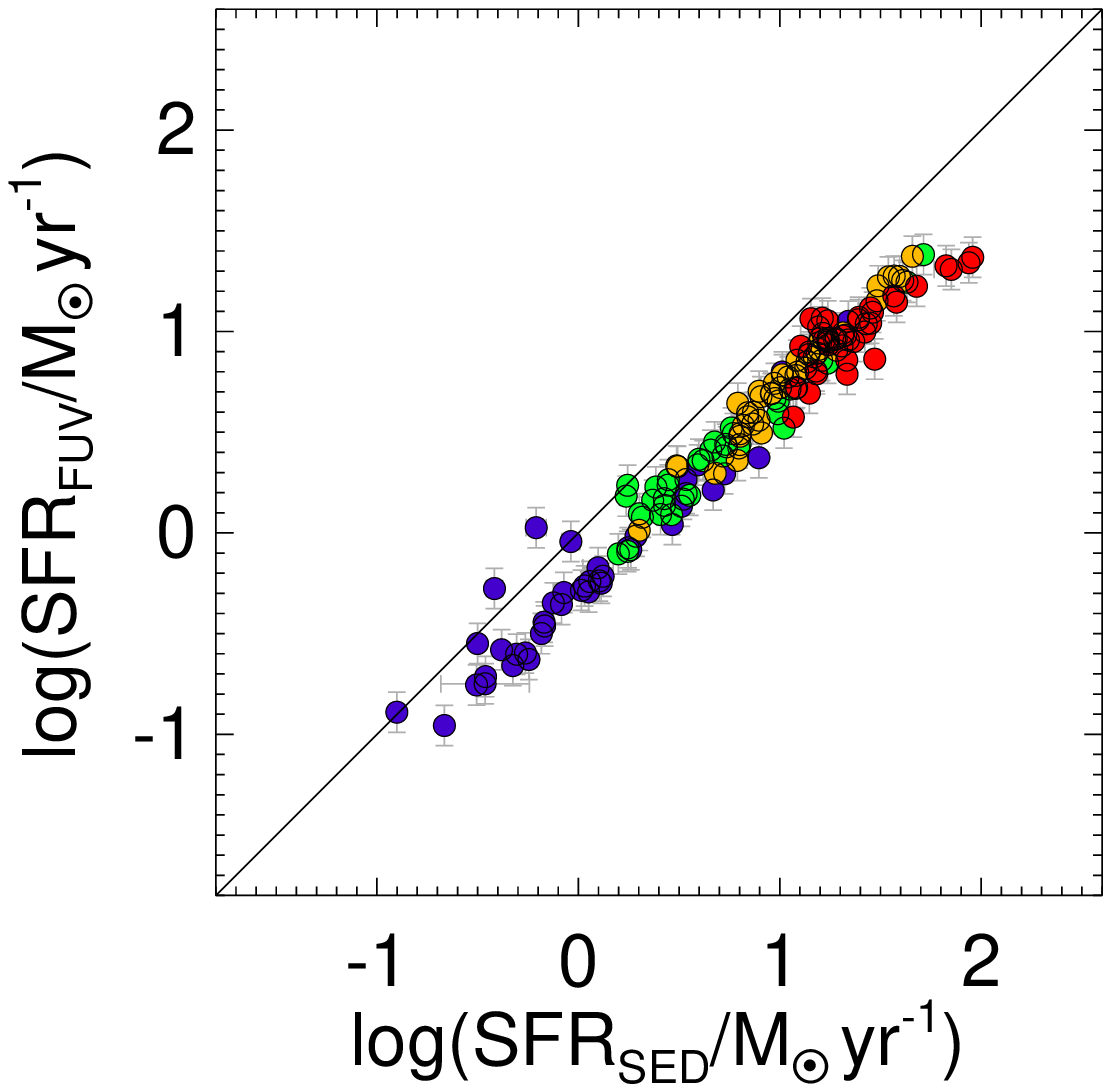}
   \caption{{Comparison between star formation rates derived from the 
   SED fitting (SFR$_{\rm SED}$), the extinction-corrected \ha\ luminosity 
   (SFR$_{\rm H\alpha}$), and the rest-frame attenuation-corrected FUV luminosity 
   (SFR$_{\rm FUV}$) for the sample of EELGs. The solid lines indicate the 
   one-to-one relation.}
              }
         \label{sfr_uv}
   \end{figure*}

Using dust-corrected FUV luminosities we also derive FUV-based SFRs 
using the calibration given by \citet{Kennicutt1998}, 
SFR$_{\rm FUV}$\,$=$\,1.4$\times$\,10$^{-28}$\,$L_{\rm FUV}$\,[erg\,s$^{-1}$\,Hz$^{-1}$], 
scaled down to a Chabrier IMF. 
{In Fig.~\ref{sfr_uv} we compare the SFRs derived 
from the SED fitting and those from \ha\ and FUV after dust-attenuation corrections.
Even though the scatter is relatively large, both SFR$_{H\alpha}$ and SFR$_{\rm FUV}$ 
are in excellent agreement.}
  
Since SFR$_{\rm FUV}$ traces massive star formation over a longer time scale than \ha\ 
\citep[typically up to a factor of 10,][]{Kennicutt2012}, this would imply that these 
galaxies are experiencing a very recent and probably the first major star 
formation episode in the last several hundreds million years. 
We note, however, that compared with the SFR derived from the SED fitting both 
SFR$_{H\alpha}$ and SFR$_{\rm FUV}$ are systematically lower by $\sim$0.2 dex, 
as shown in Fig.~\ref{sfr_uv}. 
{This offset} can be {understood as due to some of the} 
assumptions involved in the derivation of the three SFR tracers. 
{In particular, one of the most critical ones is the dust 
attenuation correction.}
   
Recent work by \citet{Castellano2014} suggests that for young, 
low-metallicity galaxies at high redshift the SFRs derived from dust-corrected UV luminosities can be underestimated by up to a factor of 2-10 regardless of 
the assumed star formation history. 
Such a discrepancy is due to the solar metallicity implied by the usual $\beta_{\rm UV}$-A$_{\rm FUV}$ conversion factor.
Since our EELGs are characterized by their strongly subsolar metallicities, 
their dust attenuations derived through the Meurer formula might be 
systematically lower than the true ones and, therefore, the derived 
SFR$_{\rm FUV}$ might be underestimated. 
{In our study} this hypothesis is supported by the fact that the 
SFR derived from the SED (which takes into account the metallicity of the 
galaxy to choose the best-fit model) is always systematically offset to higher 
values by $\sim$\,0.2 dex. 
In agreement with the results {of} \citet{Castellano2014} if this offset is 
only due to the dust attenuation correction it would imply that Meurer's 
zeropoint should be corrected upwards by this quantity for typical EELGs, 
resulting in a median dust attenuation in the FUV of about $\sim$\,10-20\% higher.  

Although a rigorous analysis of the discrepancies between different SFR
indicators is far from the scope of this paper, we caution about using dust 
attenuation corrections for low-metallicity galaxies under the
assumption of models and calibrations which are valid for solar metallicity 
environments.  

\subsection{The gas-phase metallicity of EELGs}

Given the wide redshift range of the sample, the low S/N of some faint
emission lines, and the limited wavelength coverage of the VIMOS
spectra, the derivation of  gas-phase metallicity in our sample of
EELGs cannot be addressed using a unique methodology. 
Thus, we use four different methods to derive metallicities for 
149 out of 165 EELGs ($\sim$90\% of the sample) with reliable 
measurements for  the required set of lines imposed by these 
methods, as described below. 

{Metallicities and associated uncertainties for the EELGs 
are presented in Table~\ref{Tab1}, where we also indicate the method applied 
in each case.} 
Median values for both redshift bin and morphological type are presented in 
Table~\ref{Tab2}. 

{In Figure~\ref{histo_OH} we show} the histogram 
distribution {of metallicities for the EELGs.}
{They} span a wide range of subsolar values (12$+$log(O/H)$=$\,7.3-8.5), 
with a median value {of} 12$+$log(O/H)$= 8.16$ ($\sim$\,0.18\,$Z_{\odot}$). 
{Moreover,} we do not observe a trend {in metallicity} with 
redshift. {These results are in good agreement with the typical values 
found for local star-forming galaxies \citep[e.g., H{\sc ii} galaxies and BCDs; ][]{Terlevich1991,Kniazev2004}. }

\subsubsection{Metallicity derived through the direct method}

The direct method (also known as the  $t_e$-method) is the most accurate method 
for deriving the oxygen abundance in star-forming galaxies 
\citep[e.g.,][]{Hagele08}. 
It is based on the previous determination of the electron temperature 
of the gas, using the intensity ratio of nebular-to-auroral emission lines 
(\emph{{\em e.g.,}} [\oiii]$\lambda\lambda$\,4959,5007 and [\oiii]$\lambda$\,4363) 
and the relative intensity of the strongest nebular emission lines to a 
hydrogen recombination line. 
Since for the EELGs we do not have a direct estimation of the [\oii] 
electron temperature, they have been derived using the model-dependent 
relation between the [\oiii] and [\oii] electron temperatures, $t_e$\,[\oiii] 
and $t_e$\,[\oii], proposed by \citet{Perez-Montero2003}, which 
takes into account the dependence of $t_e$\,[\oii] on the electron density. 
Then, following the expressions in \citet{Perez-Montero2009}, 
O$^+$ and O$^{2+}$ have been calculated using $t_e$\,[\oiii] 
and $t_e$\,[\oii] and the relative intensities of the corresponding bright 
emission lines, namely [\oii]$\lambda$\,3727 and [\oiii]$\lambda$\,4363, 
plus [\oiii]$\lambda\lambda$\,4959,5007.  
Finally, O$^+$ and O$^{2+}$ have been combined to estimate the total abundance 
of oxygen relative to hydrogen, O/H.

Following the direct method we have derived ionic abundances in 26 
purely star-forming EELGs ($\sim$16\%) uniformly distributed in
redshift {from $z\sim$\,0.45}\footnote{{The lower limit in redshift is due to the blue cut-off of the VIMOS grism used for the zCOSMOS bright survey, which precludes the observation of [\oii]$\lambda$3727 and [\oiii]$\lambda$4363 at lower redshift.}} and with reliable measurements 
(S/N\,$>$\,2) of all the involved emission lines, including [\oiii]$\lambda$4363. 
{ We} find direct metallicities spanning a large range of 
values, 12$+\log$(O/H)$=$7.5-8.4. 
\begin{table*}[t!]
\caption{{Derived properties of EELGs in zCOSMOS}}
\label{Tab1}
\centering
\begin{tabular}{lccccccccc}
\noalign{\smallskip}
\hline\hline
\noalign{\smallskip}
zCOSMOS\,ID & MT & $M_{B}$ & log $L_{\rm FUV}$ & $r_{50}$ & log M$_*$ & log SFR$_{H\alpha, H\beta}$ &  c(H$\beta$) & 12\,$+$\,log(O/H) & Method \\
  & & {\it mag} & $L_{\odot}$ & {\it kpc} & $M_{\odot}$ &
 M$_{\odot}$yr$^{-1}$ &  &  \\
(1) & (2) & (3) & (4) & (5) & (6) & (7) & (8) & (9) & (10)  \\
\hline
\hline
\noalign{\smallskip}
\vspace{0.15mm}
  700882&T &   -18.9   &10.23&     1.02    &8.66$\pm$ 0.12 &  0.45$\pm$  0.18& 0.28$\pm$0.18$^b$ &  ...  &  ...        \\
  701051&T &   -18.06  &9.88 &     1.48    &8.32$\pm$ 0.11 &  0.38$\pm$  0.03& 0.59$\pm$0.04$^a$ &  ...  &  ...        \\
  701741&... &   -18.94  &10.15&        ... &8.41$\pm$ 0.05 &  0.8 $\pm$  0.11& 
  0.12$\pm$0.11$^b$  &   7.46 $\pm$ 0.15   &  $T_{e}$     \\
  800984&T &   -19.04  &10.5 &     1.54    &8.76$\pm$ 0.01 &  1.03$\pm$  0.07& 0.36$\pm$0.07$^c$ &   7.95        $\pm$ 0.07   &  $T_{e}$       \\
  801094&... &   -19.27  &10.29& ...   &8.77$\pm$ 0.17 &  -0.14$\pm$0.13&
  0.17$\pm$0.13$^b$ &   8.14   $\pm$ 0.1    &  $R23$          \\
  802275&M &   -19.93  &10.7 &     2.22    &9.14$\pm$ 0.06 &  1.13$\pm$  0.08 & 0.54$\pm$0.08$^c$  &    8.18     $\pm$ 0.06   &  $R23$         \\
  803226&C &   -19.5   &10.13&     1.72    &8.86$\pm$ 0.19 &  0.41$\pm$  0.07& 0.20$\pm$0.07$^b$ &   7.87        $\pm$ 0.1    &  $T_{e}$       \\
  803892&C &   -18.96  &9.96 &     1.43    &8.55$\pm$ 0.23 &  0.19$\pm$  0.09&0.24$\pm$0.09$^b$ &  ...    &  ...    \\
  804130&M &   -20.08  &9.98 &     2.21    &9.01$\pm$ 0.07 &  0.33$\pm$  0.16& 0.45$\pm$0.09$^c$ &   8.53        $\pm$ 0.12   &  $N2$          \\
  804791&C &    -20.65  &11.05&    2.02    &9.47$\pm$ 0.05 &  0.51$\pm$  0.06& 
  0.05$\pm$0.06$^b$ &   8.29    $\pm$ 0.16   &  $R23$         \\
\noalign{\smallskip}                                                         
\hline                                                                          
\hline
\end{tabular}                                                                   
\begin{list}{}{}
\item Columns: 
(1) zCOSMOS identification number; 
(2) {Morphological type: ($R$)\,Round/Nucleated, ($C$)\,Clumpy/Chain, 
($T$)\,Cometary/Tadpole, ($M$) Merger/Interacting; }
(3) Rest-frame absolute B-band magnitude. {Median 1$\sigma$ uncertainties are $\sim$\,0.07 mag}; 
(4) Rest-frame, dust-corrected FUV luminosity. {Median 1$\sigma$ uncertainties are $\sim$\,0.11 dex};
(5) Circularized effective radius. {Median 1$\sigma$ uncertainties are $\sim$\,10\%}; 
(6) Stellar mass {from SED fitting} {(Chabrier (2003) IMF)};
(7) Star formation rate from \ha\ or \hb\ luminosity {(Chabrier (2003) IMF)};
(8) {Reddening constant derived from $(a)$ \ha/\hb\ or $(b)$ \hg/\hb\ ratios, whenever possible, or $(c)$ the SED best-fitting for those galaxies where $(a)$ and $(b)$   cannot be measured or where they produce a negative extinction (i.e., \ha/\hb$<$\,2.82 or \hg/\hb$<$\,0.47, assuming Case B recombination with  $T_e=$\,2$\times$10$^4$K, $n_e=$100\,cm$^{-3}$)}.
(9) Gas-phase metallicity. {The uncertainties quoted for 12$+$log(O/H) only take 
into account the propagation of errors from the emission line flux measurements;} 
(10) Method used for metallicity derivation ({see text for details}).
(The entire version of this table for the full sample of EELGs is available {\it \emph{online}}).
\end{list}
\end{table*}

\subsubsection{Metallicity of EELGs without [\oii] measurements: the $t_e$\,[\oiii]-Z calibration}

\begin{figure*}[t!]
   \begin{center}
   \includegraphics[angle=0,width=6.5cm]{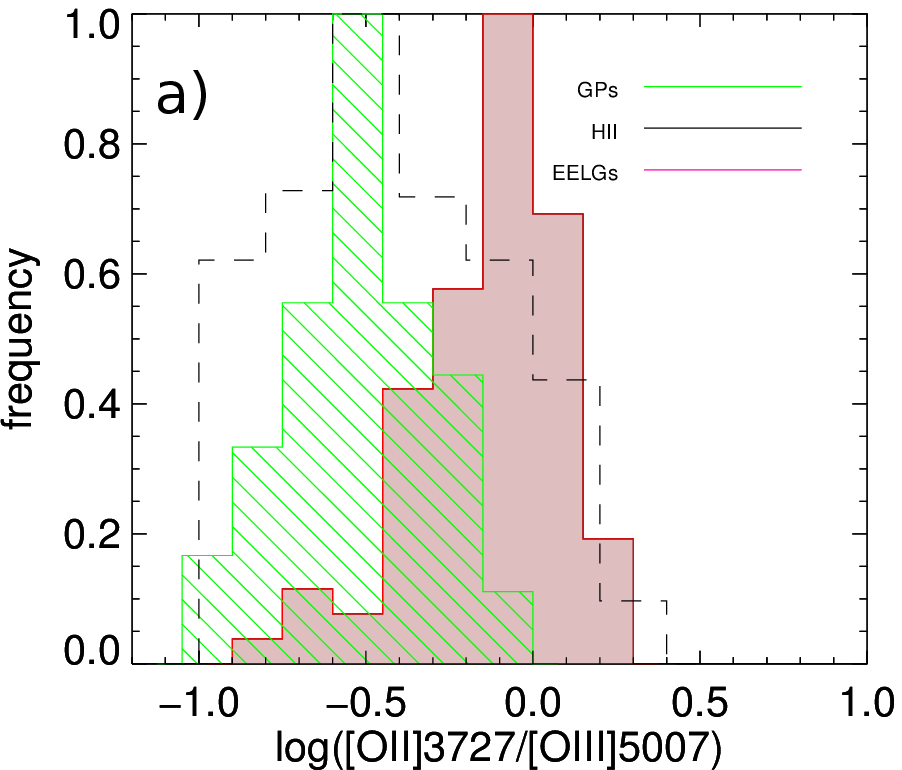}\hspace{3 mm}
   \includegraphics[angle=0,width=6.3cm]{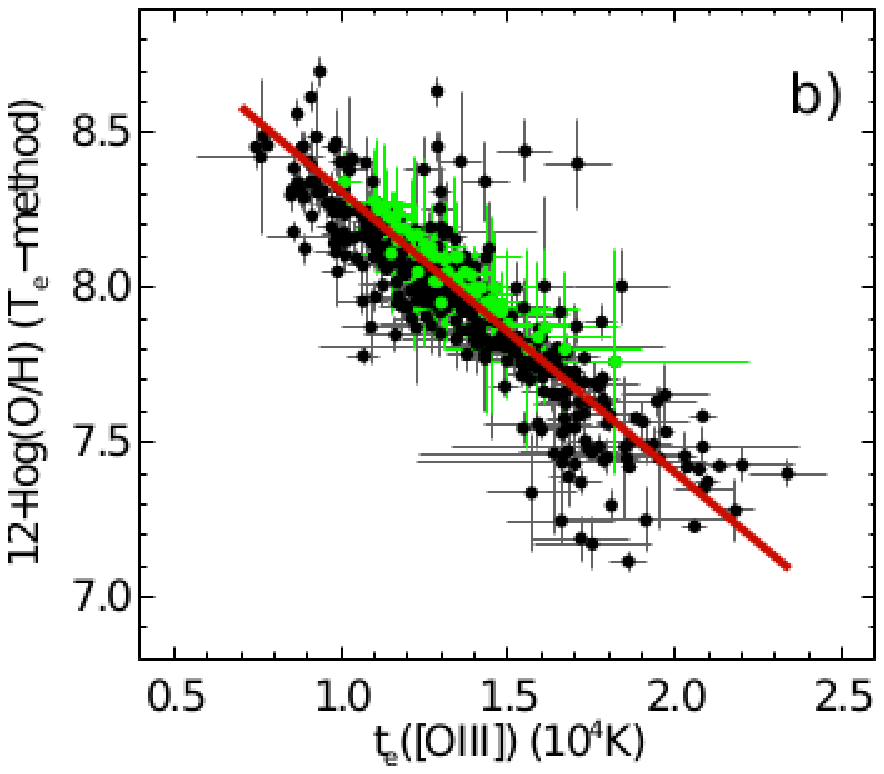}\\\vspace{2. mm}
   \includegraphics[angle=0,width=6.3cm]{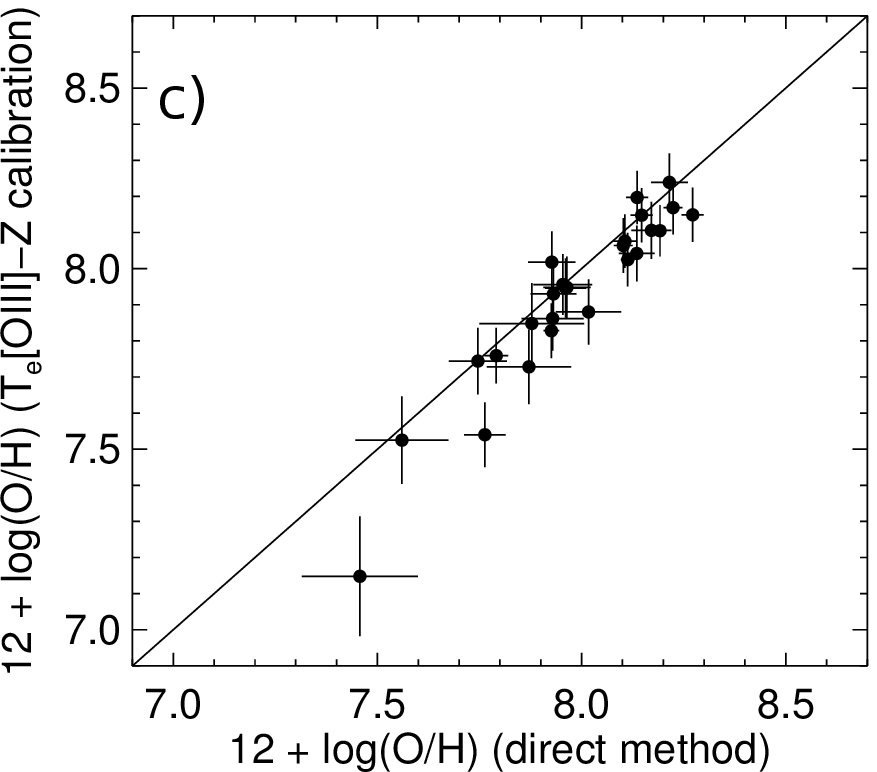}\hspace{3 mm}
   \includegraphics[angle=0,width=6.6cm]{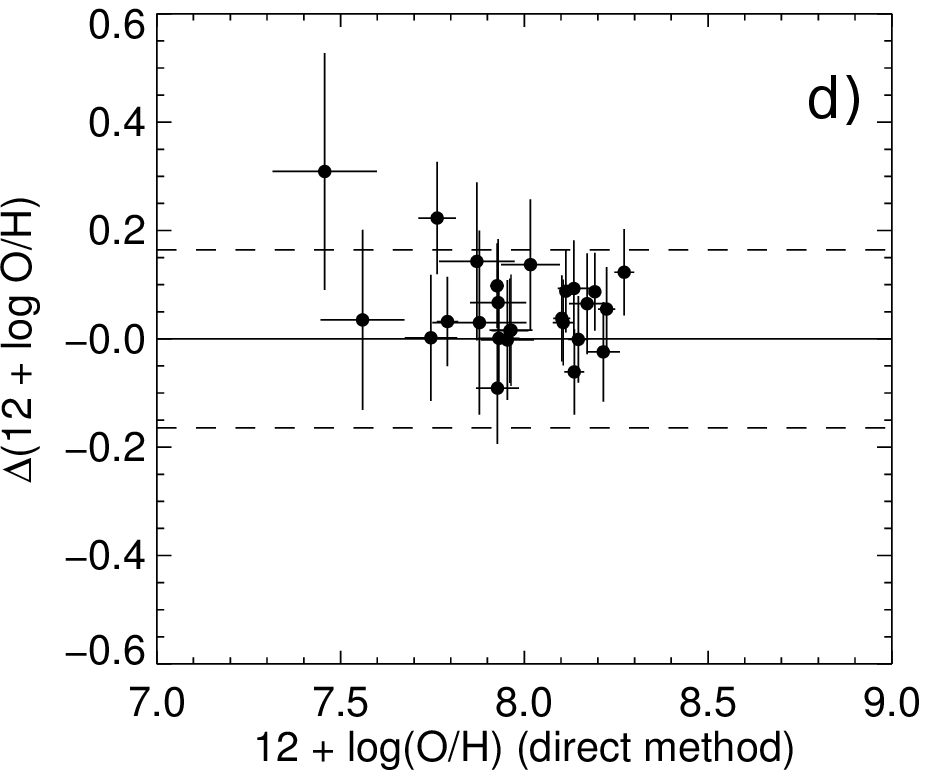}
   \end{center}   
  \caption{{{\it (a)} Histogram distributions of the ionization parameter, estimated 
  with the [\oii]/[\oiii] ratio, for EELGs, {\it green peas} (GPs), and H{\sc II}
  galaxies. 
  {\it (b)} Relation between the oxygen abundance derived from the direct method and 
  the [\oiii] electron temperature, $t_e$[\oiii], for the sample of giant H{\sc II} 
  regions and nearby H{\sc II} galaxies from 
  \citet[][\textit{black points}]{Perez-Montero2009} 
  and the sample of GPs from \citet[][\textit{green points}]{Amorin2010}. 
  The red line indicates the best-fit to the data shown in Eq.\,1. 
  {\it (c)} Comparison between the metallicity derived using the direct
  method and from the $t_e$[\oiii]$- Z$ calibration shown in Eq.\,1
  and their residuals {\it (d)}. 
  Dashed lines in {\it (d)} indicate the 2$\sigma$ limits. }
              }
         \label{Te-Z}
   \end{figure*}

For EELGs with reliable measurements of the [\oiii]\,4363,4959,5007 lines, 
but without [\oii] line measurement, i.e., those where [\oii] lines lie out of the VIMOS spectral range, we cannot derive O$^+$, so we do not have a direct measurement of the metallicity. 
In our sample, 19 galaxies ($\sim$12\%) at $z < 0.48$ { fall into this category}. 
For these galaxies, however, we can derive a semi-direct 
metallicity taking advantage of the tight relation between 
$t_e$\,[\oiii] and $Z$ expected for high-excitation environments --
like those present in EELGs -- from both observations and models 
\citep[e.g.,][]{Masegosa1994,LopezSanchez2012}. 
Thus, we calibrate the relation between $t_e$\,[\oiii] and $Z$ for
EELGs (dubbed hereafter  the $t_e$\,[\oiii]$-Z$ calibration) using a
combination of two independent datasets. 
We use metallicities derived using the direct method for 
both giant H{\sc II} regions and H{\sc II} galaxies from
\citet{Perez-Montero2009} and the sample of {\it green pea} galaxies 
from \citet{Amorin2010}. 
In order to avoid a strong dependence on the ionising 
parameter, the calibration was restricted to those objects with [\oii]/[\oiii] 
ratios in the range covered by the EELG sample (Figure~\ref{Te-Z}($a$)). 

In Figure~\ref{Te-Z}($b$) we show the $t_e$\,[\oiii]$-Z$ calibration, which 
produces the  expression 
\begin{equation}
\metal = 9.22\,(\pm 0.03) - 0.89\,(\pm 0.02)\,t_{e}[{\rm O\,III}] 
,\end{equation}
where $t_{e}$[\oiii] is the [\oiii] electron temperature {calculated from the [\oiii] 
  ($\lambda$4959$+$$\lambda$5007)$/$$\lambda$4363 ratio} in units of 10$^4$\,K.
Uncertainties in the $t_e$\,[\oiii]-Z calibration translate into 
uncertainties of $\la$0.15 dex (1$\sigma$) in metallicity.
As a consistency check, we have applied the $t_e$[\oiii]$- Z$ calibration to 
those EELGs with metallicities derived using the direct method. 

In Figure~\ref{Te-Z}($c$) we show that the two metallicity estimates based
on the electron temperature are in good agreement over the wide range of metallicity 
covered by our sample, with deviations broadly consistent within the errors. 
Only a small shift of $\sim$0.1-0.2 dex is noticed in some cases, especially at  
lower metallicities, i.e., 12$+$log(O/H)$\la 7.7$. 
In this range, the $t_e$[\oiii]$-Z$ calibration has a lower statistical 
significance and the scatter is large, probably because of an increased 
sensitivity of $t_e$[\oiii] on the ionization parameter. 
Therefore, {in extremely metal-poor 
galaxies Eq.\,1 should be applied with caution}.  
Moreover, we {emphasize} that the calibration presented in Eq.\,1 is 
well suited for determining reliable oxygen abundances only in objects with 
similar (i.e., high) ionization conditions to those shown by the EELGs. 
Thus, the $t_e$[\oiii]$-Z$ calibration may appear an alternative for 
other samples of gas-rich galaxies with similar ionization conditions where the 
limitation in the spectral coverage does not allow a proper determination of 
the [\oii] flux.
 
\subsubsection{Metallicity from strong-line methods}
%
\begin{sidewaystable*}[ht!]
\begin{minipage}[t][90mm]{0.85\textwidth}
\caption{{Median properties of zCOSMOS EELGs.}}
\label{Tab2}
\centering
\begin{tabular}{lcccccccccc}
\noalign{\smallskip}
\hline\hline
\noalign{\smallskip}
Subsample & $N$ & $z$ & $M_{\rm B}$ & $c$(H$_{\beta}$) & $r_{50}$ & $\beta_{\rm UV}$ & $\log L_{\rm FUV}$ & 
SFR$_{\rm H\alpha, H\beta}$ & $\log$ M$_*$ & 12$+\log$(O/H) \\[3pt]
  &     &     &   &     &   kpc   &  &  $L_{\odot}$   &
$M_{\odot}$\,yr$^{-1}$ & M$_{\odot}$ &   \\[3pt]
 & (2) & (3) & (4) & (5) & (6) & (7) & (8) & (9) & (10) & (11) \\
\noalign{\smallskip}
\hline
\noalign{\smallskip}
1 ($0.11 \leq z \leq 0.30$)&44&0.19&-17.6 (0.9)&0.31 (0.19)&1.6 (1.4)&-1.44 (0.51)&9.6 (0.5)&0.6 (2.6)& 8.02 (0.47)& 8.10 (0.23)\\[3pt]
2 ($0.30 < z \leq 0.50$)&43 & 0.40&-19.3 (0.7)&0.27 (0.18)& 1.3 (1.0)&-1.66 (0.32)&10.2 (0.3)&2.2 (2.4)& 8.70 (0.29)& 8.07 (0.22)\\[3pt]
3 ($0.50 < z \leq 0.70$)&42 & 0.61&-20.1 (0.7)&0.27 (0.16)& 1.5 (0.6)&-1.61 (0.31)&10.6 (0.3)&6.3 (2.1)& 9.02 (0.31)& 8.16 (0.19)\\[3pt]
4 ($0.70 < z \leq 0.93$)&36 & 0.82&-20.7 (0.4)&0.31 (0.15)& 1.0 (0.6)&-1.68 (0.43)&10.8 (0.2)&12.7 (1.7)& 9.21 (0.24)& 8.21 (0.17) \\[3pt]
\noalign{\smallskip}
\hline
\noalign{\smallskip}
All-SF                  &165& 0.48(0.23)&-19.5 (1.4)&0.27 (0.17)& 1.3 (1.0)&-1.61 (0.41)&10.4 (0.6)&3.9 (3.9)& 8.79 (1.07) & 8.16 (0.21) \\[3pt]
All-SF in groups        &48 & 0.35(0.22)&-18.9 (1.5)&0.31 (0.17)& 1.4 (1.1)&-1.55 (0.43)&10.3 (0.6)&2.4 (4.1)& 8.72 (1.08) & 8.15 (0.23) \\[3pt]
\noalign{\smallskip}                                                         
\hline  
\noalign{\smallskip}                                                         
{Round/Nucleated} & 30&0.44 (0.25)&-19.1 (1.4)&0.36 (0.20)&1.0 (0.3)&-1.54 (0.42)&10.2 (0.5)&3.2 (3.6)&8.54 (1.08)& 8.13 (0.26) \\[3pt]
{Clumpy/Chain} & 60 &0.48 (0.23)&-19.5 (1.5)&0.31 (0.15)&1.5 (0.7)&-1.61 (0.43)&10.3 (0.6)&3.9 (4.3)&8.76 (1.07)& 8.14 (0.21) \\[3pt]
{Cometary/Tadpole} & 27 &0.46 (0.20)&-19.2 (1.4)&0.27 (0.18)&1.5 (0.7)&-1.70 (0.36)&10.3 (0.6)&3.3 (3.7)&8.70 (1.07)& 8.08 (0.21) \\[3pt]
Merger/Interacting & 48&0.55 (0.23)&-20.2 (1.1)&0.27 (0.17)&1.5 (1.5)&-1.55 (0.41)&10.6 (0.5)&6.2 (3.7)&9.05 (1.05)& 8.18 (0.16) \\[3pt]
NL-AGN            & 18&0.78 (0.16)&-20.7 (0.8)&0.62 (0.45)&1.2 (0.3)&-1.32 (0.83)&10.9 (0.5)&17.2 (4.4)&9.36 (1.10)& $-$ \\[3pt]
\noalign{\smallskip}                                                         
\hline                                                                          
\hline
\end{tabular}                                                                   
\begin{list}{}{}
\item 
Columns: (2) Number of galaxies. (3) to (11) {Median (and 1\,$\sigma$ dispersion) values for: redshift, rest-frame $B-$band absolute magnitude, reddening, circularized half-light radius, UV spectral slope, dust-corrected {\it FUV} luminosity, dust-corrected star formation rate, stellar mass, and metallicity. }
\end{list}
\end{minipage}
\end{sidewaystable*}

The  two  methods explained in Sects. 3.3.1 and 3.3.2 are based on the determination of the electron
temperature and cannot be applied to the whole sample of EELGs either 
because at certain redshifts the [\oiii]$\lambda$\,4363 line is not 
included in the VIMOS spectral range, or because this line is too weak.
One alternative for the derivation of the gas-phase metallicity 
in these galaxies is the use of the so-called strong-line methods. 
{They} are based on the direct calibration of the relative 
intensity of the strongest collisionally excited emission-lines with 
grids of photoionization models or samples of objects with an accurate 
determination of the oxygen abundance, or both. 

There is a wide variety of strong-line methods used in the literature, 
which usually give different results depending on the metallicity 
range, ionization conditions, and available line ratios \citep[see][for an extensive 
discussion]{Kewley2008}. 
In order to derive metallicities consistent with those derived from  
the $t_e$-based methods, here we use three different empirical  
calibrations based on {a} sample of nearby objects with accurate determinations of 
12$+$log(O/H) via the direct ($t_e$) method.
We use the calibration proposed by \citet{Perez-Montero2009} based on
the \NDOS\ parameter, defined as the ratio of [\nii]\,$\lambda$\,6584 to \ha\ by 
\citet{Storchi-Bergmann1994}, and used by \citet{Denicolo2002} as a metallicity proxy. 
This method is our choice for EELGs at $z \la 0.30$, where [\nii] and \ha\ 
are included in the VIMOS spectra but [\oiii] is not.
Although this relation does not present any dependence  on 
reddening correction or on flux calibration uncertainties, \NDOS\ depends on 
the ionization parameter, the equivalent effective temperature 
{of the ionising cluster,} and the nitrogen-to-oxygen ratio \citep{Perez-Montero2005}. 
Taking these effects into account, in EELGs -- which show  homogeneous 
excitation properties (see Figure~\ref{Te-Z}) -- the overall uncertainty is 
$\sim$0.2 dex across their entire metallicity range. 

For EELGs at $z > 0.48$ where the [\oiii] auroral line is too
weak to derive $t_e$[\oiii] and \NDOS\ cannot be applied (i.e.,
\ha$+$[\nii] lie out of the spectral range), metallicity  
is derived using the $R_{23}$ parameter. 
This parameter is defined as the ratio between the sum of 
[\oii]$\lambda$\,3737 and [\oiii]$\lambda\lambda$\,4959,5007 
to \hb\ fluxes \citep{Pagel1979}. 
The main {drawback} of $R_{23}$ is its {degeneracy} with $Z$. 
Moreover, $R_{23}$ has a strong dependence on the ionization parameter and 
effective temperature. To minimize this dependence we use the 
calibration proposed by \citet{Kobulnicky2003}, based on the photoionization 
models from \citet{McGaugh1991}, which includes additional terms as a function of 
[\oii]/[\oiii], {a proxy of the ionization parameter}. 
Most of the EELGs are located in the {turnover region} of the $R_{23}$ 
calibration. Therefore, for galaxies with a difference of $\la$0.3 dex between 
the metallicity provided by the upper and lower branches we adopt a mean value as
the final metallicity. For galaxies where this difference is higher we adopt 
the upper branch of the $R_{23}-Z$ relation because we interpret the {weakness}  
of the [\oiii] auroral line in high S/N spectra as a possible {indication of} 
high (12$+$log(O/H)$\ga$8.2) metallicities. 
Nonetheless, we note that we do not find EELGs with 12$+$log(O/H)$>$8.5.

Finally, it is worth mentioning that important differences may arise {when using} 
different strong-line methods and/or different calibrations 
\citep[see, e.g.,][]{Kewley2008,Perez-Montero2014}. 
Here, metallicities derived using the \NDOS\ {calibration of} 
\citet{Perez-Montero2009} are consistent with those 
derived from the direct method. However, the adopted calibration of 
$R_{23}$ is not based on galaxies with direct metallicities but on grids of 
photoionization models, which may produce a systematic bias. 
To overcome these differences, we follow \citet{Perez-Montero2013} and 
convert the metallicities derived from $R_{23}$ to those derived from 
\NDOS\ using the linear relations described in \citet{Lamareille2006b}, 
which are based on models {of} \citet{Charlot2001}. 
This way, the adopted estimators find metallicities that are broadly 
consistent, within the uncertainties, with each other.   

\subsection{ Diverse morphologies of EELGs}
\subsubsection{{Visual classification}}
%
   \begin{figure*}[ht!]
   \centering
   \includegraphics[angle=0,width=6.75cm]{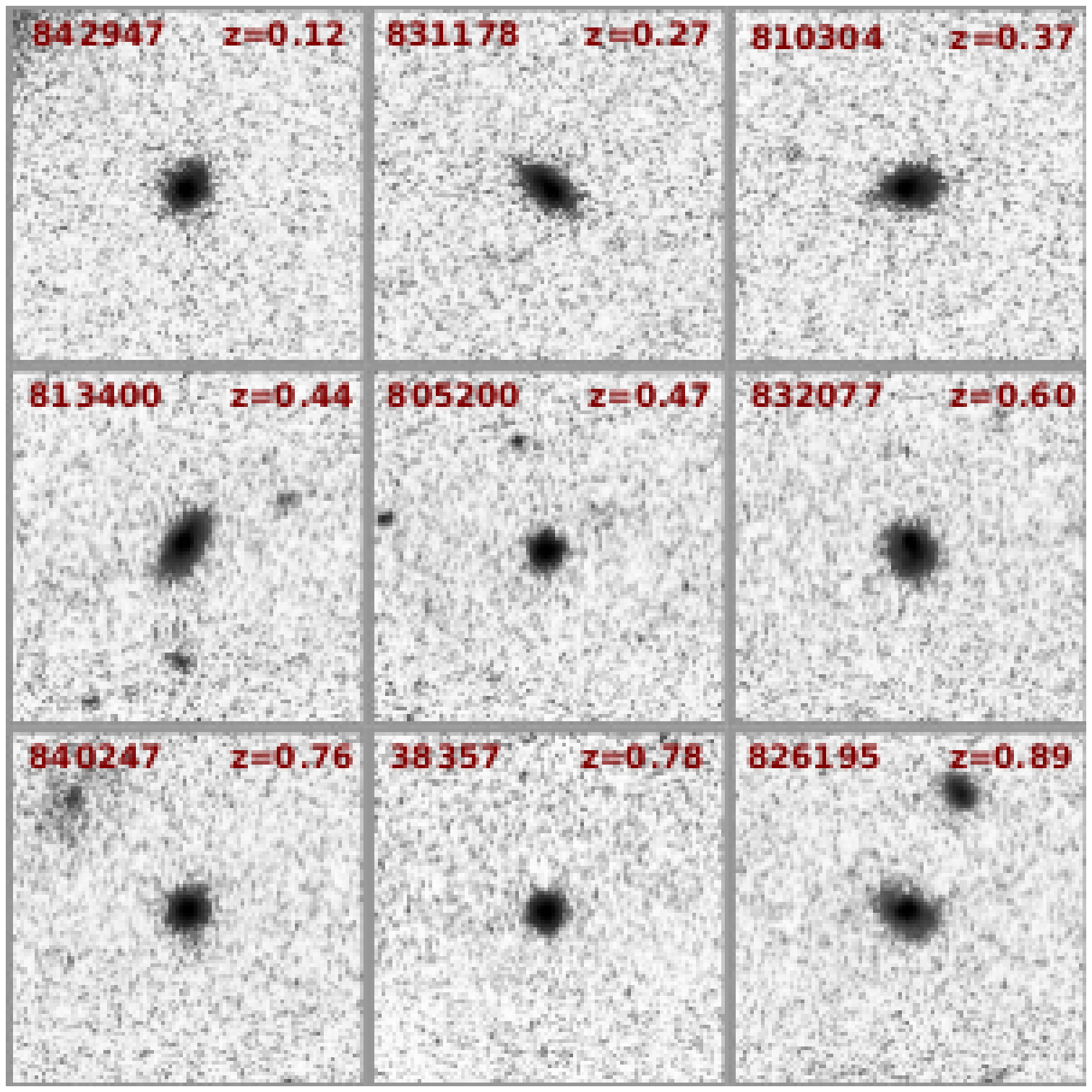}\hspace{2mm}
\includegraphics[angle=0,width=6.75cm]{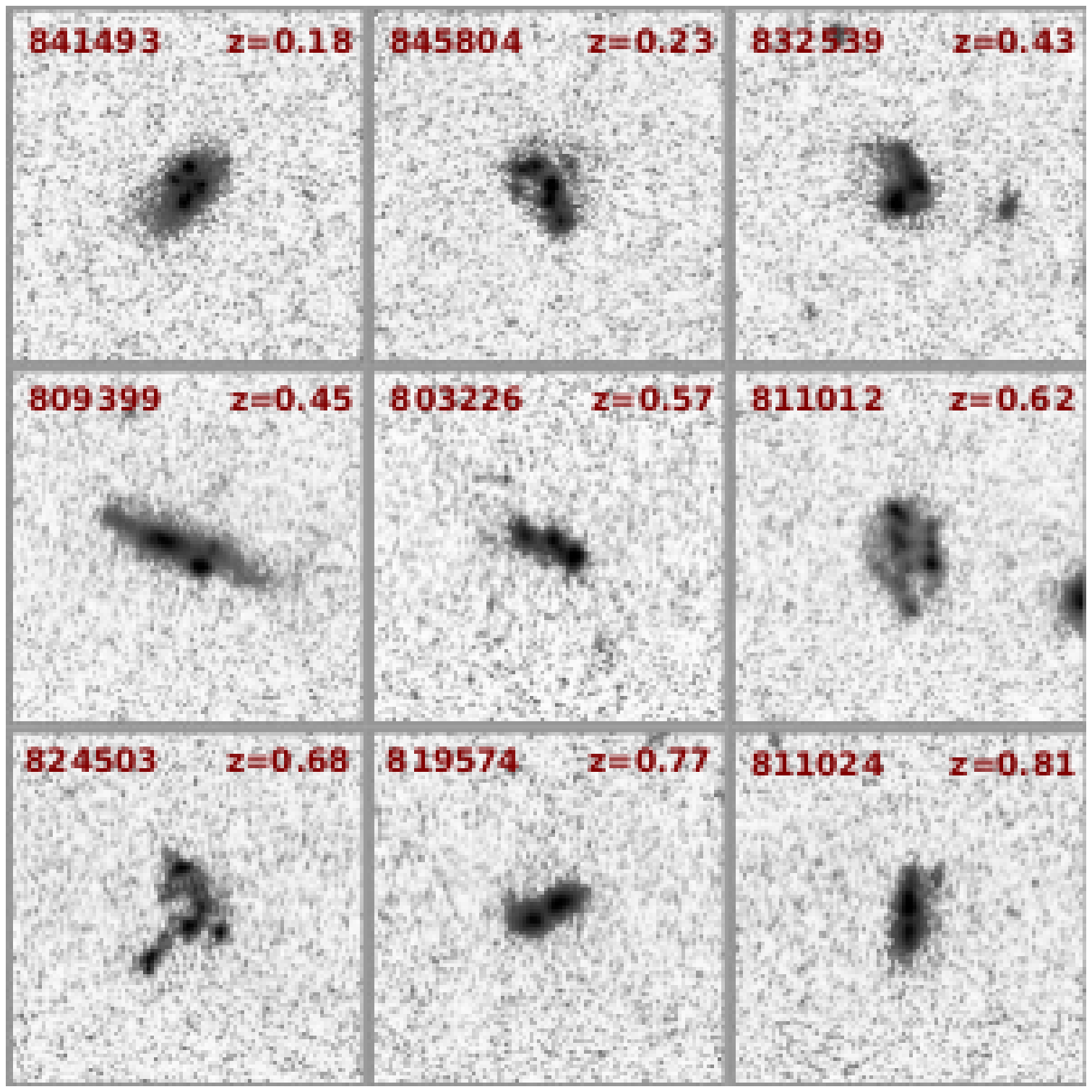}\\\vspace{2mm}
\includegraphics[angle=0,width=6.75cm]{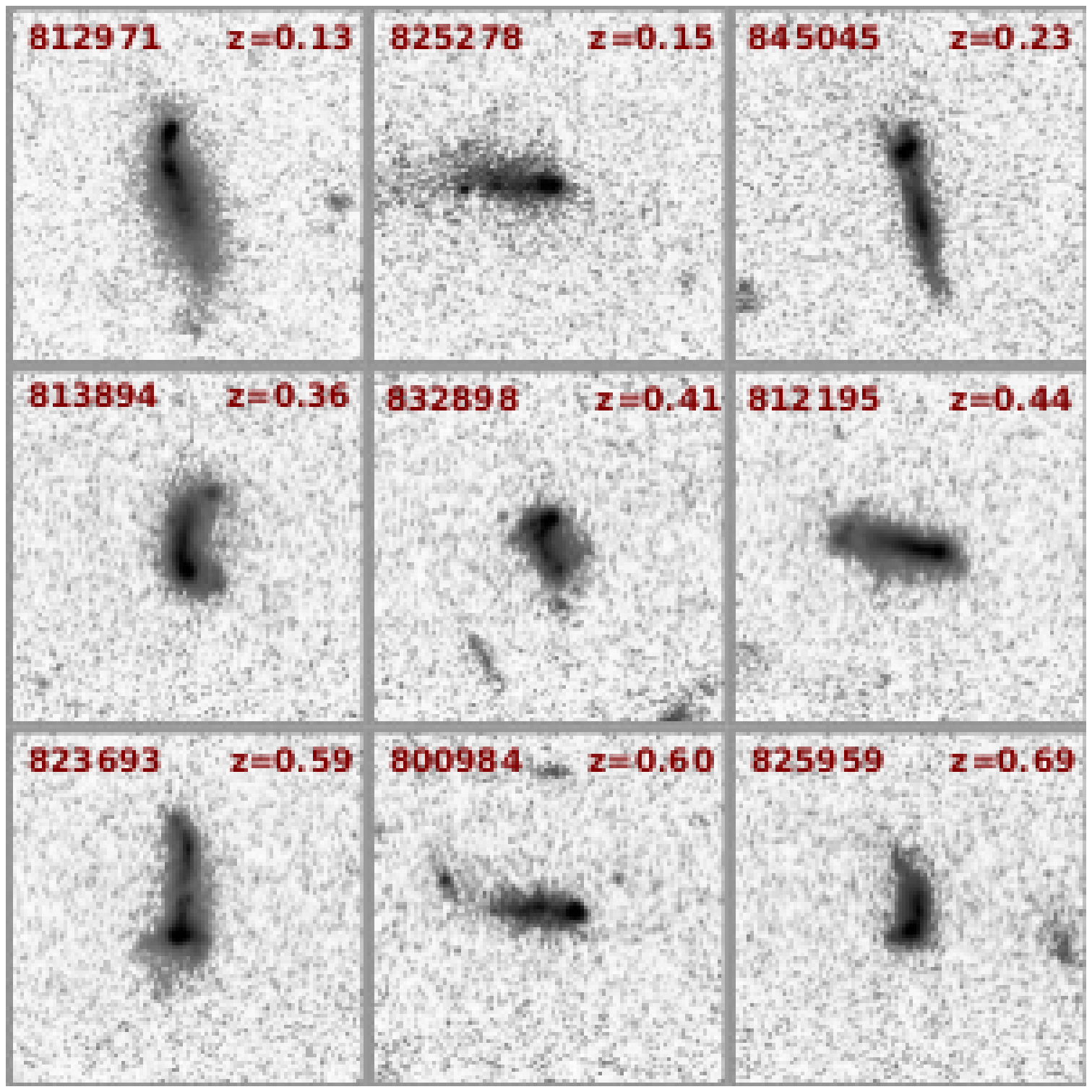}\hspace{2mm}
\includegraphics[angle=0,width=6.75cm]{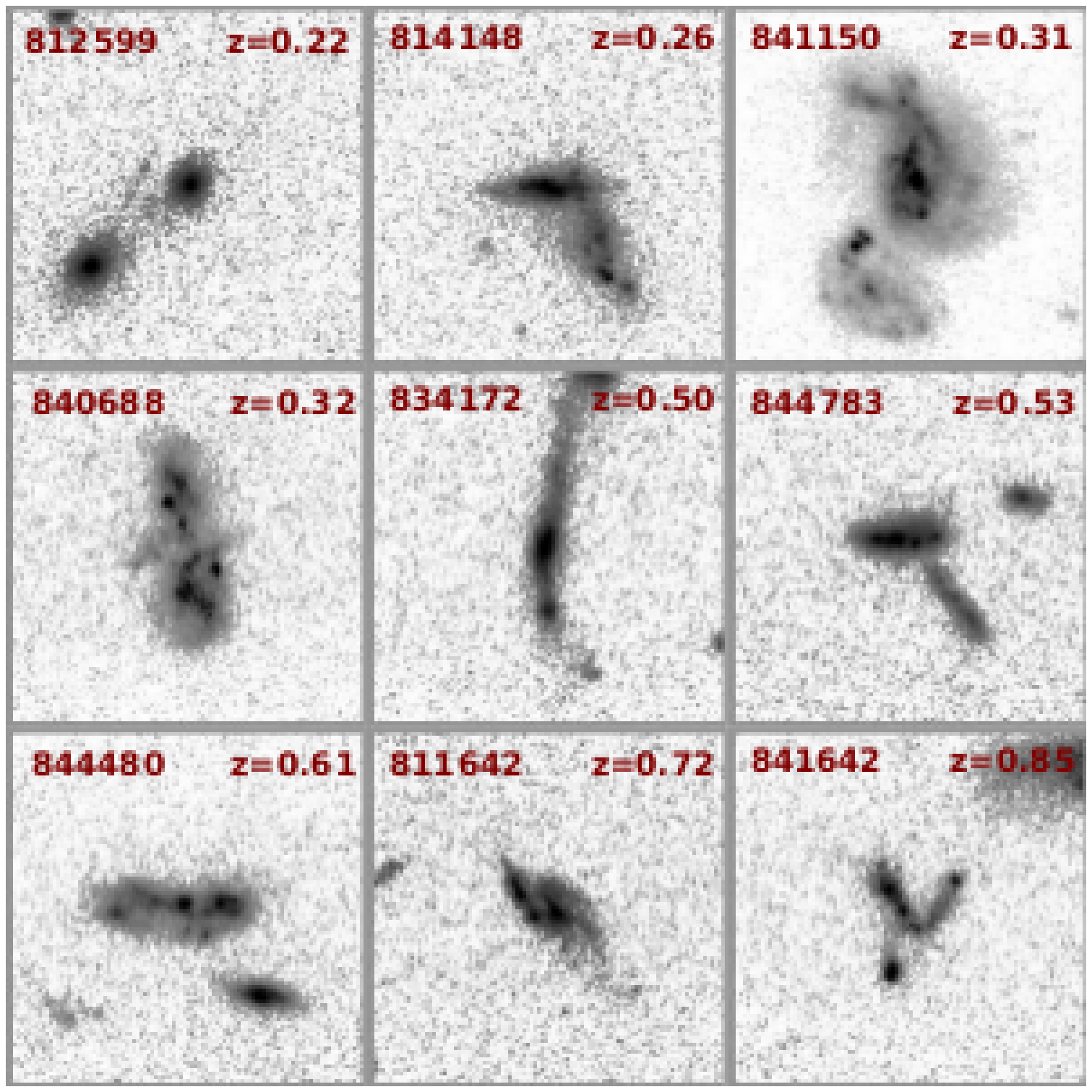}
     \caption{Morphology of star-forming EELGs. Examples of
       {round/nucleated (\textit{upper left}), clumpy/chain
       (\textit{upper right}), cometary/tadpole (\textit{bottom left}), and
       merger/interacting} (\textit{bottom right}) morphologies from their
       HST/ACS $I-$band ($F814W$) images. 
       The images are oriented to the north-east and are 6$"$ on a side. 
       The redshift for each galaxy is indicated in {the} labels. 
              }
         \label{morphologies}
   \end{figure*}

Our first approach to studying the morphological properties of the EELG
sample {was to perform} a visual classification using the {available} 
HST/ACS $F814W-$band images {from COSMOS}. 
{We excluded from the analysis five EELGs that have not been imaged with 
ACS and are nearly unresolved in 
ground-based images. }
{Inspired by} classical visual classifications of
  BCDs \citep[e.g.,][]{Cairos2001}  
 we distinguish here four major {morphological} classes 
 {of EELGs} according to the distribution and shape of {their} high- 
 and low-surface-brightness components: 
\begin{enumerate}
 \item {\it {Round/Nucleated}}: galaxies showing one bright 
 star-forming knot {embedded} in a nearly symmetric low surface 
 brightness envelope {and galaxies with point-like/unresolved appearance}. 
 About 18\% of EELGs are in this class.
\item {\it {Clumpy/Chain}}: galaxies with two or more high-surface-brightness knots spread out over a diffuse or asymmetric low-surface-brightness 
component. These EELGs represent $\sim$\,37\% of the sample galaxies.
 \item {\it Cometary/Tadpole}: galaxies with {head-tail shape}, where 
 {a main} bright star-forming {clump} is located {at the 
 head and a low-surface-brightness tail is off to one side}. 
 About 16\% of the EELGs are in this class.
\item {\it Merger/Interacting}: galaxies with a distorted low-surface-brightness component, features {potentially} associated with past 
  or current interaction with very close companions, e.g., tails, bridges, etc. 
  These EELGs are about 29\% of the sample.
\end{enumerate}
{Morphological classes for the sample of EELGs are given in 
Table~\ref{Tab1}, while several illustrative examples are shown in 
Fig.~\ref{morphologies}.}
None of the {above} classes {appears} to be biased to any 
redshift bin;  all of them show nearly the same median redshift {(Table~\ref{Tab2})}.  
{We note} that EELGs belonging to the last three {morphological} 
classes can be simply considered as ``irregular'' galaxies.  
Although there might be an inevitable overlap between them 
(e.g., some clumpy/chain or cometary/tadpole EELGs 
may be also interpreted in terms of interactions) 
a more detailed description of both the distribution of the star-forming 
regions and the shape of the underlying diffuse component in broadband 
images remains interesting.  
In particular, it may be useful to study the possible mechanisms 
responsible for the origin of the starburst and chemical enrichment  
in these galaxies 
\citep[e.g.,][]{Papaderos2008,Filho2013,SanchezAlmeida2013,SanchezAlmeida2014} 
and to compare them with galaxies of similar morphologies identified 
at higher redshifts \citep[e.g.,][]{Elmegreen2012,Elmegreen2013}. 

\subsubsection{{Quantitative analysis}}

In addition to our visual classification, we {adopt a quantitative classification 
scheme based on different non-parametric diagnostics of galaxy structures 
\citep[e.g.,][]{Abraham1996,Conselice2003,Lotz2004,Huertas-Company2008}. 
In short, we use high-resolution imaging in the $F814W$ filter 
from the HST/ACS and the fully automated method developed by \citet{Tasca2009} to 
derive standard morphological parameters\footnote{\label{note1}See \citet{Tasca2009} and \citet{Huertas-Company2008} for details on the definition of morphological parameters 
and on the algorithms adopted for the morphological classification.}, such as half-light 
radius $R_{50}$ and axial ratio $q=b/a$, concentration index ($C$), asymmetry ($A$), 
and the Gini coefficient ($G$). }

\begin{figure}[t!]
   \centering
   \includegraphics[angle=0,width=4.35cm]{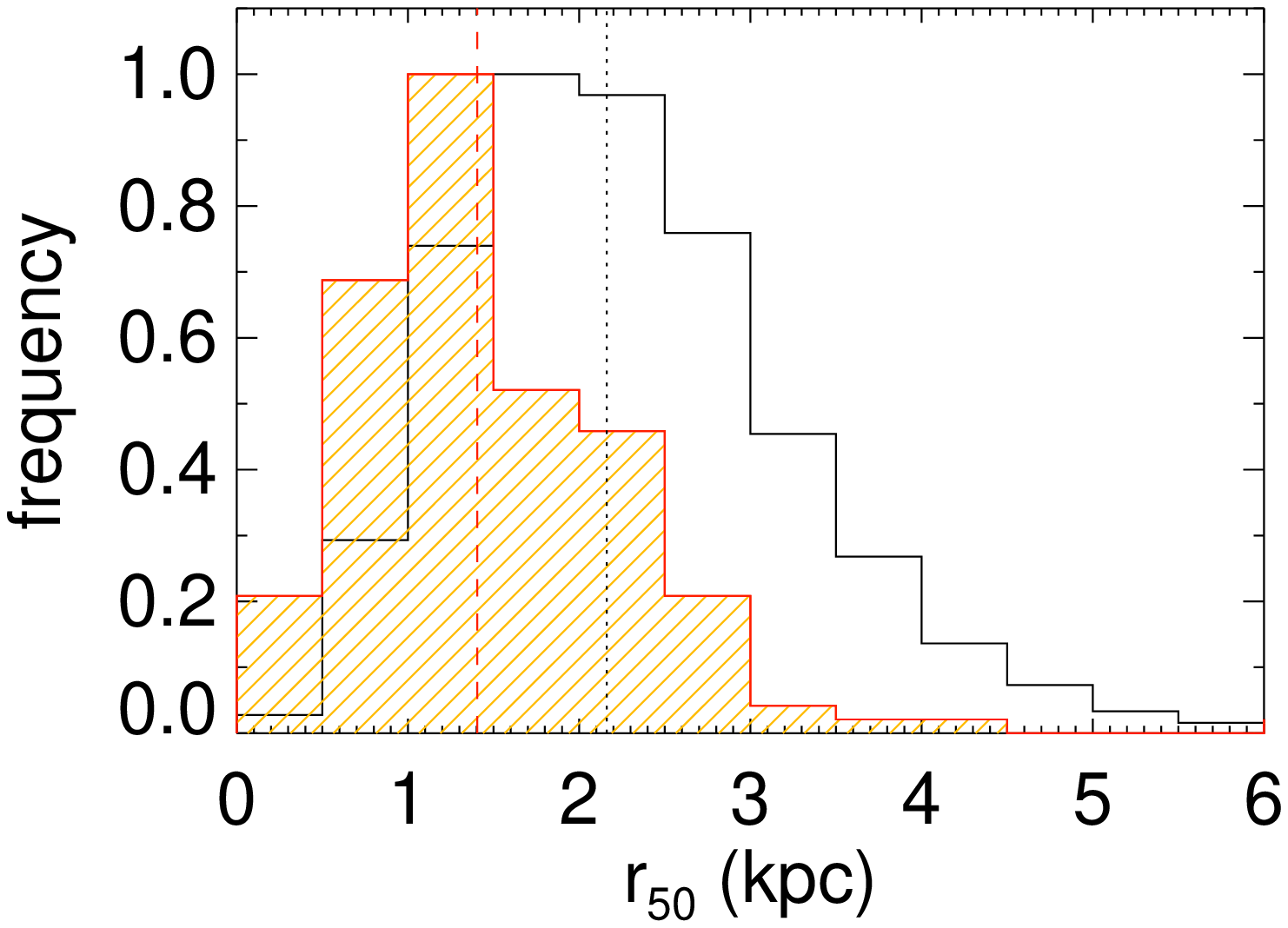}\hspace{0.5mm}
   \includegraphics[angle=0,width=4.5cm]{FIG8b.eps}\\\vspace{2.mm}
   \includegraphics[angle=0,width=4.40cm]{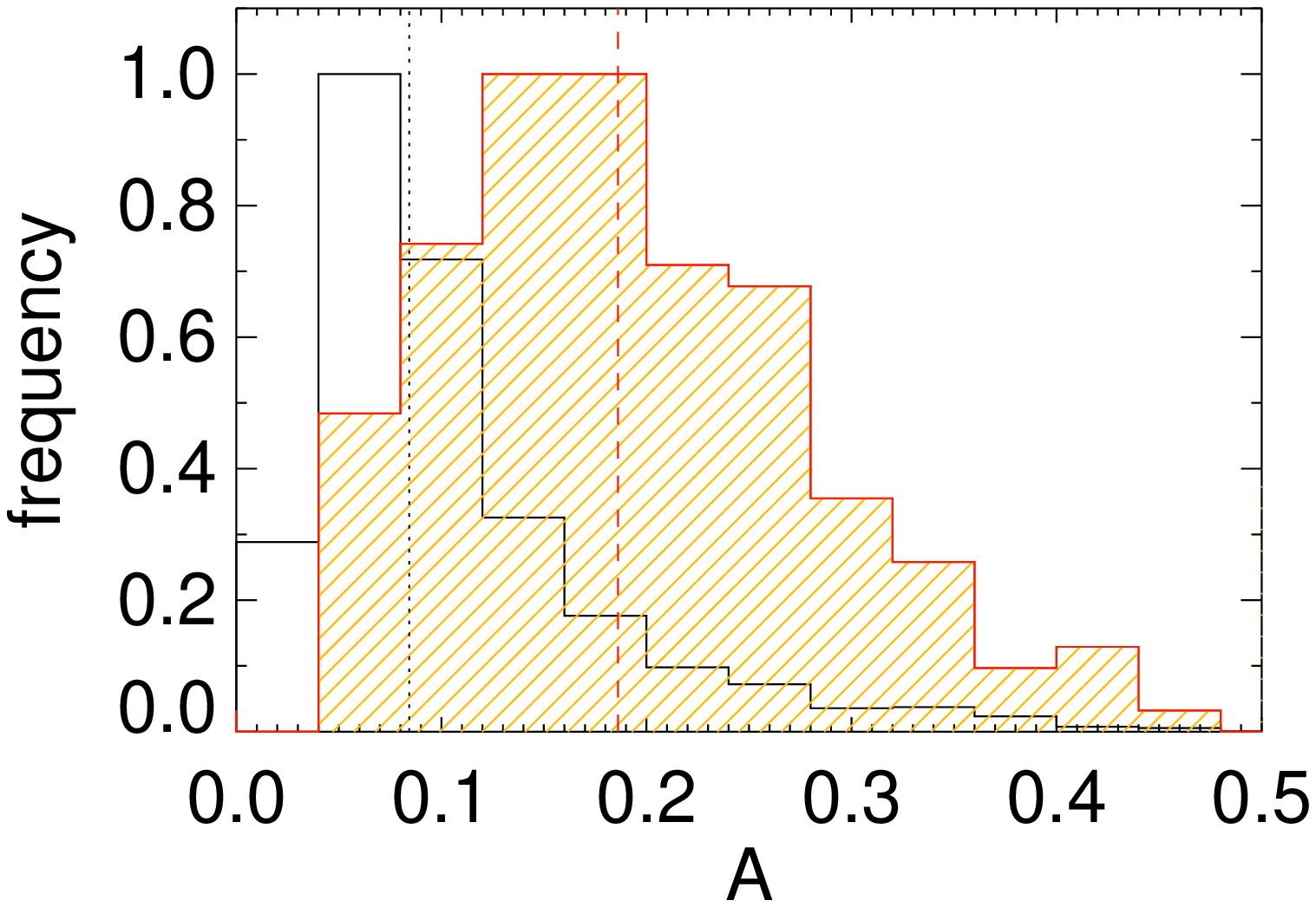}
   \includegraphics[angle=0,width=4.30cm]{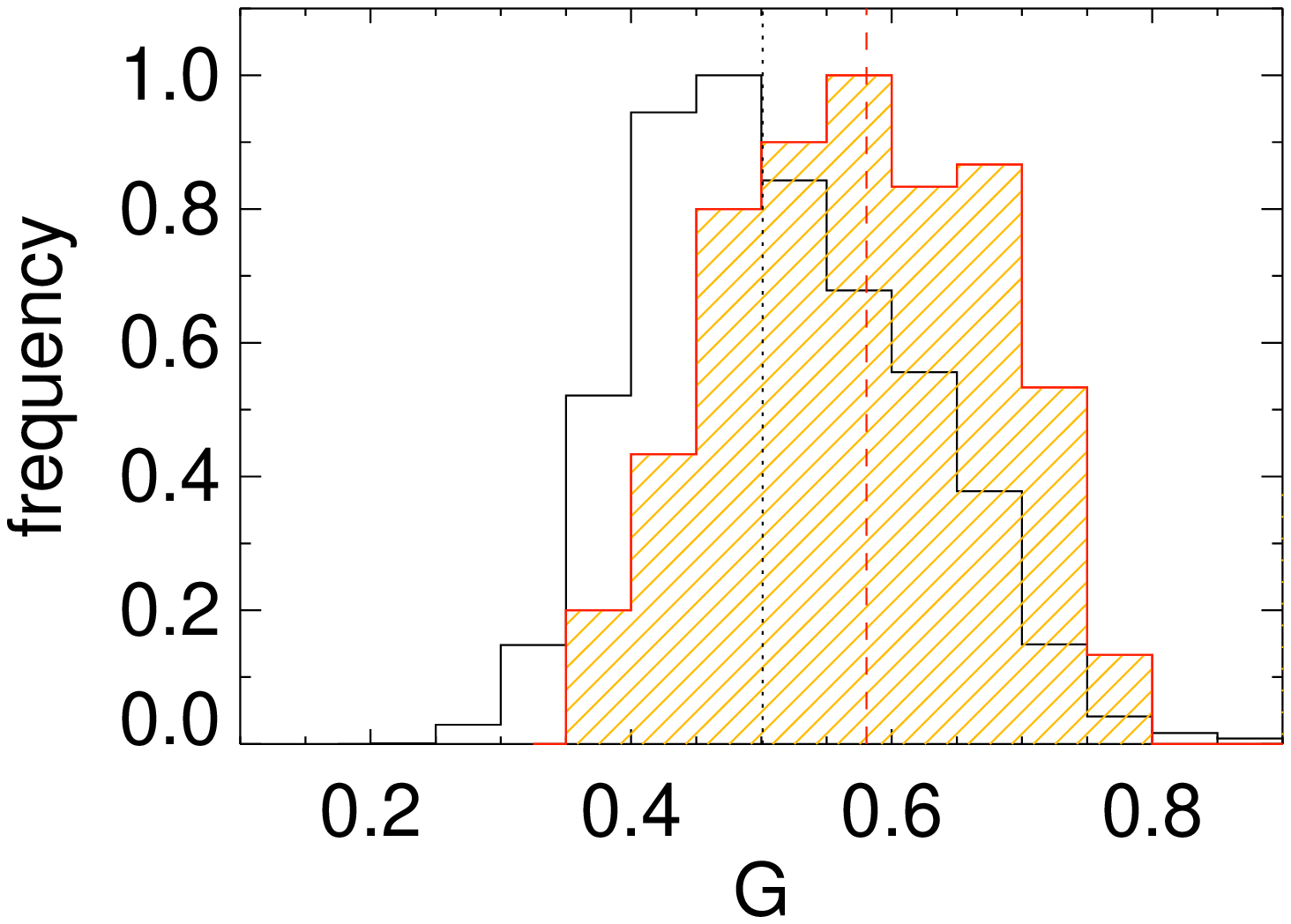}\\\vspace{2.mm}
   \includegraphics[angle=0,width=4.25cm]{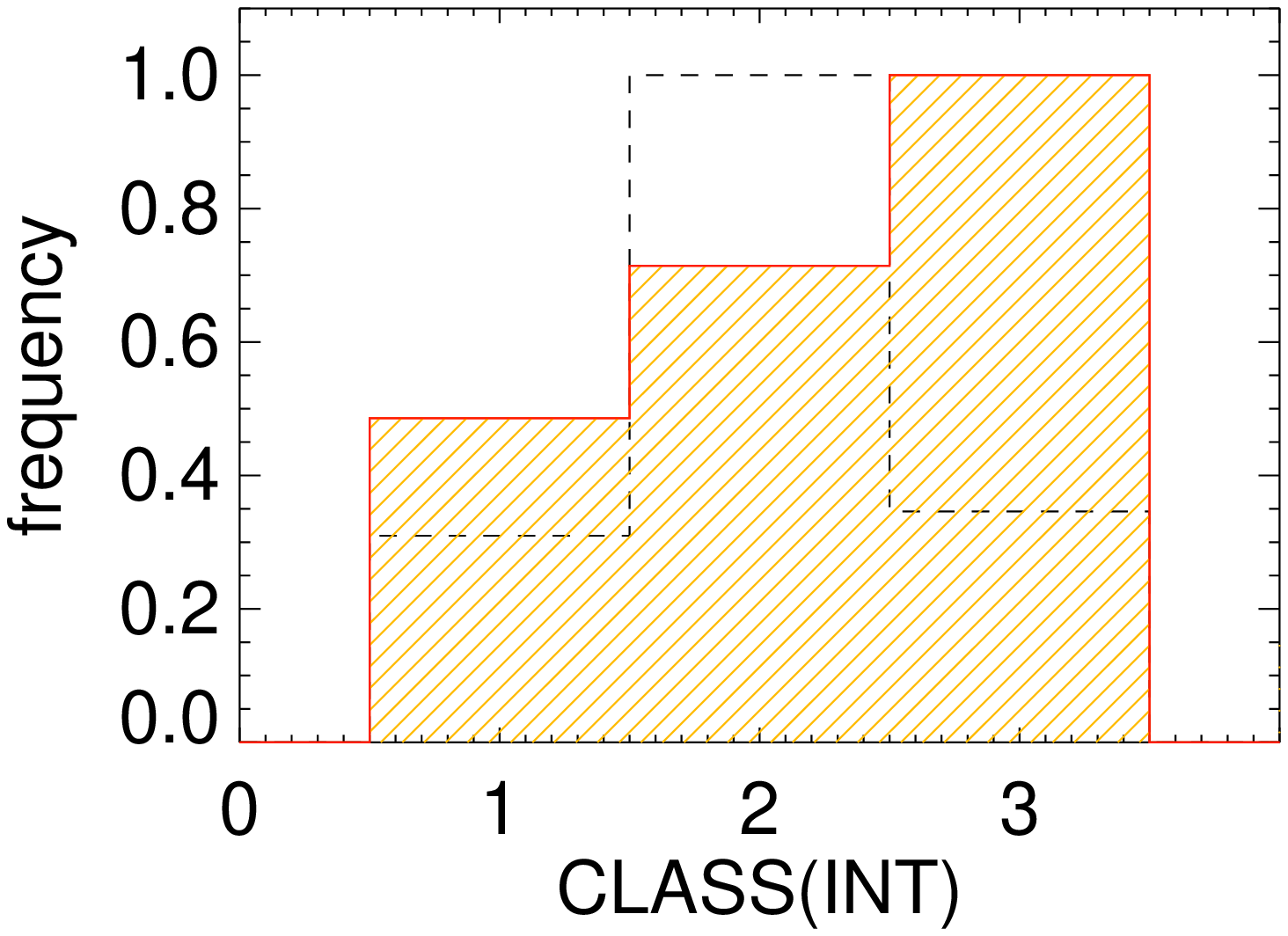}\hspace{2mm}
   \includegraphics[angle=0,width=4.35cm]{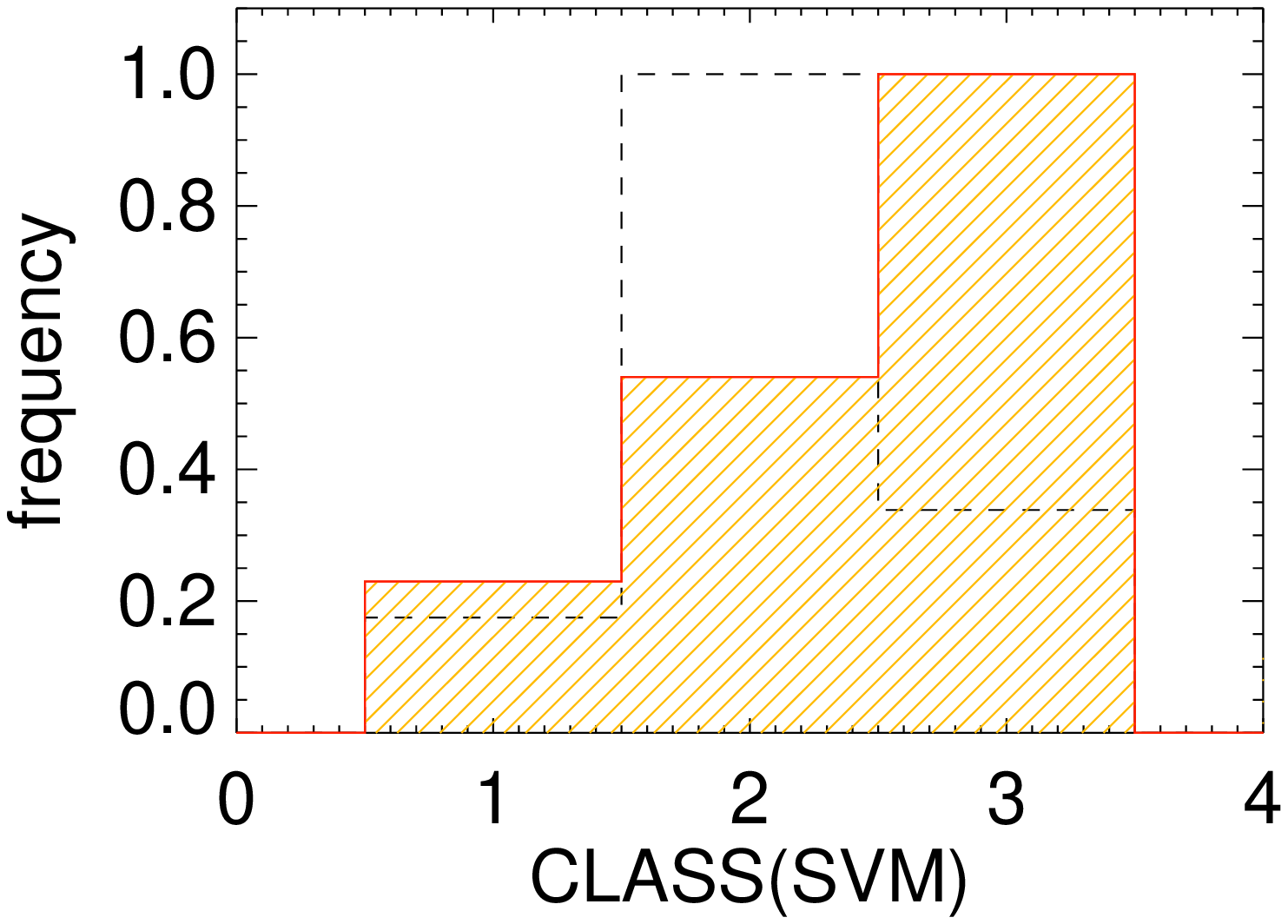}
    \caption{{Normalized distribution of morphological parameters and classes for 
    EELG and SFG-20k samples. 
    Panels show, from the upper left to the bottom right side,  
    the half-light radius ($r_{50}$), the concentration index (C), the asymmetry (A), and the
    Gini (G) coefficients, and the morphological classes INT and SVM proposed by 
    \citet{Tasca2009}, where $1 = elliptical$, $2=spiral$, and $3= irregular$. 
    Lines and colors are as in Fig.~\ref{histograms}. }}
         \label{morphology2}
   \end{figure}

{The above parameters are used simultaneously by two different optimized 
algorithms\textasciicircum7\ referred to as INT \citep{Tasca2009} and SVM 
\citep{Huertas-Company2008}, to define the boundaries between three predefined 
morphological classes, {\it early}, {\it spiral}, and {\it irregular}, in an 
automated and objective way}. 
In Fig.~\ref{morphology2} {(bottom panels)} we show the normalized 
distributions of SFGs and EELGs in zCOSMOS over these three classes for the 
two methods. We find that both INT and SVM algorithms provide similar results. 
While most SFGs in zCOSMOS are classified as {\it spiral}, the EELG sample 
contain a significantly higher fraction of galaxies classified as {\it irregular.}

We find a {qualitative} agreement between the results of 
the classification schemes shown in Fig.~\ref{morphology2} and 
our {visual} classification, in the sense that almost all the 
EELGs {in the} cometary/tadpole, clumpy/chain,
and merger/interacting {classes} are automatically classified 
as irregular or spiral, whereas most EELGs classified as
round/nucleated are classified as elliptical. 
Taken together, we conclude that at least $\sim$80\% of the sample 
presents non-axisymmetric morphologies.

In Figure~\ref{morphology2} we {also} show the normalized distribution of 
morphological parameters for the {EELG and SFG-20k} samples {in the 
same stellar mass range}. 
The EELGs are small systems, with half-light radii\footnote{{We have circularized 
the measured half-light radii $R_{50}$ as $r_{50}$$=$$R_{50}$\,$q^{0.5}$, where 
$q$ is the axial ratio $b/a$. Both quantities have been measured from the $I$-band
  ($F814W$) HST-ACS images.}} {$r_{50}$ in a range between 
  $\sim$\,0.3 and $\sim$\,4\,kpc, with a median value of 1.3\,kpc.} 
  Similarly to most galaxies in the {SFG-20k} sample, the $C$, $A$, and 
$G$ parameters for the EELGs {are} spread over a wide range of values. 
However, there is a clear tendency {for} the EELGs to show larger 
asymmetry and also higher concentration and Gini parameters than normal SFGs. 

\begin{figure}[t!]
   \centering
   \includegraphics[angle=0,width=8.5cm]{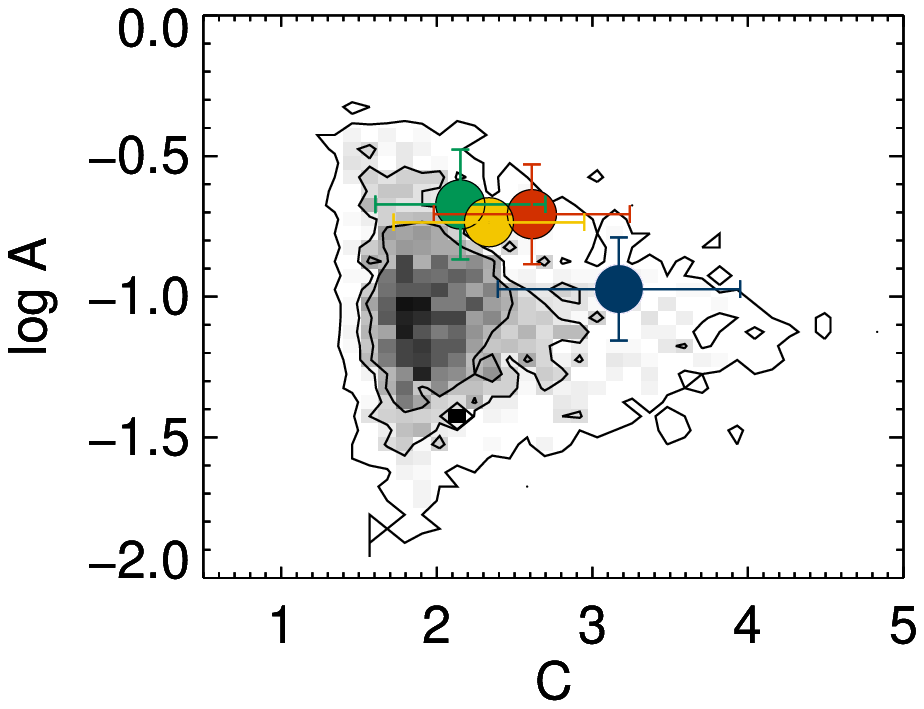} \\\vspace{5mm}
   \includegraphics[angle=0,width=8.5cm]{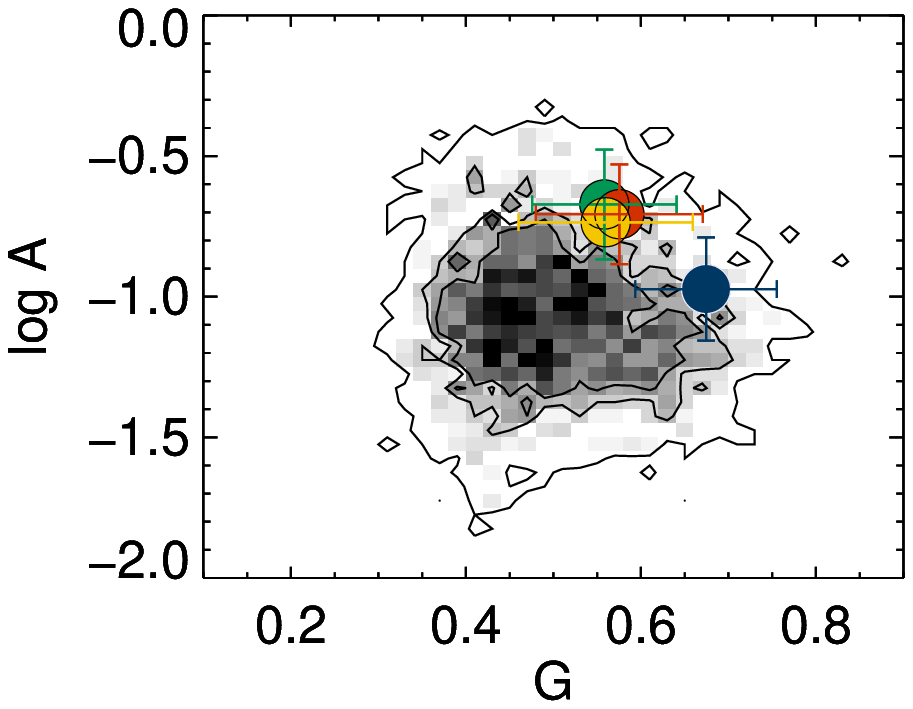}
\caption{{Asymmetry-concentration (\textit{top}) and asymmetry-Gini (\textit{bottom}) diagrams. The gray density plot show the location of the SFG-20k sample. The inner and outer contours enclose 50, 68 and 95\% of the sample, respectively. 
Median values and dispersions for the EELGs are indicated by large color circles and error bars. Yellow, red, green and blue colors indicate EELGs visually classified as "merger/interaction", "cometary/tadpole", "clumpy/chain" and "round/nucleated", respectively. }}  
\label{morphology3}
   \end{figure}

{The same conclusion arises from Fig.~\ref{morphology3}}, where 
we {show} the asymmetry-concentration and asymmetry-Gini 
diagrams for both zCOSMOS SFGs and EELGs. 
Especially at {low values of C}, the EELGs show larger 
{values of A compared to those of} normal SFGs at a given $C$ or $G$. 
From Fig.~\ref{morphology3} we also test the consistency between {the}  
visual and quantitative classifications. 
Those EELGs visually classified as round/nucleated show higher concentration 
and Gini parameters than those included in irregular classes. 
Figure~\ref{morphology3} {also highlights} the difficulty of 
distinguishing between galaxies with different irregular 
morphologies, such as cometary or clumpy galaxies on the basis 
of such quantitative diagrams only.

{Among our EELGs we do not find any clear correlation between the morphological properties and other galaxy-averaged properties such as redshift, absolute magnitudes, SFRs, stellar masses, extinction, or gas-phase metallicity. In particular, we do not find significant differences between the median properties of {rounded} and irregular EELGs (see Table~\ref{Tab2}). }
%

\subsection{The environment of EELGs}

We study the large-scale environment of the EELG sample using the zCOSMOS 20k group catalog \citep{Knobel2012}. This catalog includes about 16500 galaxies between 0.1\,$\la z \la$\,1, and contains 1498 groups in total, of which 192 have more than five members. Full details about the catalog can be found in \citet{Knobel2009,Knobel2012}. 

{We cross-match the group catalog and both the SFG-20k and the EELG samples. 
We find that $\sim 26$\% of the galaxies in the SFG-20k sample are in groups of two or more spectroscopic members.  
Similarly, we find 48 out of 165 EELGs ($\sim$\,29\%) {classified} as  group  members with a probability $\geq$\,50\% and $\geq$\,90\% in 46 and 34 of them, respectively. 
Out of these 48 galaxies, 27 EELGs belong to pairs of galaxies, 11 belong to triplets, and only 10 of them belong to groups of four or more spectroscopic members.}
The probability that these EELGs are the most massive {galaxies} of their group is $\leq$\,10\% for all but five EELGs, all of them in pairs.  
Only two galaxies in our sample, zCOSMOS\,ID\#823693 and ID\#823694, constitute on their own a spectroscopic pair of EELGs. 
The median properties of EELGs in groups are shown in Table~\ref{Tab2}. 

{We find that only $\sim$\,29\% of EELGs are in groups with one
  or more spectroscopic companions. 
Thus, we conclude that EELGs are located in relative isolation, in agreement 
with previous findings for local star-forming dwarf galaxies \citep[e.g.,][]{Vilchez1995,TellesTerlevich1995,Lee2000,Noeske2001,Pustilnik2001,Koulouridis2013}. 
It should be noted, however, that the fraction of EELGs members of groups is nearly the 
same as in the SFR-20k sample. Moreover, most of these groups show a non-negligible number 
of additional photometric members, which may constitute in most cases neighbors of lower luminosity, so the above numbers should be considered as lower limits.
Because of spectroscopic incompleteness we may miss, in these and in the remaining 70\% of the EELG sample, possible faint companions that are often seen projected closely to the EELGs.} 

{Faint neighbors can be important, for example, to evaluate  the role of interactions in the triggering of star formation. Some observational evidence shows that local star-forming 
dwarfs are usually found with low-surface-brightness companions \citep{Brosch2004,Sanchez-Janssen2013}. If these neighbors are located in the very close environment ($<<$\,1 Mpc) 
of the galaxies they may have an influence on the star formation triggering processes and subsequent evolution \citep{Pustilnik2001}. In the case of our EELGs this will be a topic for a future, more detailed investigation. 
 }

\begin{table*}[ht!]
\caption{The properties of very metal-poor EELGs in zCOSMOS.}
\label{Tab3}
\centering
\begin{tabular}{l c c c c c c c c c}
\noalign{\smallskip}
\hline\hline
\noalign{\smallskip}
zCOSMOS\,ID & $z$ & MT & $c$($H_{\beta}$) & $EW$(\hb) & $t_e$([\oiii]) & log(\oiii/\hb) & log([\nii]/\ha) & log(N/O) & 12$+\log$(O/H)  \\[3pt]
(1) & (2) & (3) & (4) & (5) & (6) & (7) & (8) & (9) & (10) \\
\noalign{\smallskip}
\hline
\noalign{\smallskip}
701741 &0.504 &2&0.12$\pm$0.05& 113$\pm$20 & 2.28$\pm$0.11 & 0.50$\pm$0.05&...& ... & 7.46$\pm$0.15$^{b}$ \\[3pt]
809215 &0.124 &2&0.20$\pm$0.07$^{a}$ &...&...&...& -1.79$\pm$0.18 & -1.83$\pm$0.15$^{e}$ & 7.65$\pm$0.07$^{d}$ \\[3pt]
825959 &0.690 &3  &0.00$\pm$0.08$^{a}$& 66$\pm$13 & 1.87$\pm$0.06 & 0.52$\pm$0.06&...& ... & 7.56$\pm$0.12$^{b}$ \\[3pt]
836108 &0.351 &3  &0.30$\pm$0.13& 74$\pm$16 & 1.92$\pm$0.04  & 0.77$\pm$0.04& ... & ... & 7.47$\pm$0.10$^{c}$ \\[3pt]
840051 &0.250 &1  &0.28$\pm$0.04& 84$\pm$10 &...& 0.61$\pm$0.02 & -1.75$\pm$0.11 & -2.00$\pm$0.26$^{e}$ & 7.69$\pm$0.08$^{d}$ \\[3pt]
840962 &0.121 &1  &0.31$\pm$0.06$^{a}$ & ... & ... & ... & -2.18$\pm$0.13 & -1.75$\pm$0.24$^{e}$ & 7.35$\pm$0.11$^{d}$ \\
\noalign{\smallskip}                                                         
\hline                                                                          
\hline
\end{tabular}                                                                   
\begin{list}{}{}
\item Columns: (1)zCOSMOS identification number; (2) Redshift; (3)
  Morphological type: (1)\,Regular, (2)\,Clumpy/Irregular,
  (3)\,Cometary/Tadpole; (4) Reddening constant from the \ha/\hb\
  ratio, {except} for those with the superscript $(a)$, which are taken
  from SED fitting; (5) \hb\ equivalent width in \AA; (6) [\oiii]
  electron temperature in units of 10$^4$K; (7) [\oiii]5007/\hb\
  ratio; (8) \NDOS\ parameter; (9) Nitrogen-to-oxygen ratio; (10)
  Gas-phase metallicity; Method used to derive metallicity and N/O:
  $(b)=$Direct method, $(c)=T$([\oiii])$-Z$, $(d)=$\NDOS, $(e)=${\it
    N2S2}.
\end{list}
\end{table*}

\section{Discovery of extremely metal-poor EELGs}
\label{sect:discussion1}

Extremely metal-poor (XMP) galaxies are the least evolved systems in 
the Universe and, therefore, they provide a unique environment in which to
study the first stages of galaxy evolution and
chemical enrichment \citep{KunthOstlin2000}. 
{However, XMPs are extremely scarce \citep[$\sim$0.01\% of galaxies in the 
Local Universe; ][]{Morales-Luis2011} and only a handful of 
{\it \emph{bona fide}} XMPs have been discovered so far at $z>0.3$  
\citep[e.g.,][]{Hu2009,Ly2014,Maseda2014,Amorin2014a}}.  

\begin{figure}[t!]
   \centering
   \includegraphics[angle=0,width=8.5cm]{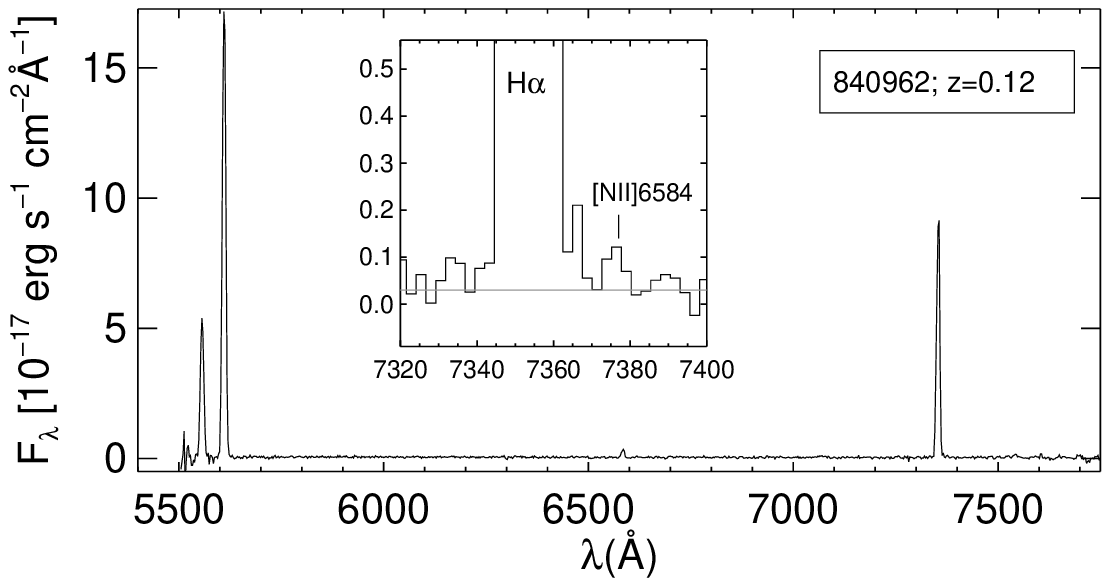}\\\vspace{5mm}
    \includegraphics[angle=0,width=8.5cm]{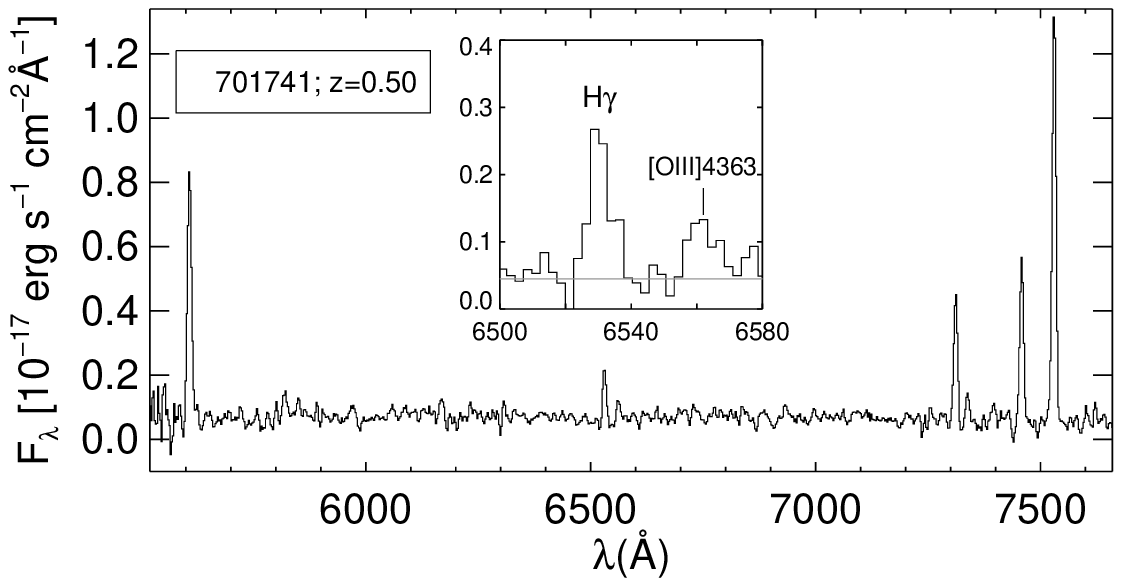}
 \caption{VIMOS spectra of the very metal-poor EELGs zCOSMOS\,ID\#840962
   and ID\#701741 at $z=0.12$ and $z=0.50$, respectively. 
The spectra have been smoothed by a two-{pixel boxcar} filter. 
The insets show a close-up view of the \ha$+$[\nii] and \hg$+$[\oiii]$\lambda$4363 lines. 
              }
         \label{spectrum_xmp}
   \end{figure}

In {the subset} of 149 EELGs with reliable metallicities we 
find six objects ($\sim$4\%) with metallicities {below} 
the limit for XMPs \citep[$\sim 1/10 Z_{\odot}$, e.g.,][]{Kniazev2004,Ekta2010}.
{We show two examples in Figure~\ref{spectrum_xmp} and summarize 
their main properties in Table~\ref{Tab3}.}
{In three EELGs, zCOSMOS\,ID\#836108, ID\#701741, 
and ID\#825959, at $z \ga$\,0.3 the metallicity 
has been derived using the electron temperature. 
However, we are cautious about the metallicity of zCOSMOS\,ID\#836108 because 
it was derived using the $T$([\oiii])$-Z$ calibration, which at very low 
metallicities may underestimate the true metallicity (see Fig.~\ref{Te-Z}). 
The remaining three EELGs have $z \la 0.3$ and their metallicities 
have been derived using the \NDOS\ parameter. 
One of these, zCOSMOS\,ID\#840952, is the most metal-poor galaxy in our sample ($Z\sim$\,0.04$Z_{\odot}$), and is comparable to the most metal-poor 
H{\sc II} galaxies known \citep[e.g., I\,Zw\,18 and SBS\,0335-052;][]{Papaderos2006,Izotov2009}}. 
{Its [\nii] line is extremely weak (S/N$\sim$\,2.5; 
Fig.~\ref{spectrum_xmp}), so its flux might be considered as an upper limit. }
{For these three EELGs we have derived the nitrogen-to-oxygen ratio 
(N/O) using the {\it N2S2} calibration \citep{Perez-Montero2009}. 
Their very low nitrogen abundance, $\log$(N/O)$\la-1.7$, is typical 
of chemically unevolved systems, where the nitrogen is still 
primarily produced by massive stars \citep[e.g.,][]{Molla2006}.  } 
{Deeper, high S/N spectroscopy covering the entire spectral 
range should provide a definitive confirmation of the extremely low 
oxygen abundance for these EELGs.}

An intriguing aspect of XMPs is that over 60\% of them\ turn out to have cometary/tadpole morphologies \citep[][]{Papaderos2008,SanchezAlmeida2013,Filho2013}. 
While cometary/tadpole morphologies are rather common at high redshift 
\citep[$\sim$6-10\% of all galaxies in the Hubble Ultra Deep Field,][]{Straughn2006,Elmegreen2010}, this percentage  decreases at 
lower redshifts \citep[$<$1\% in the local Universe;][]{Elmegreen2012}. 
{Here we find a large number of tadpoles among EELGs and 
at least half of the most metal-poor EELGs are indeed tadpoles. 
This can provide additional clues about their nature.  \citet{SanchezAlmeida2013}  
studied their 
morphological and dynamical properties and suggested that XMPs with tadpole morphologies are in early stages of 
their disk assembly. In this scenario, a massive accretion of 
external metal-poor gas feeds the starburst, leading to large inhomogeneities or gradients 
of metallicity from head to tail \citep{SanchezAlmeida2014}, 
closely resembling  recent findings at higher redshift \citep[e.g.,][]{Cresci2010,Queyrel2012,Troncoso2014}. }
{Future studies of very metal-poor EELGs using high-quality 3D 
spectroscopy will be used to test this scenario.}

\section{Lyman-alpha emission in EELGs}
\label{sect:discussion2}

High-redshift star-forming galaxies are generally recognized in 
surveys by their high UV luminosity and/or by their strong 
Ly$\alpha$ emission, with equivalent widths 
EW(Ly$\alpha$)$\geq$20 \citep[e.g.,][]{Shapley2003,Mallery2012}. 
Although Ly$\alpha$ selection may  systematically trace different
galaxies at different redshifts \citep{Nilsson2011} and a small
fraction of Ly$\alpha$ emitters (LAEs) may be evolved galaxies
\citep{Pentericci2009}, most of them  typically show
low metallicity, blue colors, small sizes, and low dust attenuation,
{indicating} an early stage of galaxy formation 
\citep[e.g.,][]{Pirzkal2007,Cassata2011,Cowie2011,Finkelstein2011}. 
{A significant fraction of their low redshift ($z\sim$\,0.3) 
analogues are also found to be EELGs \citep[e.g.,][]{Cowie2011}.} 

\begin{figure*}[t!]
   \centering
   \includegraphics[angle=0,width=16.cm]{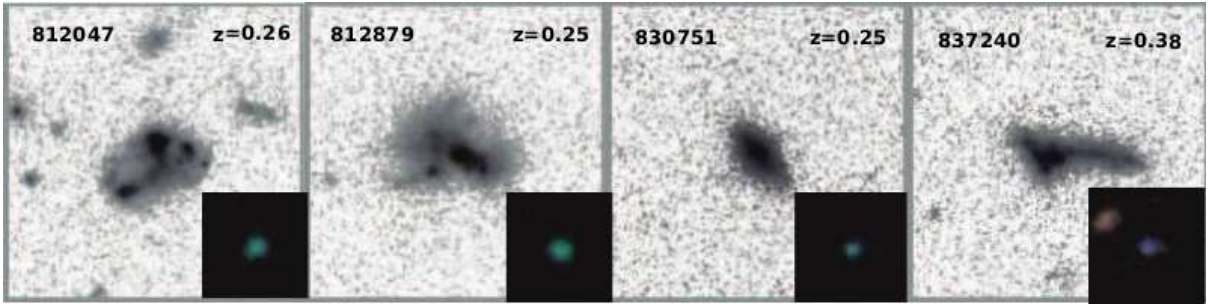}
     \caption{HST/ACS $I-$band images of EELGs with {detected Ly$\alpha$ emission. 
     The insets show {\textit{ugriz} color-composite} SDSS-$DR9$ 
     postage-stamps ($FWHM \sim$\,1$''$) for the same galaxies. 
     Each {ACS and SDSS cutouts have 6$''$ and 30$''$ on a side, respectively}. 
     Labels indicate zCOSMOS ID number and redshift. }
                   }
         \label{LAEs}
   \end{figure*}

{Although} they were not selected for their UV properties, our 
sample EELGs are very luminous in the UV continuum, so it is interesting 
to {investigate} whether some of these galaxies have been identified as 
LAEs in the literature. 
{To this end}, we cross-correlate our sample with GALEX grism 
spectroscopy surveys. {We find} that only four zCOSMOS EELGs at 
$z=0.25-0.38$ have been observed so far, {and they are included 
in the catalog of low$-z$ GALEX LAEs of \citet{Cowie2010}. 
We identify and show these four EELGs in Figure~\ref{LAEs}.}
Remarkably, \textit{\emph{all}} of them show prominent Ly$\alpha$ emission 
lines, with luminosities $\log$(L$_{Ly\alpha}$)$=$41.8-42.4 erg s$^{-1}$, 
and high equivalent widths of EW(Ly$\alpha$)$=$22-45\AA\ {(rest-frame)}. 

The observed Ly$\alpha$/H$\alpha$ ratios of these EELGs are between 0.5-2, 
well below the Case B recombination value, even after correction {for} dust extinction. 
Comparing their Ly$\alpha$ and H$\alpha$ equivalent widths 
(EW(H$\alpha$)$=$320-580\AA), these galaxies are among those
with larger EW(H$\alpha$) low$-z$ LAEs in the catalog {of} \citet{Cowie2010}. 
Their EW(H$\alpha$)/EW(Ly$\alpha$) ratios ($\sim$14) are instead comparable 
with some high-excitation LAEs at higher redshift
\citep[e.g., $z\sim$2][]{Finkelstein2011,Nakajima2013}. Compared with
model predictions \citep[e.g.,][]{Schaerer2003}, these
EW(H$\alpha$)/EW(Ly$\alpha$) ratios are in good agreement with 
tracks for instantaneous burst models with young ages ($\sim$10$^{7}$yr) 
and low metallicities \citep[see][their Figure\,10]{Nakajima2013}, probably 
superposed with a more constant (e.g., exponentially declining) underlying 
star formation history \citep[e.g.,][]{Amorin2012a}. 

As shown in Figure~\ref{LAEs}, zCOSMOS EELGs with Ly$\alpha$ 
emission display {a variety of} morphologies in HST-ACS imaging, 
{while in SDSS images they appear nearly unresolved.}  
{Although considered} as {\it green pea} galaxies owing to their 
large EW([\oiii]) and green colors {in the SDSS thumbnails}, these 
galaxies were not included in the {\it green pea} sample {of} 
\citet{Cardamone2009} {because of their low luminosity, which is 
$\ga$\,2 mag} fainter than the SDSS spectroscopic limits. 
The connection between LAEs, {\it green peas}, and our sample of EELGs
is not entirely surprising. 
{\citet{Cowie2011} have shown that $\sim$75\%  of 
low-$z$ LAEs have EW(H$\alpha$)$>$100\AA, while only $\sim$30\% of 
UV-continuum selected galaxies with EW(H$\alpha$)$>$100\AA\ are LAEs. 
Moreover, recent studies have found evidence for high-ionization
state and low metallicities in LAEs out to $z\ga2$ \citep[][see also
Cassata et al. 2012, for HeII\,$\lambda$1640 detection 
in LAEs]{Xia2012,Nakajima2013}. These two properties are 
an imprint of {\it green peas} \citep{Amorin2012a,Jaskot2013} 
and EELGs in general (see  Figure~\ref{histograms}).
Thus, LAEs could be ubiquitous among  
low-mass galaxies selected for their unusually large equivalent 
widths. }

%
\section{Comparison with other EELG studies}
\label{sect:discussion3}

{Star-forming galaxies with very high [\oiii] EWs have been identified at lower and higher redshifts in previous studies, as we mentioned in our introduction. 
At lower redshift luminous H{\sc ii} galaxies and \textit{green peas} \citep{Cardamone2009,Amorin2010,Amorin2012a} show very similar properties to our EELGs, e.g.,  very large [\oiii] EWs up to $\sim$\,1500\AA, low-metallicity, extreme compactness, and high sSFR. 
At low to intermediate redshift ($z\la$\,1), samples of EELGs selected from narrowband imaging \citep[e.g.,][]{Kakazu2007,Hu2009} and from very deep spectroscopic surveys for their strong [\oiii] lines \citep[e.g.,][]{Hoyos2005,Ly2014,Amorin2014a} also show strong similarities to our EELG sample. 
Similar conclusions can be obtained by comparing the properties of zCOSMOS EELGs with 
EELGs at higher redshift (i.e., $z>$\,1), which are typically selected by their unusually strong [\oiii] EWs in low-resolution HST NIR spectroscopy \citep[e.g.,][]{Atek2011,Atek2014,vanderWel2011,Xia2012,Maseda2014,Masters2014}. 
As many of these studies have shown, the overall morphology, size, stellar mass, sSFR, and metallicity properties suggest that EELGs are distributed in the same parameter space.  }

\begin{figure*}[t!]
   \centering
  \includegraphics[angle=0,width=13cm]{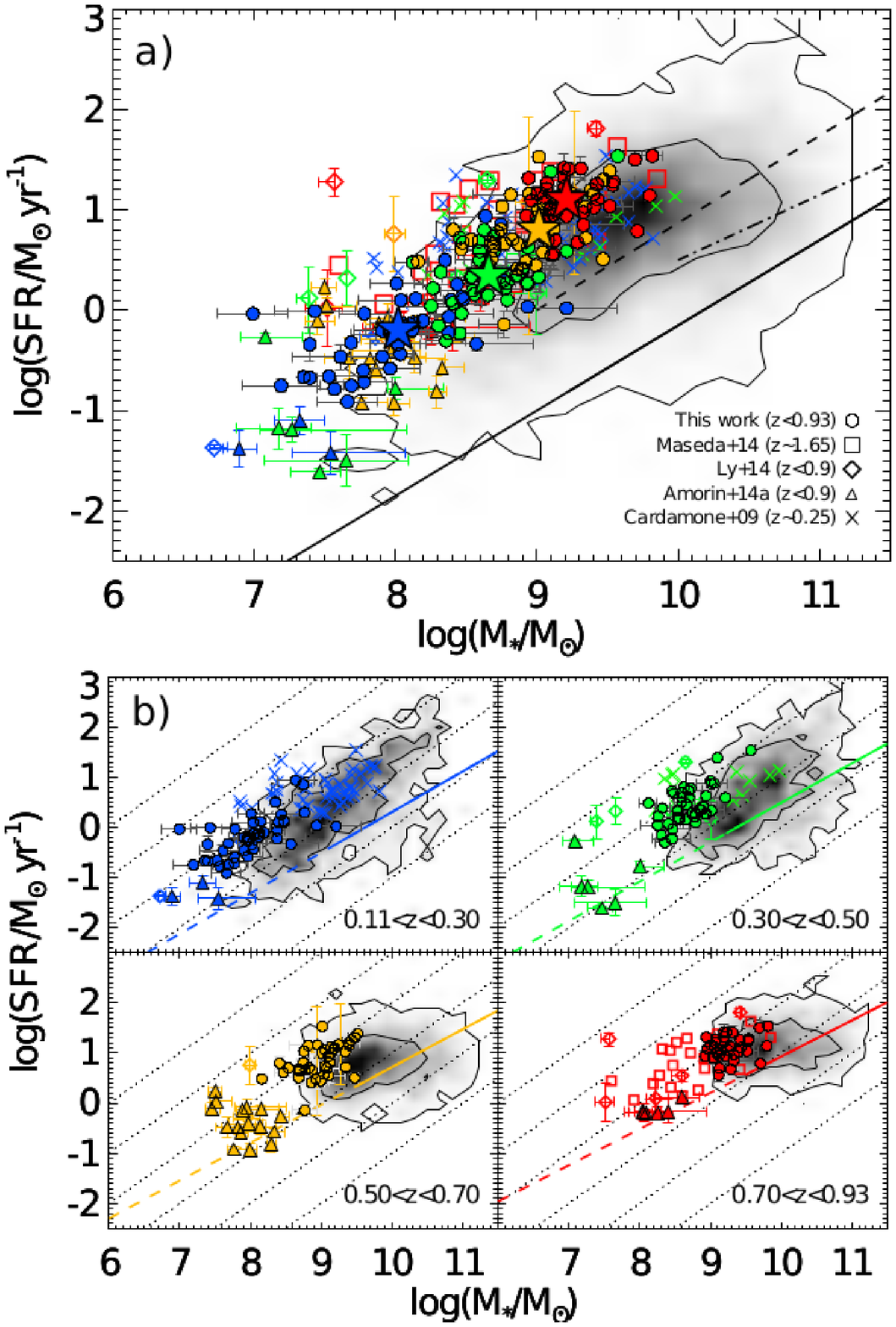} 
    \caption{{($a$) The location of EELGs and SFGs in the SFR-M$_*$ plane. 
    The gray density contours show the position of the SFG-20k sample. 
    The inner and outer contours enclose  68\% and 99\% of the sample, respectively. 
    Big stars show the median values for EELGs (see Table~\ref{Tab2}). 
    Solid, dot-dashed, and dashed lines show the  main sequence (MS) 
    of galaxies at $z=0$ \citep{Whitaker2012}, $z\sim 0.7$  \citep{Noeske2007}, 
    and $z\sim 1$  \citep{Elbaz2007}, respectively. 
    ($b$) Same as ($a$) but divided into four redshift bins. Solid and dashed 
    colored lines show the MS of galaxies at each redshift and their extrapolation 
    to the low-mass regime, respectively, according to \citet{Whitaker2012}.
    Dotted lines indicate constant sSFR from 10$^{-11}$\,yr$^{-1}$ 
    (\textit{bottom left}) to 10$^{-6}$\,yr$^{-1}$ (\textit{upper right}). 
    At all redshifts the EELGs follow nearly the same relation, which is offset 
    by $\gtrsim$\,1 dex from the local MS}.   }
  \label{M-SFR}
   \end{figure*}

{Scaling relations are useful tools for comparing the observed properties of galaxies and the predictions of models. A thorough analysis of different scaling relations including  size, mass, metallicity, and SFR, will be the subject of the second paper of this series. However, we anticipate in Fig.~\ref{M-SFR} the relation between SFR and stellar mass 
for the EELGs and SFGs in zCOSMOS, comparing them to other EELGs from the literature. 
In the SFR-$M_{\star}$ plane, nearly all EELGs follow a well-defined relation. 
However, this trend is  $\sim$\,1 dex above the extrapolation to low stellar mass of 
the main sequence followed by normal SFGs at a 
given redshift \citep[e.g.,][]{Noeske2007,Elbaz2007,Whitaker2012}. 
This means that EELGs have enhanced sSFR at a given stellar mass compared to typical SFGs. 
Their values, in the range $\sim$\,10$^{-9}$--10$^{-7}$\,yr$^{-1}$, imply short stellar mass doubling times $<$\,1 Gyr, which clearly suggest that EELGs are forming stars in strong bursts.} 

{This result is in excellent agreement with similar studies, which have also 
shown that most EELGs are typically more metal-poor than predicted by the 
mass-metallicity relation at a given redshift \citep[e.g.,][]{Amorin2014a,Ly2014}. 
One possible interpretation for the offset position of EELGs in scaling relations 
involving mass, metallicity, and SFR is that strong gas inflows (e.g., due to 
interactions or mergers) and outflows (e.g., due to SNe feedback) may play a 
significant role in regulating their chemical abundances and mass growth 
\citep[e.g.,][]{Amorin2010,Xia2012}.  
Finally, another common point among EELGs at low and high redshift is the high-ionization state of their ISM. In our sample we find a common range of high-ionization parameters (as measured by the [\oiii]/[\oii] ratio, see Fig.~\ref{Te-Z}) for EELGs, GPs,   
and local H{\sc ii} galaxies. Recently \citet{Nakajima2014} show that, for a given stellar mass and SFR, the ionization of GPs and other low-$z$ EELGs is much higher than in typical 
SDSS SFGs, being only comparable to the ionization found in some high-$z$ LBGs and LAEs.}    

{Overall, in terms of scaling relations involving size, stellar mass, SFR, metallicity, and also the ionization parameter, EELGs seem to be a relatively homogeneous class 
regardless of redshift \citep[e.g.,][]{Xia2012,Ly2014,Nakajima2014,Maseda2014,Amorin2014a}.    
If true,  all these low-mass galaxies are probably being caught in a similar, transient 
and extreme stage of their formation history, where they are efficiently building up a
significant fraction of their present-day stellar mass in a young, galaxy-wide starburst. }

{Nevertheless, a more detailed and quantitative comparison of the galaxy-averaged 
EELG properties and number densities through cosmic time is needed to reach firm 
conclusions on the physical properties behind the strong star-formation activity at different cosmic epochs. 
In particular, this comparison should be made using complete samples studied with 
homogeneous  methodologies and high-quality datasets over a wide redshift range. 
Studies like these, which are currently ongoing, will strongly benefit from the statistical value and wealth of data products provided in this study. 
}

\section{Summary and conclusions}
\label{sect:conclusions}

{Using the zCOSMOS 20k bright survey we have selected a large sample 
of 183 extreme emission-line galaxies (EELGs) at $0.11 \leq z \leq 0.93$ 
showing unusually high [\oiii]$\lambda$5007 rest-frame equivalent widths (EW([\oiii])$\geq$100\AA). 
We have used zCOSMOS optical spectroscopy and multiwavelength COSMOS photometry 
and HST-ACS {\it I}-band imaging to characterize the main properties of EELGs, 
such as sizes, stellar masses, SFR, and metallicity, as well as morphology 
and large-scale environment. }
We summarize our main findings as follows:


\begin{enumerate}
\item {The adopted} selection criterion based on EW([\oiii]) 
{lead to a sample of galaxies with the highest EWs in all 
the observed strong emission lines, e.g., H$\beta$ ($\ga$\,20\AA) and H$\alpha$ ($\ga$\,100\AA), suggesting galaxies dominated by young ($\la$\,10 Myr) star-forming regions}. {The EELGs constitute} 3.4\% of {SFGs} in our parent 
zCOSMOS sample. 
Using emission-line diagnostic diagrams we divided the 
sample {into} 165 purely star-forming galaxies plus 18 NL-AGN candidates ($\sim$\,10\%). {Only four of them are detected as bright X-ray sources.}

\item EELGs form the low-end of stellar mass and {the} high-end of sSFR distributions of SFGs in zCOSMOS up to $z\sim$1. 
Stellar masses of EELGs, as derived from multiband SED fitting, lie in the range
$7 \la \log$\,(M$_*$/M$_{\odot}) \la 10$. {Our sample, however, 
is not complete in mass below $\sim$\,10$^9$\,M$_{\odot}$ in the considered 
redshift range.} 
Star formation rates from both \ha\ and FUV luminosities after
corrections for dust attenuation and extinction are consistent with
each other and range 0.1\,$\la$\,SFR\,$\la$\,35\,M${_\odot}$\,yr$^{-1}$ 
(Chabrier IMF). 
Both quantities increase similarly with redshift, so this results {in} a
tight range of {\it specific} SFRs (median
sSFR$=$\,10$^{-8.18}$\,yr$^{-1}$) and stellar mass doubling times 
0.01\,Gyr$<$M$_{*}/$SFR$<$1\,Gyr. 

\item EELGs are characterized by their low metallicities,
  7.3\,$\la$\,12$+\log$(O/H)\,$\la$\,8.5 
(0.05-0.6\,$Z_{\odot}$), as derived using both the direct measurements 
based on electron temperature ($t_{\rm e}$) and strong-line methods 
calibrated consistently with galaxies with $t_{\rm e}$ measurements. 
Therefore, the chemical abundances of EELGs at $0.11 \leq z \leq 0.93$ 
are very similar to those of nearby H{\sc II} galaxies and BCDs. 
{We find six ($\sim$4\%) extremely metal-poor ($Z
  <$\,0.1\,$Z_{\odot}$) galaxies in our sample. }

\item EELGs are moderately low-dust, very compact UV-luminous galaxies, as
  evidenced by their typically blue colors ($\beta \sim$\,$-1.6$), high
  FUV luminosities ($L_{\rm FUV}$\,$\sim$\,10$^{10.4} L_{\odot}$) and
  high surface brightnesses $\mu_{\rm  FUV}$\,$\ga$\,10$^{9} L_{\odot}$\,kpc$^{-2}$. 
We find only four EELGs with GALEX-UV spectroscopic
observations. All these galaxies are strong Ly$\alpha$
  emitters, with large equivalent widths and luminosities in the
ranges EW(Ly$\alpha$)$=$22-45\AA\ and $\log(L_{\rm Ly\alpha})$\,$=$\,41.8-42.4 
erg s$^{-1}$, respectively. 

\item Using HST-ACS $I-$band COSMOS images we classify star-forming
  EELGs {into} four morphological classes according to the distribution
  and shape of their high- and low-surface-brightness components 
(i.e., SF knots and {underlying} galaxy, respectively).  
{We} show that 18\% have {round/nucleated} morphologies, {most of which 
are barely resolved, {while} the remaining 82\% have 
irregular morphologies. These irregular morphologies are visually classified 
as} {\it clumpy/chain} (37\%), {\it cometary/tadpole} (16\%), and 
{\it merger/interacting} (29\%). 
Therefore, we conclude that at least $\sim$80\% of the EELG sample 
shows non-axisymmetric morphologies. 
{Using quantitative morphological parameters} we find that 
EELGs show smaller half-light radii ($r_{50} \sim$1.3\,kpc in 
the median) and larger concentration, asymmetry, and Gini parameters 
than other SFGs in zCOSMOS,  most of them being classified as {\it irregular} 
galaxies {by} automated {algorithms}. 
{Among the defined morphological classes we do not find 
any significant difference in the redshift distribution or physical properties. } 

\item {As star-forming dwarfs in the Local Universe, EELGs are usually 
found in relative isolation. 
While only very few EELGs belong to compact groups, almost one third of 
them are found in spectroscopically confirmed {loose} pairs or triplets. 
Comparing isolated and grouped EELGs we do not find any significant differences in the 
redshift distributions or physical properties.  }
\end{enumerate}

In conclusion, we have shown that galaxies selected by their extreme
strength of optical emission lines led us to a homogeneous,
representative sample of compact, low-mass, low-metallicity, vigorously 
star-forming systems identifiable with luminous, higher-$z$ versions of 
nearby H{\sc II} galaxies and blue compact dwarfs. 
The extreme properties of some of these rare systems closely resemble 
those of luminous compact galaxies, such as the {\it green peas}  
\citep{Cardamone2009,Amorin2010} and other samples of emission 
line galaxies with very high equivalent widths recently found 
at similar and higher redshift \citep[e.g.,][]{Hoyos2005,
Kakazu2007,Salzer2009,Izotov2011,Atek2011,vanderWel2011,vanderWel2013,Xia2012,
Shim2013,Henry2013,Ly2014,Maseda2014,Amorin2014b}. 
The EELGs are galaxies likely caught in a transient and early period
of their evolution, where they are efficiently building up a
significant fraction of their present-day stellar mass in a young,
galaxy-wide starburst. 
Therefore, they constitute an ideal benchmark for comparative 
studies with samples of high redshift Ly$\alpha$ emitters and
Lyman-break galaxies of similar mass and high-ionization state. 

\begin{acknowledgements}
We thank the referee for her/his deep and thorough reports  
which significantly contributed to improving this manuscript. 
We also gratefully acknowledge Polychronis Papaderos, Marco Castellano, 
Veronica Sommariva, Jorge S\'anchez Almeida and Casiana Mu\~noz-Tu\~n\'on 
for a careful reading of the paper and the helpful comments and suggestions
provided. We also thank M. Maseda and A. van der Wel for kindly providing us 
assistance with their data. 
This work was partially funded by the Spanish MICINN under the
Consolider-Ingenio 2010 Program grant CSD2006-00070: First Science
with the GTC\,\footnote{\tt http://www.iac.es/consolider-ingenio-gtc},
and by projects AYA2007-67965-C03-02 and AYA2010-21887-C04-01 of the
Spanish National Plan for Astronomy and Astrophysics, and by the
project TIC114  {\em Galaxias y Cosmolog\'\i a} of the  Junta de
Andaluc\'\i a (Spain). \\
R.A. acknowledges the contribution of the FP7 SPACE project “ASTRODEEP”
(Ref.No: 312725), supported by the European Commission. \\
This work has also been partially supported by the CNRS-INSU and its 
Programmes Nationaux de Galaxies et de Cosmologie (France).\\
The VLT-VIMOS observations have been carried out on guaranteed time (GTO) allocated by the European Southern Observatory (ESO) to the VIRMOS consortium, under a contractual agreement between the Centre National de la Recherche Scientifique of France, heading a consortium of French and Italian institutes, and ESO, to design, manufacture and test the VIMOS instrument.\\
Based on observations obtained with MegaPrime/MegaCam, a joint project of CFHT and CEA/DAPNIA, at the Canada-France-Hawaii Telescope (CFHT) which is operated by the National Research Council (NRC) of Canada, the Institut National des Science de l'Univers of the Centre National de la Recherche Scientifique (CNRS) of France and
the University of Hawaii. This work is based in part on data products produced at TERAPIX and the Canadian Astronomy Data Centre as part of the Canada-France-Hawaii Telescope Legacy Survey, a collaboration project of NRC and CNRS.
\end{acknowledgements} 

\bibliographystyle{aa}
\bibliography{22786_hk_RA.bbl}
\end{document}